\documentclass[a4paper,11pt]{article}
\pdfoutput=1 
\usepackage{jcappub} 
\usepackage[T1]{fontenc} 
\usepackage{graphicx}   
\usepackage{changepage}
\usepackage{subcaption}
\usepackage{orcidlink}
\usepackage{subcaption}
\usepackage{academicons}
\usepackage{natbib}
\usepackage{hyperref}
\usepackage{aas_macros}
\definecolor{orcidlogocol}{HTML}{A6CE39}

\title{\boldmath{A power spectrum approach to the search for Axion-like Particles from resolved galaxy clusters using CMB as a backlight}}

\author{Harsh Mehta\orcidlink{0009-0007-4664-4820},}
\author{and Suvodip Mukherjee\orcidlink{0000-0002-3373-5236}}

\affiliation{Department of Astronomy and Astrophysics, Tata Institute of Fundamental Research, Homi Bhabha Road, Mumbai, 400005, India}

\emailAdd{harsh.mehta@tifr.res.in}
\emailAdd{suvodip@tifr.res.in}

\begin{document}

\abstract{Axions or axion-like particles (ALPs) are hypothetical particles predicted by beyond standard model theories, which make one of the dark matter candidates. These particles can convert into photons and vice-versa in the presence of a magnetic field, with a probability decided by its coupling strength $\mathrm{g_{a\gamma}}$. One of the ways to detect these particles is by using th e Cosmic Microwave Background (CMB) as a backlight. 
As the CMB photons pass through a galaxy cluster, they can get converted into ALPs in the mass range $10^{-15}$ eV to $10^{-11}$ eV through resonant conversion in the presence of cluster magnetic fields. This leads to a polarized spectral distortion ($\alpha$-distortion) in the CMB as the photon polarization parallel to the magnetic field in the galaxy cluster is involved in the conversion. The fluctuations in the magnetic field and electron density in a galaxy cluster lead to spatially varying $\alpha$-distortion around the cluster, with a power spectrum that is different from the lensed E-mode and B-mode CMB polarization power spectrum for the standard model of cosmology. By measuring the difference in the polarization power spectrum around a galaxy cluster from the all-sky signal, one can find new $\alpha$-distortion in the sky. For the resolved galaxy clusters, if the redshift, electron density, and magnetic field profiles of the cluster can be constrained using optical, X-ray, and radio observations, one can measure the coupling strength $\mathrm{g_{a\gamma}}$ from the ALP power spectrum. The contamination from CMB and galactic foregrounds such as synchrotron and dust can be mitigated by using multiple frequency bands by leveraging on the difference in the spectral shape of the signal from foregrounds. Using the ILC technique to clean the foregrounds, we show that the new power spectrum-based approach of the resolved galaxy clusters from upcoming CMB experiments such as Simons Observatory and CMB-S4 can detect (or put constraints) on the ALP-photon coupling strength of $\mathrm{g_{a\gamma} < 5.2 \times 10^{-12} \, GeV^{-1}}$ and $\mathrm{g_{a\gamma} < 3.6 \times 10^{-12} \, GeV^{-1}}$ at 95\% C.I. respectively for ALPs of masses $10^{-13}$ eV or for smaller $\mathrm{g_{a\gamma}}$ for lighter ALP masses.}
\maketitle
\section{Introduction}
\label{sec:intro}
Axions or axion-like particles (ALPs) are hypothetical particles whose existence is predicted by beyond-Standard model theories. These are pseudo-Nambu Goldstone bosons that possess mass, making them
one of the dark matter candidates \cite{dine1983not,abbott1983cosmological,preskill1983cosmology,Ghosh:2022rta,1992SvJNP..55.1063B,khlopov1999nonlinear,sakharov1994nonhomogeneity,sakharov1996large}.
Their production in particle-based experiments requires high energies, thus probing them using the weak interactions they may be having with photons provides a way for us to detect them \cite{Carosi:2013rla,Mukherjee_2020,Ghosh:2023xhs}.

The Cosmic Microwave Background (CMB) is the nearly isotropic black-body radiation \citep{Fixsen_2009} surrounding us with tiny spatial anisotropies of one part in $10^{5}$ as shown by various space and ground-based experiments like the Wilkinson Microwave Anisotropy Probe (WMAP) \citep{Bennett_2013}, Planck \citep{2020}, Atacama Cosmology Telescope (ACT) \citep{Das_2011}, South Pole Telescope (SPT) \citep{benson2014spt}, etc. Along with temperature fluctuations, CMB also exhibits around $5$\% polarization measured using E-mode and B-mode polarization (or Stokes parameters Q and U) \citep{2020_planck,Austermann_2012}. These are generated during the epoch of decoupling of the CMB photons (known as primary anisotropies: such as the baryon acoustic oscillations, Sachs-Wolfe effect, etc.\citep{Dodelson:2003ft,Hu_2002}) and anisotropies after decoupling (known as secondary anisotropies such as lensing \cite{Smith_2007}, Sunyaev Zeldovich (SZ) effects \citep{1972CoASP...4..173S}, reionization \citep{adam2016planck}, etc.). The fluctuations in the CMB temperature and polarization field are primarily captured by the power spectrum in terms of angular frequency $\ell$ in the spherical harmonic basis. The power spectrum of the CMB is very well known from the temperature and polarization power spectra ($\mathrm{C_{\ell}^{TT},  C_{\ell}^{EE}}$), and the correlation between these spectra  ($\mathrm{C_{\ell}^{TE}}$) \cite{PhysRevD.104.022003}.  
The measurement of the temperature and polarization power spectrum has led to the inference of cosmological parameters and has played a key role in building the standard model of cosmology known so far. 

Along with inference of cosmological parameters, CMB photons can also probe signatures of physics beyond the standard model of particle physics, by exploring the signature of any spatial and/or spectral anisotropy on the CMB. One such signature is the interaction of axions (or axion-like particles) with the CMB photons in the presence of magnetic fields in astrophysical systems such as galaxy clusters \cite{osti_22525054,Mukherjee_2018,Mukherjee_2020,Mukherjee:2021jgv}. The CMB photons as they pass through a galaxy cluster, can undergo resonant conversion in the presence of a magnetic field leading to spatially fluctuating polarized spectral distortion signal around galaxy clusters, which we call the $\alpha$-distortion\cite{Mukherjee_2020}. This interaction does not require axions or ALPs to be dark matter constituents and depends on the mass and coupling strength of the ALPs with photons. The conversion is resonant if the effective mass of the photon is equal to that of the ALP being formed.  These conversions dominate and the probability of conversion depends on the electron density and magnetic field profiles of the cluster, its redshift, frequency of observation, and the strength of the ALP-photon
coupling, quantified by the coupling constant $\mathrm{g_{a\gamma}}$. The photon-ALP resonant conversion will polarize the CMB photons passing through galaxy clusters along their line of sight, leading to additional fluctuations in the CMB. This polarization will depend on the magnetic field direction at the resonant location. Also, the 2D-projected shape and strength of the signal in the sky will be different for different mass ALPs that may be forming in the clusters. Thus the polarization measurements of the CMB along the cluster line of sight may be used to probe ALPs of different masses using their weak coupling with photons.  The additional fluctuations of the ALP distortion signal generated in galaxy clusters will affect the CMB at small angular scales, which clusters occupy. These small angular scales correspond to high multipoles in the spherical harmonics domain.

In this paper, we explore how the measurement of the residual power spectrum of the polarization signal in multiple frequency bands of CMB can explore the $\alpha$-distortion. This signal in the polarization power spectrum around a galaxy cluster depends on the cluster redshift, electron density, and cluster magnetic field. Observational measurements of these quantities are possible for clusters that are resolved (jointly observed) in multi-wavelengths such as radio, microwave, optical, and X-ray observations. Optical surveys can be useful to identify the redshift of the galaxy cluster\cite{2014ApJS..210....9B}. The measurement of the Sunyaev-Zeldovich signal \cite{Birkinshaw_1999} from CMB temperature or the X-ray observation can probe the electron density, and radio observations can probe the magnetic field in galaxy clusters \cite{GOVONI_2004}.  A new analysis pipeline for multi-band analysis of galaxy clusters is under preparation \citep{Mehta:2024:new3}.

The ALP signal is contaminated by galactic foregrounds and even CMB, which can affect the constraints on the coupling constant. Thus a cleaning is required to isolate the ALP signal. This can be done using the distinct spectral and spatial variation of the ALP distortion spectrum compared to the spectrum of the contaminants. This is implemented using ILC \cite{Eriksen_2004,ilc2008internal} or via a template matching of foregrounds \cite{Hansen_2006}. We show that the experiments with improved angular resolution, low instrument noise, and operating in multiple bands such as the Simons Observatory (SO) \cite{Ade_2019} and CMB-S4 \cite{abazajian2016cmbs4} will be capable of probing the $\alpha$-distortion, by isolating the signal from the galactic foregrounds. 

The paper is organized as follows, the motivation for the power spectrum-based analysis is explained in Sec. \ref{sec:motive}, followed by a brief review of axions in Sec.\ref{sec:axions}. The resonant conversion of CMB photons to ALPs in galaxy clusters is described in Sec.\ref{sec:ALPs from CMB}, followed by the ALP distortion power spectrum and its sources of variation in Sec.\ref{sec:alpha_l}. The estimator for the ALP signal and the changes induced due to partial sky observation is explained in Sec.\ref{sec:estimator}. 
The foregrounds affecting the ALP signal and their cleaning using ILC are studied in Sec.\ref{sec:contaminants}.
Then we discuss the method for Bayesian estimation of the ALP coupling constant in Sec.\ref{sec: bayesian}. The results using a Bayesian framework are described in Sec.\ref{sec: results} and the conclusion is discussed in Sec.\ref{sec: conclusion}. Mostly, natural units ($\mathrm{\hbar = 1, c = 1, k_B = 1}$) have been used in this paper, unless explicitly stated. The cosmological parameters have been used from Planck 2015 results \cite{2016}.

\section{Motivation} \label{sec:motive}
 The angular power spectrum contains information about the variation of a sky-projected signal at different angular scales in the harmonic space. The use of harmonic space imparts a way of distinguishing various signals based on their expected spatial variation. 
The photon-ALP resonant conversion in galaxy clusters will generate additional fluctuations in the CMB at small angular scales, which clusters occupy. These small angular scales correspond to high multipoles (or angular frequencies "$\ell$") at which the ALP distortion power spectrum may become dominant, based on the photon-ALP coupling strength. This will result in additional fluctuations along the cluster line of sight with the ALP power spectrum scaling as $\ell^2$. Also, the ALP power spectrum increases with the frequency of observation in the microwave and radio spectra following a dependence that goes as $\mathrm{\propto \nu^2 I_{cmb}(\nu)^2}$, where $\mathrm{I_{cmb}(\nu)}$ is the Planck black-body intensity. This spatial and spectral dependence marks the signature for the ALP distortion signal. 

\begin{figure}[h!]
     \centering
\includegraphics[height=8cm,width=15cm]{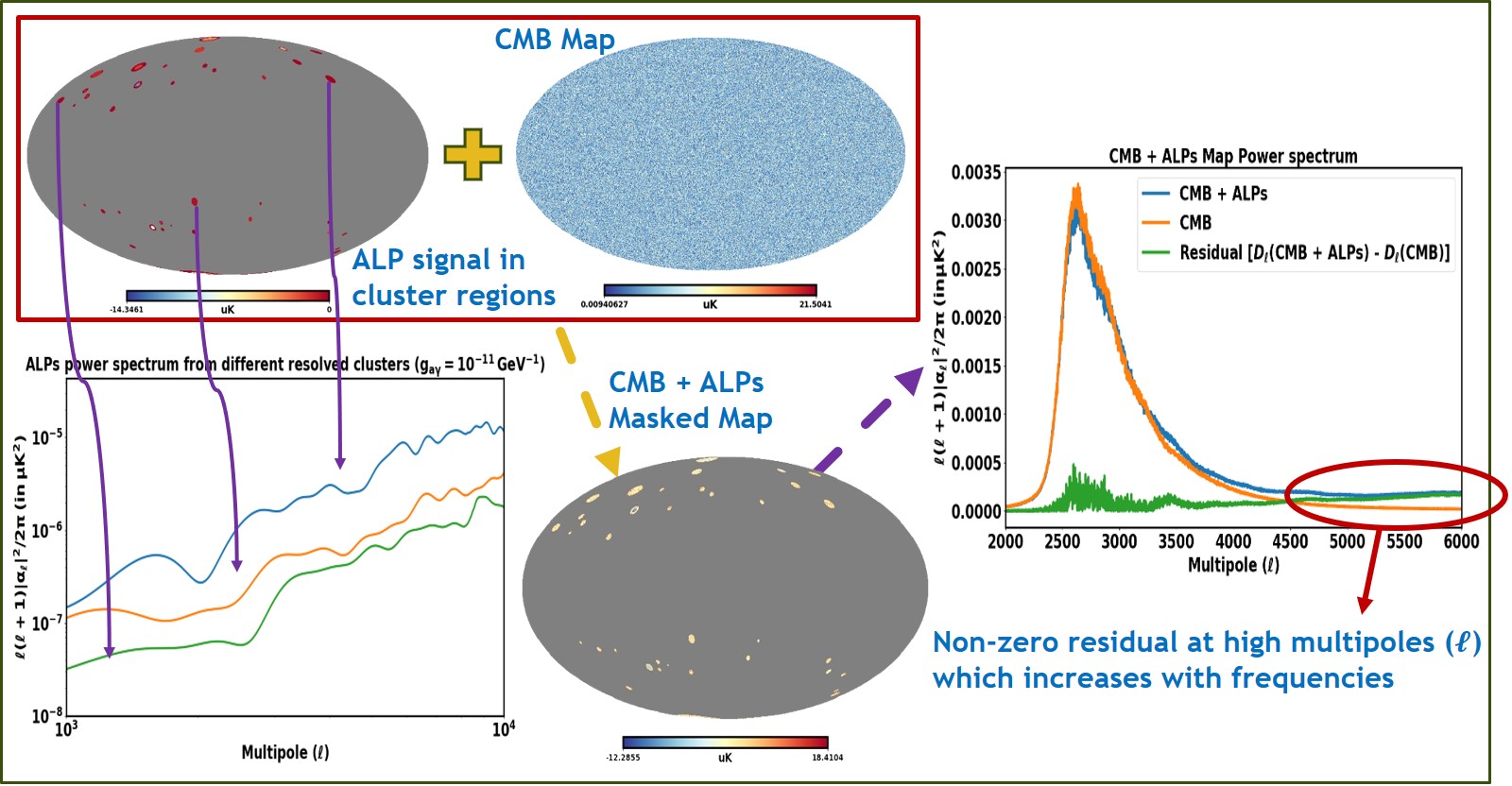}
    \caption{The following figure shows the contribution of ALP distortion signals ($\mathrm{g_{a\gamma} = 10^{-11} \, GeV^{-1}}$) from different galaxy clusters to the partial sky power spectrum from the cluster regions. The ALP power spectra from three of the cluster regions are shown in the figure. Such spectra from different galaxy cluster regions add up to give the observed ALP power spectrum from all cluster regions.  The non-zero residual, obtained by subtracting the theoretical or fiducial (non-ALPs case) CMB power spectrum in cluster regions from the observed power spectrum of these regions would increase with multipoles and frequencies, which will provide a signature for the existence of ALPs. }
       \label{fig:motivcmbax}
\end{figure}

The ALP distortion angular power spectrum generated in various clusters will result in fluctuations in the primordial CMB spectrum, which is well studied (refer to Fig.\ref{fig:motivcmbax}) using polarization and temperature spectra at various frequencies and angular scales. The ALP distortion spectrum from different clusters will contribute to the power spectrum in the cluster region. This spectrum can be probed using its dependence at small angular scales or high multipoles, at which the residual spectrum, obtained after subtracting the theoretical CMB power spectrum in partial sky cluster regions from the observed power spectrum of these regions, would show non-zero increasing values at high multipoles, which increase with frequencies of observation in microwave and radio spectra. Based on the covariance of the fiducial (no-ALPs case) power spectrum, we can set bounds on the contribution of the ALP distortion to the observed power spectrum. These bounds will be related to the strength of the photon-ALP coupling, enabling us to obtain constraints on the coupling constant $\mathrm{g_{a\gamma}}$.  

The ALP distortion is a weak signal and requires precise measurements across different EM bands. Experiments such as the Square Kilometer Array (SKA) \cite{Carilli_2004}, e-Rosita \cite{merloni2012erosita}, JWST \cite{Gardner_2006}, etc., will provide loads of data on galaxy clusters in different EM bands and mapping of their evolution to high redshifts would provide further insight into their properties. These observations can be combined to study the variations in the CMB due to the ALP distortion signal and to obtain bounds on ALP coupling constants from resolved clusters. 
 
Experiments such as COBE, WMAP, and Planck have provided substantial information about the CMB and other signals such as the Sunyaev Zeldovich effects (inverse Compton scattering in galaxy clusters) \cite{2014A&A...571A..21P}. The polarization spectrum of the CMB is not yet very well probed though and contains invaluable information about phenomena such as lensing, synchrotron emission, etc. With the upcoming high-resolution detectors such as Simons Observatory (SO) \cite{Ade_2019}, and CMB-S4 \cite{abazajian2016cmbs4}, the spectral distortions \cite{2014PTEP.2014fB107T} in the CMB have only recently attracted the attention of cosmologists, ever since constraints were put on the $\mu$ and $y$-type distortions by COBE at $2 \sigma$ level as: $|\mu| \leq 9 \times 10^{-5}$ and $|y| \leq 1.5 \times 10^{-5}$ \cite{Fixsen_1996}.

The spatial and spectral dependence of the ALP distortion spectrum can thus be compared to the deviations in the CMB power spectrum, and thus be used to detect or rule out ALPs, while constraining their coupling constants.
 The phenomenon does not require axions to be dark matter and acts as an independent way to search for one of the beyond standard model particles.
 
\section{Axion-Like Particles} \label{sec:axions}
Axions or ALPs (axion-like particles) are hypothetical beyond Standard model particles and one of the  dark matter candidates. Axion is a pseudo Nambu-Goldstone boson and is formed due to global symmetry breaking (the Peccei-Quinn chiral U(1) symmetry ) by the vacuum expectation value of a scalar field and solves the strong CP problem \cite{berezhiani1991cosmology}. Since the underlying symmetry is not exact at low energies, axions carry a small mass. Hence they are pseudoscalar bosons. 
The ALP-photon interaction lagrangian with axion field 'a' is given as \cite{Raffelt:1996wa}:

\begin{equation}\label{eq:lagr}
 \mathrm{ \mathcal{L}_{int} = -\frac{g_{a \gamma} F_{\mu \nu}\tilde{F}^{\mu \nu} a}{4} = g_{a \gamma} E \cdot B_{ext} a,}  
\end{equation}
here $\mathrm{F_{\mu \nu}}$ is electromagnetic field tensor and $\mathrm{\tilde{F}^{\mu \nu}}$ is the dual tensor. The variation of the coupling constant $\mathrm{g_{a\gamma}}$ for axion masses is uniquely determined by the model, but for ALPs, there can be a large range of values.
The $\mathrm{E} \cdot \mathrm{B}$ takes into account the interaction of the photon's electric field $\mathrm{E}$ component along the transverse magnetic field $\mathrm{B}$. There is no conversion in the presence of a longitudinal magnetic field. This leads to a polarized distortion in the photon as an ALP is produced. 

The photon-ALP conversion is one of the speculated  weak dark matter interactions that will cause a polarized spectral distortion in the CMB black-body. It has a weak coupling, with  bounds from COBE FIRAS placing the product of magnetic field and coupling constant at $\mathrm{ g_{a\gamma} B < 10^{-13} \, GeV^{-1} \, nG }$ \cite{Mirizzi_2009}. The CAST bound on the ALP coupling constant up to 95\% confidence interval (C.I.) is $\mathrm{\leq 6.6 \times 10^{-11} \, GeV^{-1}}$ \cite{2017}. The upper bound on the RMS fluctuation due to ALP distortions from Planck data is $18.5 \times 10^{-6}$ \cite{Mukherjee_2019}.

\section{{Axion detection using Cosmic Microwave Background as a backlight}} \label{sec:ALPs from CMB}
The CMB peaks in the microwave region of the electromagnetic spectrum at a frequency of about $160.2$ GHz and is also the most ideal black-body in the universe. The power associated is aptly described by the Planck black-body function:
\begin{equation}
\label{eq:cmb}
\mathrm{I_{cmb}(\nu) = \left( \frac{2h \nu^3}{c^2}  \right) \frac{1}{e^{h\nu /kT_{cmb}} - 1} .}
\end{equation}
There are spectral distortions in the CMB which refer to tiny departures of the CMB from the Planck black-body spectrum like the  $\mu$ and $y$ distortions caused due to the inefficiency of photon number and energy redistributing processes to maintain thermal equilibrium, owing to the Hubble expansion. Late-time spectral distortions include those caused due to the inverse-Compton effects in galaxy clusters, also called the Sunyaev-Zeldovich (SZ) effects \cite{Birkinshaw_1999,Staniszewski_2009}. 

If ALPs exist in the universe, the CMB photons will undergo conversion to ALPs in galaxy clusters. 
The photon-ALP conversion leads to a new type of spectral distortion ($\alpha$-distortion) in the CMB black-body, as the photon-ALP conversion is a frequency-dependent phenomenon. This distortion introduces a new kind of polarized fluctuations in the CMB. The resonant conversion takes place between photons and ALPs \cite{Mukherjee_2020}, which can be measured without an absolute temperature calibrator. 
The resonant conversion takes place when the mass of ALP being formed is equal to the effective mass of the photon given by 
\begin{equation}
\mathrm{  m_a = m_{\gamma} = \frac{\hbar \omega_p}{c^2} \approx \frac{\hbar}{c^2}\sqrt{n_e e^2 / m_e \epsilon_{0}} . }
\label{eq:resonance mass}
\end{equation}
Here $\mathrm{\omega_p}$ is the plasma frequency, that depends on the $\mathrm{n_e}$ electron density at the conversion location. Higher mass ALPs are formed in regions of high electron density and vice-versa, if the conversion is resonant.
There can be non-resonant conversions, but the resonant ones are dominating. The resonant conversions lead to polarized distortions in the CMB. For an incoming CMB photon, the polarization will be transverse to our line of sight. From the Lagrangian for the interaction in Eq.\ref{eq:lagr}, it is understood that the transverse magnetic field plays a role in these conversions as for an unpolarized CMB photon passing through a galaxy cluster, the polarization parallel to the transverse magnetic field is lost. This results in the photon getting polarized, with an ALP being formed, whose mass depends on the electron density at the conversion location. 
Based on observed cluster electron densities, we expect ALPs of masses $10^{-15} - 10^{-11}$ eV to be forming in them.

\begin{figure}[h!]
     \centering
\includegraphics[height=5cm,width=10cm]{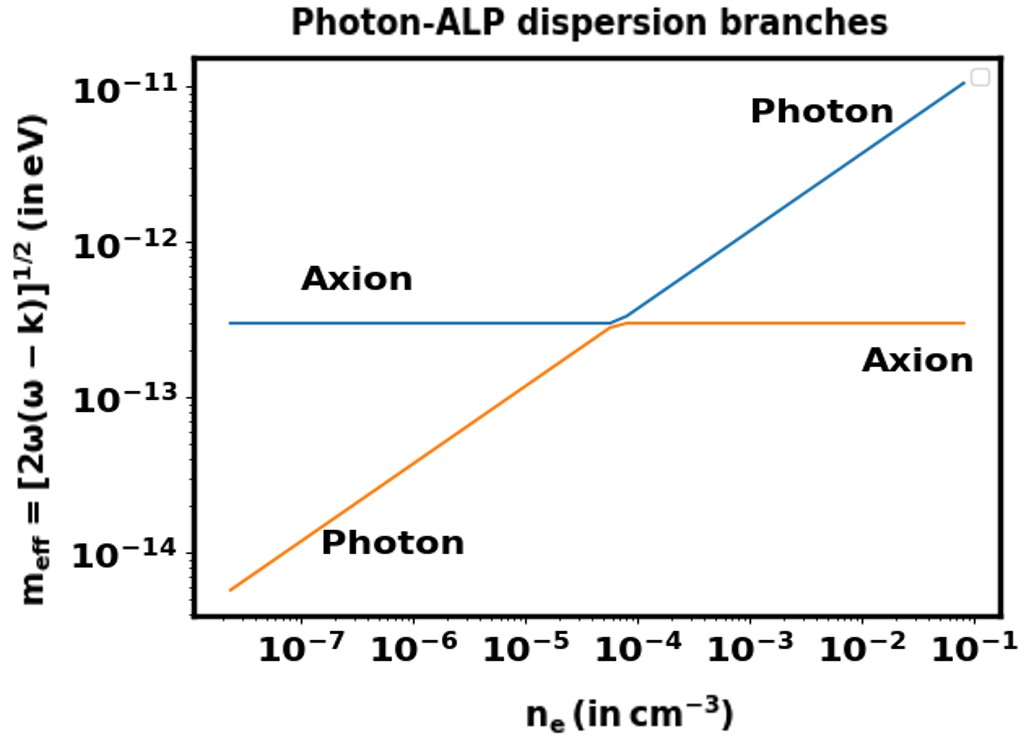}
    \caption{The dispersion branches for the photon-ALP resonant conversion (ALP mass = $3 \times 10^{-13}$ eV)}
       \label{fig:disp_branches}
\end{figure}

We neglect the Faraday rotation caused by the longitudinal magnetic field. This will cause a change in polarization but will be negligible in the range of microwave frequencies being considered. The conversion can be related to a branch shift (see Fig.\ref{fig:disp_branches}) in the dispersion relation of the photon-ALP conversion. Defining the following set of parameters: 
\begin{equation}
\label{eq:x}
\begin{split}
\mathrm{\Delta_a} &= \mathrm{- m_a^2 / 2\omega } \,,
\qquad
\mathrm{\Delta_e \approx \omega_p^2 / 2\omega} \,,
\\
\mathrm{\Delta_{a\gamma}} &= \mathrm{g_{a\gamma}B_{t} / 2 }\,,
\qquad
\mathrm{\Delta_{osc}^2 = (\Delta_a - \Delta_e)^2 + 4\Delta_{a\gamma}^2 },
\end{split}
\end{equation}
the dispersion relation can be written as 
\begin{equation}
\label{eq:disper}
\mathrm{m_{eff}^2 = \omega^2 - k^2 \approx 2\omega (\omega - k) = - \omega(\Delta_e + \Delta_a) \pm \omega \Delta_{osc} = \frac{m_a^2 + m_{\gamma}^2}{2} \pm \left[ \left( \frac{m_a^2 - m_{\gamma}^2}{2} \right)^2 + 
\omega^2 g_{\gamma a}^2 B_t^2  \right]^{1/2}}.
\end{equation}

From this dispersion relation, we get two branches for a particular mass ALP (see Fig.\ref{fig:disp_branches}). The probability of branch shift at resonance ($\mathrm{m_a = m_{\gamma}}$) is given using the adiabaticity parameter that compares the photon-ALP oscillation scale to the scale of variation of electron density :
\begin{equation}
\mathrm{\gamma_{ad} = \frac{\Delta_{osc}^2}{|\nabla \Delta_{e}|} = \left| \frac{2g_{\gamma a}^2 B_t^2 \nu (1 + z)}{\nabla \omega_p^2} \right|}. 
\label{eq:gamma}
\end{equation}
The gradient is with respect to the conversion length scale along the path that the CMB photon is treading. The probability of branch shift when passing from a high-density to low-density region or vice-versa is given as \cite{Mukherjee_2018}:
\begin{equation}
\mathrm{p = e^{-\pi \gamma_{ad}/2}}.    
\label{eq:prob}
\end{equation}

For adiabatic resonance ($\gamma >> 1$), the oscillation length scale is much smaller than the length scale over which electron density varies. Thus, the branch shift will not occur leading to a conversion. But if $\gamma << 1$, the resonance is non-adiabatic, and the probability for conversion is given as: $1 - $p .  
Based on the cluster profiles and frequencies we are considering, the resonance will be in the non-adiabatic regime. Thus, the ALP signal strength will depend on the conversion probability, if the resonance condition Eq. \ref{eq:resonance mass} is satisfied.

For ALPs of masses within a narrow range, the resonance will occur in a spherical shell within the cluster, if the cluster is spherically symmetric, with electron density decreasing away from the cluster center.
Thus, for an unpolarized CMB photon moving through such a cluster, there will be two resonances, once when entering the resonance shell and once when leaving it. The electron density change for the first resonance will be from low to high, while it will be from high to low for the second one. To obtain a polarized photon (which is the ALP distortion signal), the conversion has to take place at one of the two resonances \cite{Mukherjee_2020}. This gives the ultimate conversion probability  for the case of CMB photons passing through a galaxy cluster:
\begin{equation}
   \mathrm{ P(\gamma \rightarrow a) = 2p(1 - p) = 2e^{-\pi \gamma_{ad} / 2}(1 - e^{-\pi \gamma_{ad} / 2}) \approx \pi \gamma_{ad} .}
   \label{eq: Probab}
\end{equation}
This changes the CMB intensity along the cluster line of sight at a particular frequency as:

\begin{equation}
    \mathrm{
    \Delta I_{\nu} = P(\gamma \rightarrow a) I_{CMB,\nu}, \approx \pi \gamma_{ad} \left( \frac{2h \nu^3}{c^2}  \right) \frac{1}{e^{h\nu /kT_{cmb}} - 1} .}
    \label{eq:Distort}
\end{equation}

\section{The ALP distortion power spectrum  from a resolved cluster} \label{sec:alpha_l}
\begin{figure}[h!]
     \centering
\includegraphics[height=6cm,width=14cm]{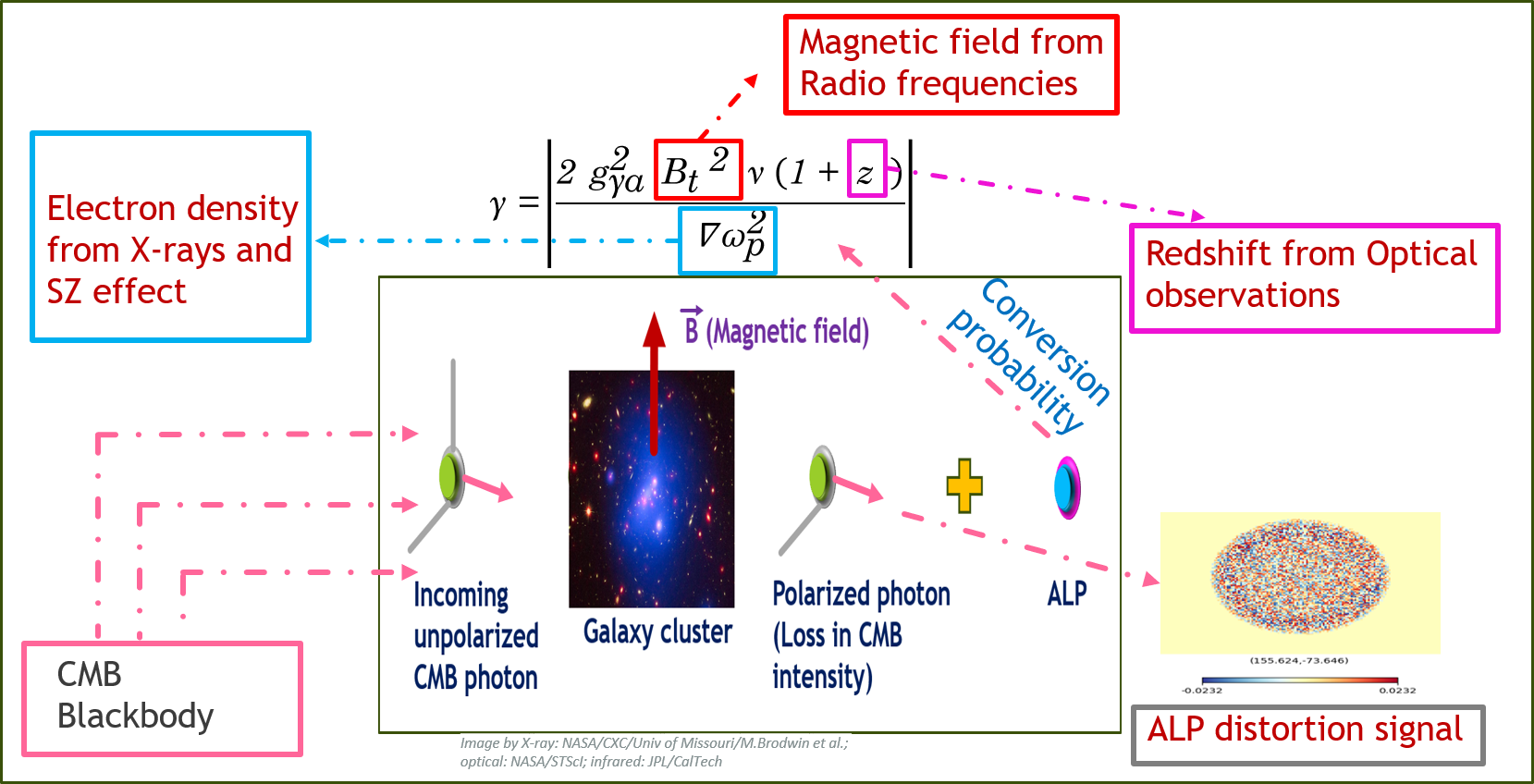}
    \caption{The following figure shows the generation of ALP distortion signal in the CMB black-body spectrum due to resonant conversion in galaxy clusters in the presence of the transverse magnetic field. The conversion probability for such a conversion depends on the magnetic field profile, electron density (via the plasma frequency), and redshift of the cluster. The magnetic field profile can be inferred from the emission by these clusters at radio frequencies. The electron density profile can be obtained using SZ observation or X-rays, while the redshifts from optical wavelengths. These can then be used to model the distortion caused in the CMB black-body due to photon-ALP resonant conversion.}
       \label{fig:resolvedclust}
\end{figure}

We consider those clusters as resolved clusters which can be observed at various wavelengths of the electromagnetic spectrum, namely in radio, microwave, optical and X-rays from which we can have a measurement of the cluster magnetic field, redshift and electron density. As shown in Fig. \ref{fig:resolvedclust}, observing the clusters at radio frequencies can provide the magnetic field profile of galaxy clusters using synchrotron emission, Faraday rotation, etc \cite{GOVONI_2004,carilli2002cluster,bonafede2010galaxy}. From the SZ effect in CMB temperature and X-ray emission in galaxy clusters, one can infer the electron density profile \cite{Birkinshaw_1999,cavaliere1978distribution,sarazin1986x}. Also, optical surveys can provide the redshifts of these clusters \cite{2014ApJS..210....9B,yee2001optical,zehavi2011galaxy,burenin2018optical}. Constraining the electron density, redshift and magnetic field profiles using multi-band observations will be discussed in a follow-up work \citep{Mehta:2024:new3}.
The resolved clusters generally have large radii ($\sim$ 3 to 5 Mpc) with masses of $\mathrm{10^{14} - 10^{15} \, M_{\odot}}$ and  particularly occupy low redshifts ($\mathrm{z \lesssim 1}$). 

The ALP distortion signal depends on the astrophysics of the galaxy clusters that affect its profiles.
We model the clusters as spherically symmetric, with the profiles depending only on the distance from the cluster center. 
We neglect any turbulence on the scales over which conversion is happening and assume smooth profiles which otherwise would lead to some fraction of depolarization of the ALP polarized distortion signal. This is valid when the coherence scale of the ordered magnetic field is larger than the scale over which resonant conversion is taking place (or the scale of the electron density variation). 
We use a modified $\beta$-model for electron density profile \citep{Vikhlinin_2006,mcdonald2013growth}:
\begin{equation}
\mathrm{  n_e^2 = Z\left[ n_0^2 \frac{(r/r_{c1})^{-\alpha}}{(1 + r^2/r_{c1}^2)^{3\beta - \alpha /2}}\frac{1}{(1 + r^{\gamma}/r_s^{\gamma})^{\epsilon / \gamma}} + \frac{n_{02}^2}{(1 + r^2/r_{c2}^2)^{3\beta_2}}\right].}
\label{eq:elec dens}\end{equation}
This model takes into account certain features of the observed cluster profiles like a cusp core following a power law, an inner core with high electron density, and the slope at a large distance from the center \cite{Vikhlinin_2006}. 

The magnetic field in galaxy clusters is difficult to model due to their coherence length being smaller than the beam size for most experiments. We consider this will not be the case with upcoming high-resolution detectors such as the SKA \cite{Carilli_2004}, and use a power law profile that fits well for clusters at low redshifts \cite{bonafede2010galaxy,bohringer2016cosmic,carilli2002cluster}:
\begin{equation}
    \mathrm{B(r) = B_0 r^{-s}}.
\label{eq:mag prof}
\end{equation}
We assume a random orientation of the magnetic field at various locations in the cluster. The significance of various parameters and their values are mentioned in Table \ref{tab:params}. 

\begin{table}[h!]
\caption{A list of parameters used in estimating the ALPs signal and their significance}
\label{tab:params}
\hspace{-2cm}
\begin{tabular}{|c|c|c|c|}

\hline
\textit{\textbf{Notation}} &
\textit{\textbf{Description}} &
\textit{\textbf{Typical Estimate}} &
\textit{\textbf{Range}} 

\tabularnewline \hline
s & Magnetic field steepness  & 0.5 & $\mathrm{0.5 < s < 2}$ \\
 \hline
 $\mathrm{B_0}$ & Order of magnetic field at 1 Mpc & 3 & $\mathrm{0.01 < B_0 < 0.5}$  \\ 
 \hline
$\mathrm{n_{02}}$ & Inner region electron density order & $\mathrm{10^{-1} \, cm^{-3}}$ & $\mathrm{5\times 10^{-2} \, cm^{-3} < n_{02}  < 1.5 \times 10^{-1} \, cm^{-3}}$  \\
 \hline
$\mathrm{n_{0}}$ & Outer region electron density order & $\mathrm{10^{-3} \, cm^{-3}}$ & $\mathrm{5\times10^{-4}\, cm^{-3} < n_{0}  < 1.5 \times 10^{-3} \, cm^{-3}}$  \\ 
 \hline
$\gamma$ & Transition width parameter & 3 & $ 2 < \gamma < 4$  \\
 \hline
$\alpha$ & Cusp slope parameter & 2 & $1 < \alpha < 3$ \\ 
 \hline
$\beta_1$ & Outer $ \beta$ exponent & 0.64 & $0.5 < \beta_1 < 0.8$  \\ 
\hline
$\beta_2$ & Inner $ \beta$ exponent & 1 & $0.8 < \beta_2 < 1.2$ \\ 
 \hline
$\mathrm{r_{s}}$ & Scaling radius & $\mathrm{1 \, Mpc}$ & $\mathrm{0.5 \, Mpc < r_s < 1.5 \, Mpc}$  \\ 
 \hline
$\mathrm{r_{c1}}$ & Outer core radius & $\mathrm{0.1 \, Mpc}$ & $\mathrm{0.05 \, Mpc < r_{c1} < 0.7 \, Mpc}$  \\ 
 \hline
$\mathrm{r_{c2}}$ & Inner core radius & 
  $\mathrm{0.01 \, Mpc}$ & $\mathrm{0.008 \, Mpc < r_{c2} < 0.05 \, Mpc }$ \\ 
 \hline
$\epsilon$ & Knee slope & 4 & $2 < \epsilon < 5$ \\ 
 \hline
 $\mathrm{Z}$ & Metallicity & 1.12 & $ 0.8 < \mathrm{Z} < 1.5$\\ 
 \hline
$\mathrm{g_{a\gamma}}$ & ALP coupling constant & $ < \mathrm{10^{-11} \, GeV^{-1}}$ & $\mathrm{10^{-14}\, GeV^{-1} < \mathrm{g_{a\gamma}} < 10^{-11} \, GeV^{-1}}$\\ [1ex] 
 \hline
\end{tabular} 
\end{table}
ALPs that are produced from resonant conversion have masses that depend on the electron density at that location. Thus, in the case of spherically symmetric clusters, ALPs of a particular mass range will be produced in a spherical shell within the cluster, if the resonant condition is met $\mathrm{(m_a = m_{\gamma}})$. With a higher electron density near the cluster center, higher mass ALPs are produced near the center in a smaller spherical shell as compared to low mass ALPs, which form away from the center in regions of low electron density. These shells appear as disks in the sky with smaller disks for high mass ALPs and larger disks corresponding to low mass ALPs. For resolved clusters, these signal disks can be probed to obtain coupling constants for ALPs of a particular mass range. However, in reality, clusters are not spherically symmetric. So, the signal will exhibit some distorted disc-like shape in the galaxy clusters. 

For our analysis, if not mentioned otherwise, we assume that ALPs of mass $10^{-13}$ eV with 30\% tolerance are being produced in the galaxy clusters if the resonant condition is satisfied ($\mathrm{m_a = m_{\gamma}} $). The coupling constant has been assumed to be constant for all ALP masses in this range: $\mathrm{g_{a\gamma} = 10^{-12} \, GeV^{-1}}$. 

We consider a galaxy cluster at a redshift of z $= 0.3$ with the following fiducial set of electron density profile parameters based on the observations of galaxy clusters: $$\mathrm{n_0 = 10^{-3} \, cm^{-3},r_{c1} = 100 \, kpc, r_{s} = 1 \, Mpc, \beta_{1} = 0.64,}$$ 
$$\mathrm{\beta_2 = 1, r_{c2} = 10 \, kpc, n_{02} = 10^{-1} cm^{-3}, \alpha = 2, \epsilon = 4, \gamma = 3, Z = 1.099},$$
while the magnetic field profile parameters are set as: $\mathrm{B_0 = 0.3 \, \mu G \, Mpc^{1/2} , s = 1/2 }$. 

\begin{figure}[h!]
     \centering
     \begin{subfigure}[h]{0.45\textwidth}
         \centering    
\includegraphics[height=5cm,width=7.5cm]
         {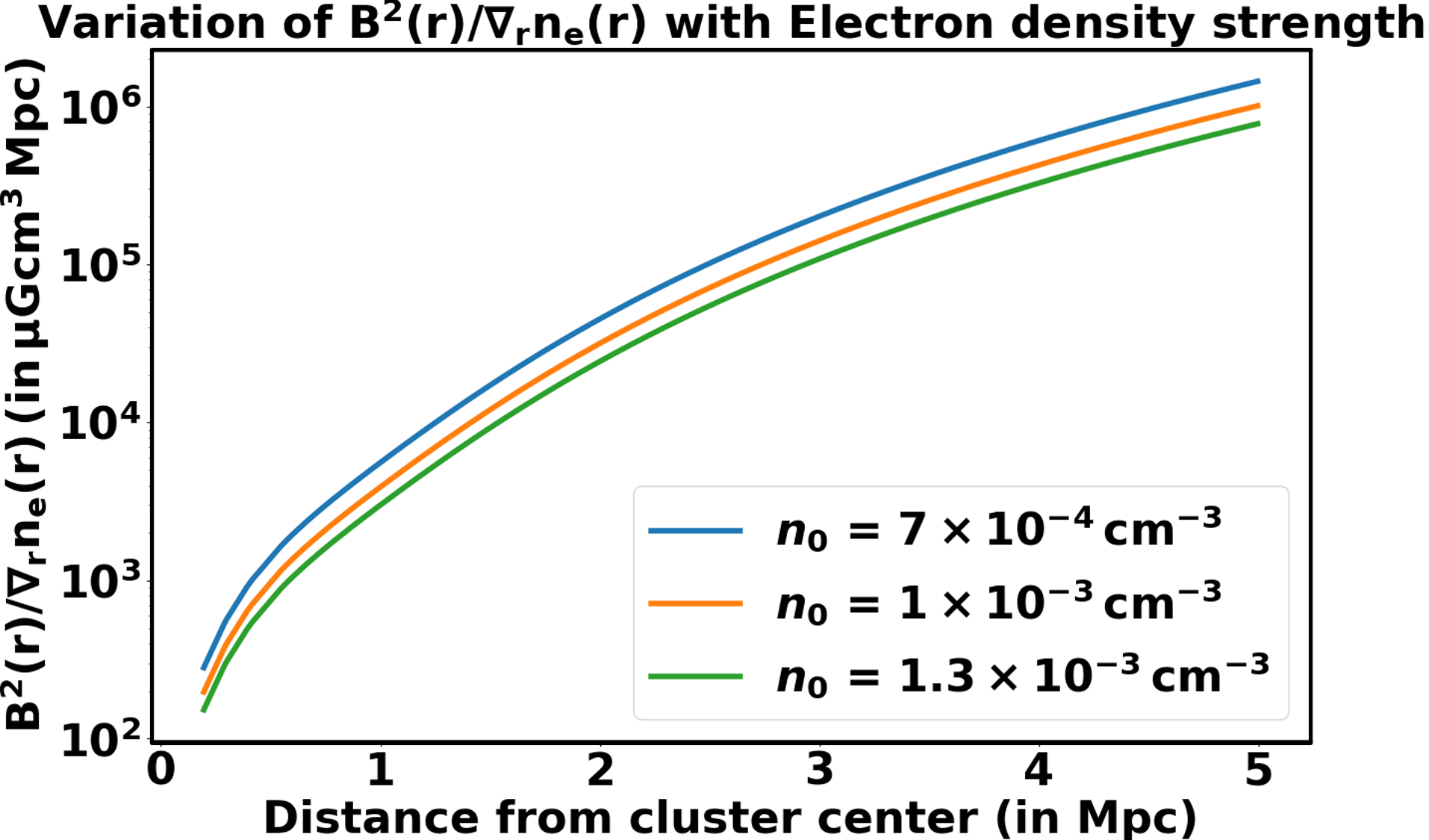}
         \caption{\textbf{Variation in $\mathrm{B(r)^2 / \nabla_r n_e(r) }$ with Electron density strength}}
         \label{fig:dne_n0_vary}
     \end{subfigure} 
     \hspace{0.01cm} 
     \begin{subfigure}[h]{0.45\textwidth}
         \centering
    \includegraphics[height=5cm,width=7.5cm]
         {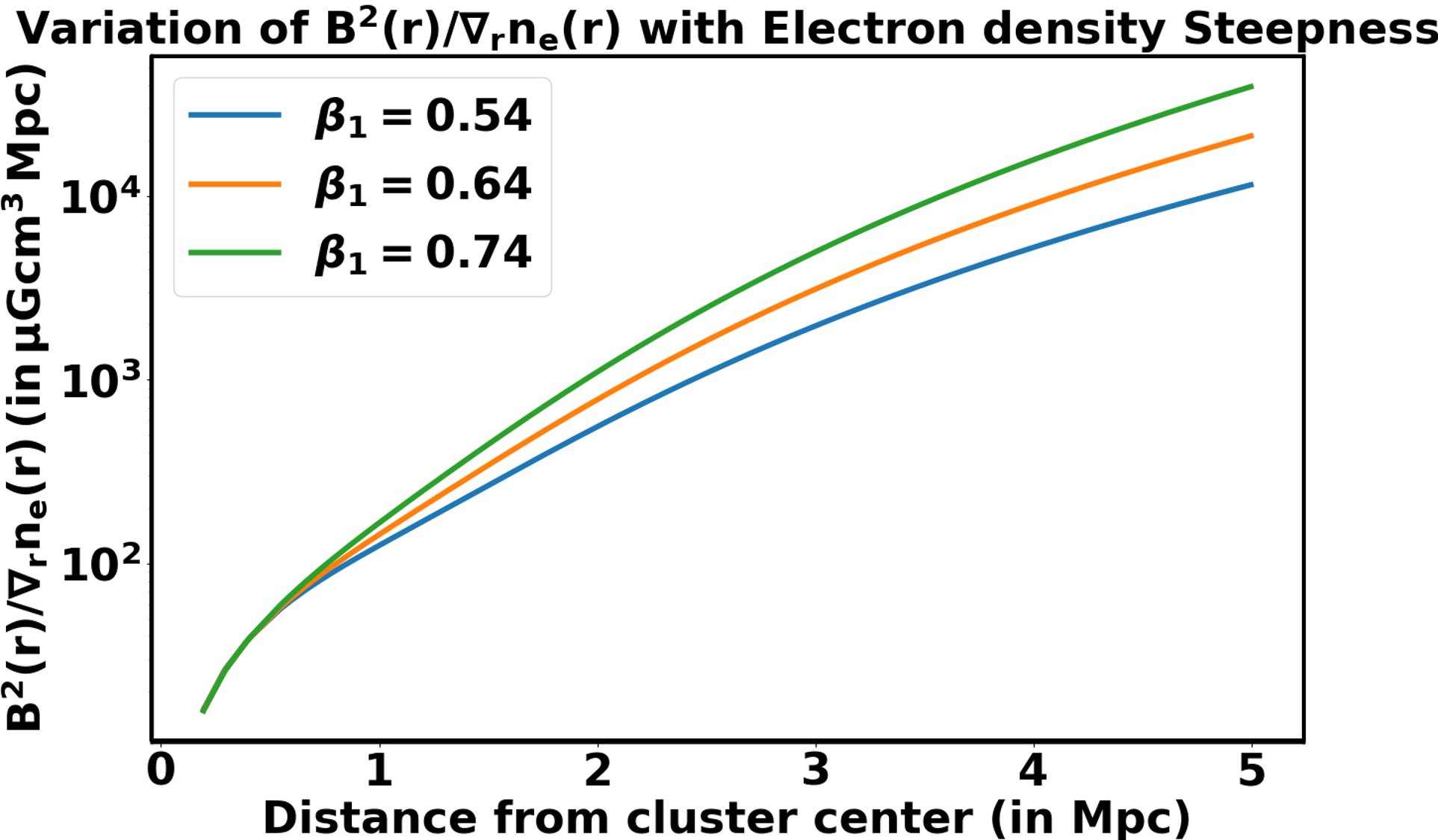}
         \caption{\textbf{Variation in $\mathrm{B(r)^2 / \nabla_r n_e(r) }$ with Electron density steepening}}
         \label{fig:dne_b1_vary}
     \end{subfigure}
     \begin{subfigure}[h]{0.45\textwidth}
         \centering
    \includegraphics[height=5cm,width=7cm]
         {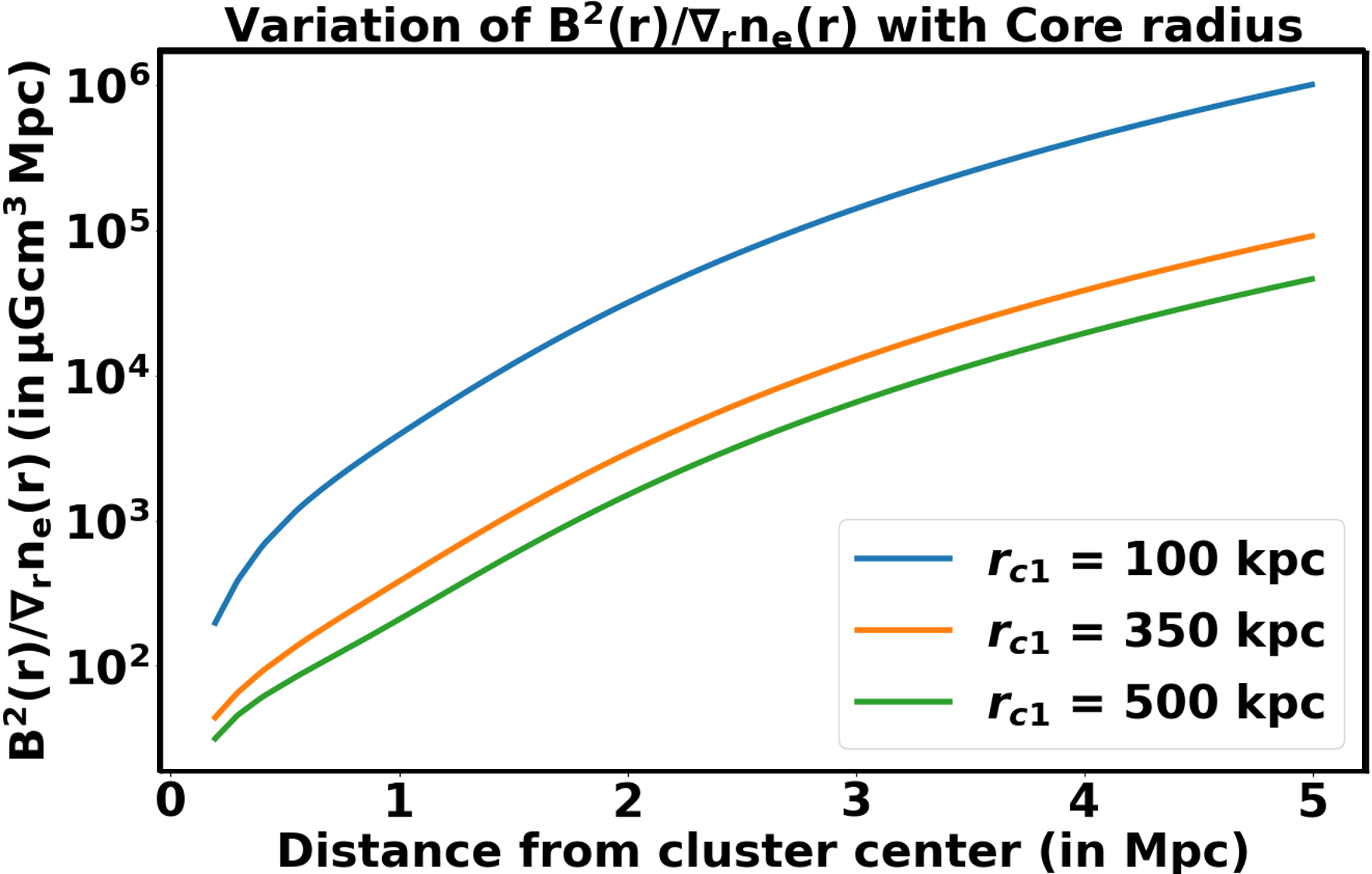}
         \caption{\textbf{Variation in $\mathrm{B(r)^2 / \nabla_r n_e(r) }$ with Core radius}}
         \label{fig:dne_rc1_vary}
     \end{subfigure}
     \hspace{0.01cm} 
     \begin{subfigure}[h]{0.45\textwidth}
         \centering
    \includegraphics[height=5cm,width=7cm]{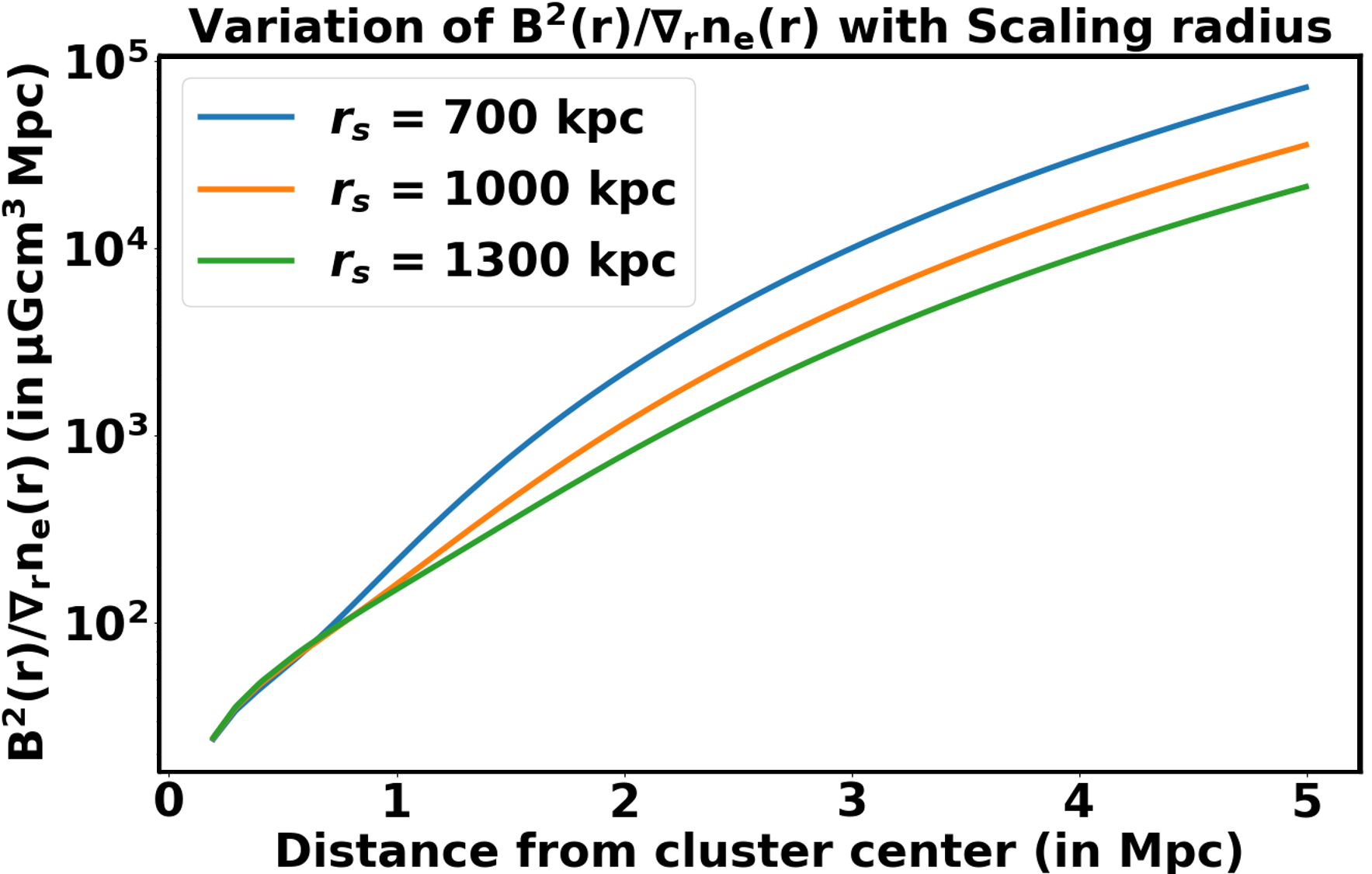}
         \caption{\textbf{Variation in $\mathrm{B(r)^2 / \nabla_r n_e(r) }$ with Scaling radius}}
         \label{fig:dne_rs_vary}
     \end{subfigure}
\caption{Variation in $\mathrm{B(r)^2 / \nabla_r n_e(r)}$ due to  electron density profile parameters}
\label{fig: dne_vary}
\end{figure}

The ALP distortion spectrum can be probed using information from the power spectrum at high multipoles. The strength of this spectrum at various multipoles depends on the distance from the cluster center where ALPs are formed and the conversion probability at the conversion location. The conversion probability is proportional to the quantity $\mathrm{B(r)^2 / |\nabla n_e(r)|}$. This quantity depends on the angle between the line of sight and the radial vector from the cluster center to the conversion location. We plot the quantity $\mathrm{B(r)^2 / |\nabla_r n_e(r)|}$, where $\mathrm{\nabla_r n_e(r)}$ denotes the radial derivative of electron density profile for some of the parameters, which affect it the most in Fig.\ref{fig: dne_vary}. 

\begin{figure}[h!]
     \centering
\includegraphics[height=8cm,width=11cm]{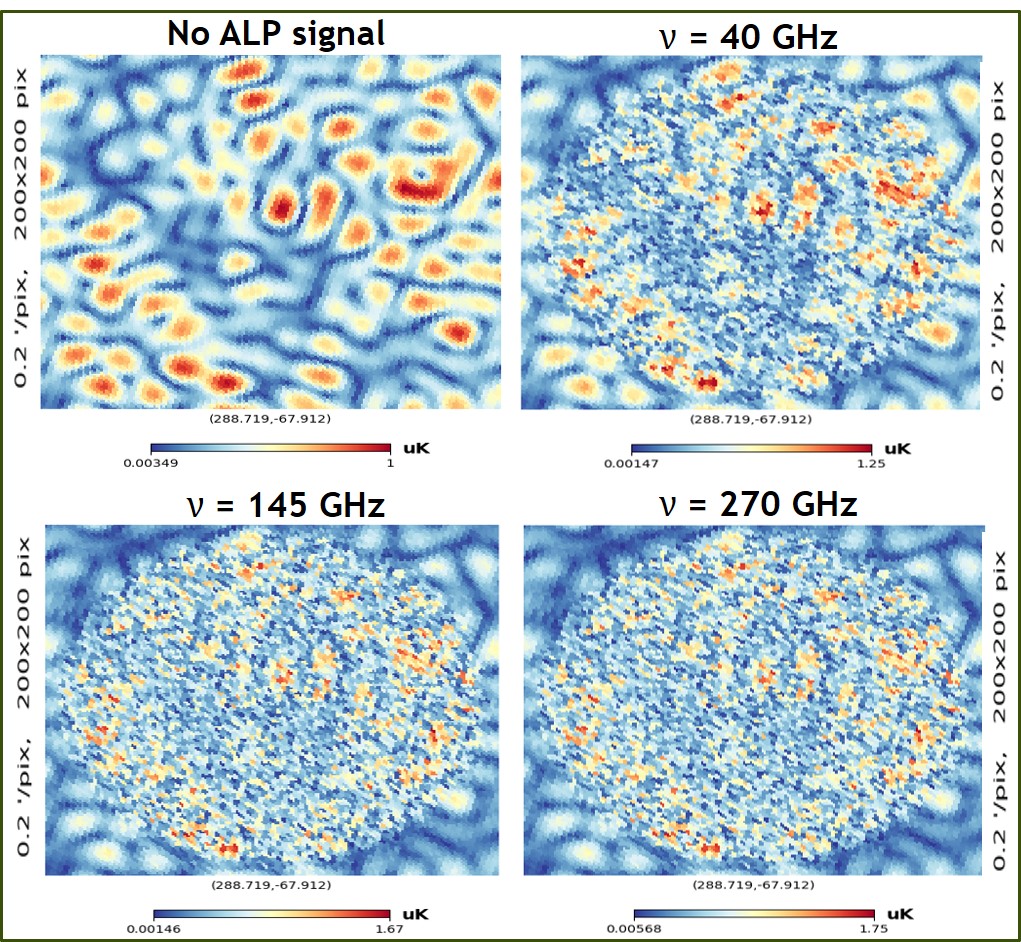}
\caption{This figure depicts the change in CMB polarization fluctuations at small angular scales (HEALPix NSIDE = 8192) in the cluster region in the absence (a) and presence (b, c, d) of ALP distortion signal. The increase in fluctuations with an increase in frequency from 40 to 270 GHz is apparent. The region shows fluctuations in 37 $\times$ 37 sq. arcminutes area on the sky. The cluster occupies the circular region in the sky, where fluctuations increase, while the surrounding region retains the characteristics of the smooth CMB. }
\label{fig: cmbaxmap}
\end{figure}
\begin{figure}[h!]
     \centering
\includegraphics[height=6cm,width=12cm]{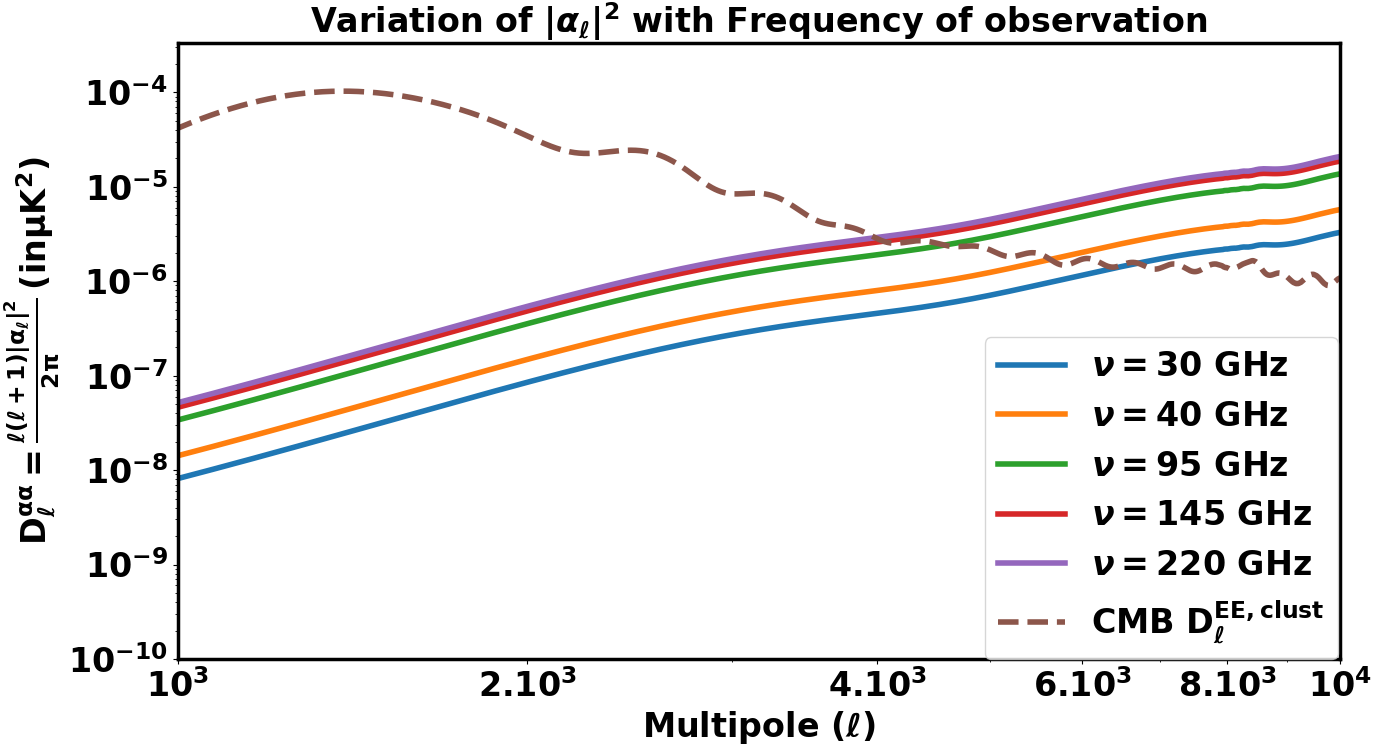}
\caption{Variation in $\mathrm{|\alpha_{\ell}|^2}$ with frequency. The ALP distortion increases with frequency in the radio and microwave spectra. Here $\mathrm{g_{a\gamma} = 10^{-11} \, GeV^{-1}}$. Small scales CMB power spectrum is shown around the cluster region.}
\label{fig: f_clust_vary}
\end{figure}

The ALP signal will affect the CMB at small angular scales in the cluster region.
Using the parameters mentioned before, we generate the ALP signal for a galaxy cluster at redshift $\mathrm{z= 0.2}$ and show its effect in Fig.\ref{fig: cmbaxmap} for a map with a HEALPix NSIDE = 8192 \cite{2005ApJ...622..759G, Zonca2019}. The angular scale corresponds to the angular size of the cluster in the sky of about 37 arcminutes. As can be seen in  Fig.\ref{fig: cmbaxmap}, the CMB-only map is very smooth at very small angular scales. The presence of an ALP signal in a cluster (which occupies the circular area in between), causes additional fluctuations in the CMB in the cluster region and these fluctuations amplify with an increase in frequency of observation in the microwave and radio spectra. The smooth CMB will retain its characteristics in the cluster-less regions in the sky. This can be explained via the variation of ALP distortion power spectrum at small angular scales which correspond to high multipoles in the spherical harmonics domain.
The ALP conversion probability increases linearly with the frequency of observation in the microwave parts of the electromagnetic spectrum. Thus, the power spectrum scales as $\mathrm{\nu^2 I_{cmb}(\nu)^2} $ (Fig. \ref{fig: f_clust_vary}). Since the power spectrum increases with an increase in fluctuations, there is an increase in power at high multipoles which increases with frequency. Also, the CMB at small angular scales around the cluster region is shown in the plot. It is obtained by filtering out the contributions of the low multipoles (or large angular scales) to the CMB power spectrum. This power spectrum is referred to as $\mathrm{D_{\ell}^{EE, \, clust}}$ in the plots. 

\begin{figure}[h!]
     \centering
\includegraphics[height=6cm,width=12cm]{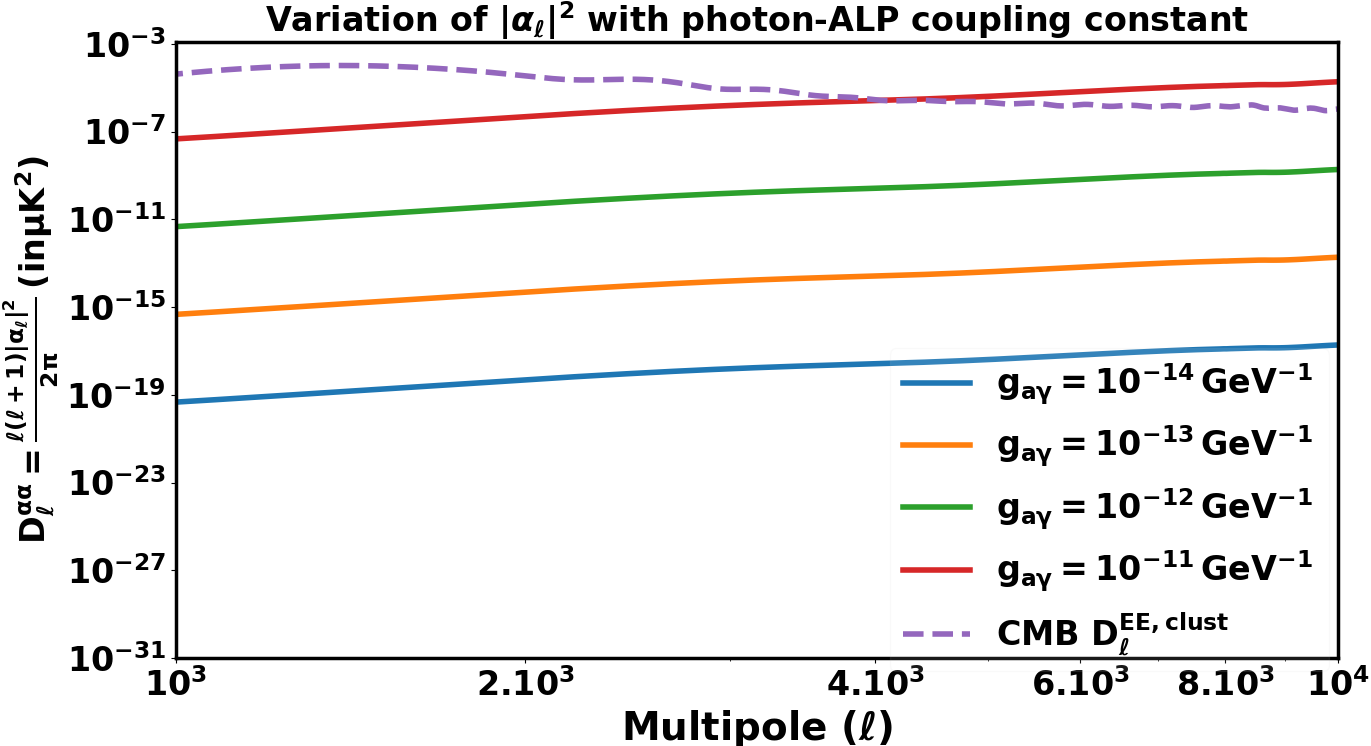}

\caption{Variation in $\mathrm{|\alpha_{\ell}|^2}$ with photon-ALP coupling constant ($\mathrm{g_{a\gamma}}$). The power spectrum varies as $\mathrm{g_{a\gamma}^4}$. Also, small angular scales CMB power spectrum is shown around the cluster region.}
\label{fig: g_clust_vary}
\end{figure}

\begin{figure}[h!]
     \centering
     \begin{subfigure}[h]{0.45\textwidth}
         \centering    
\includegraphics[height=5cm,width=7cm]
         {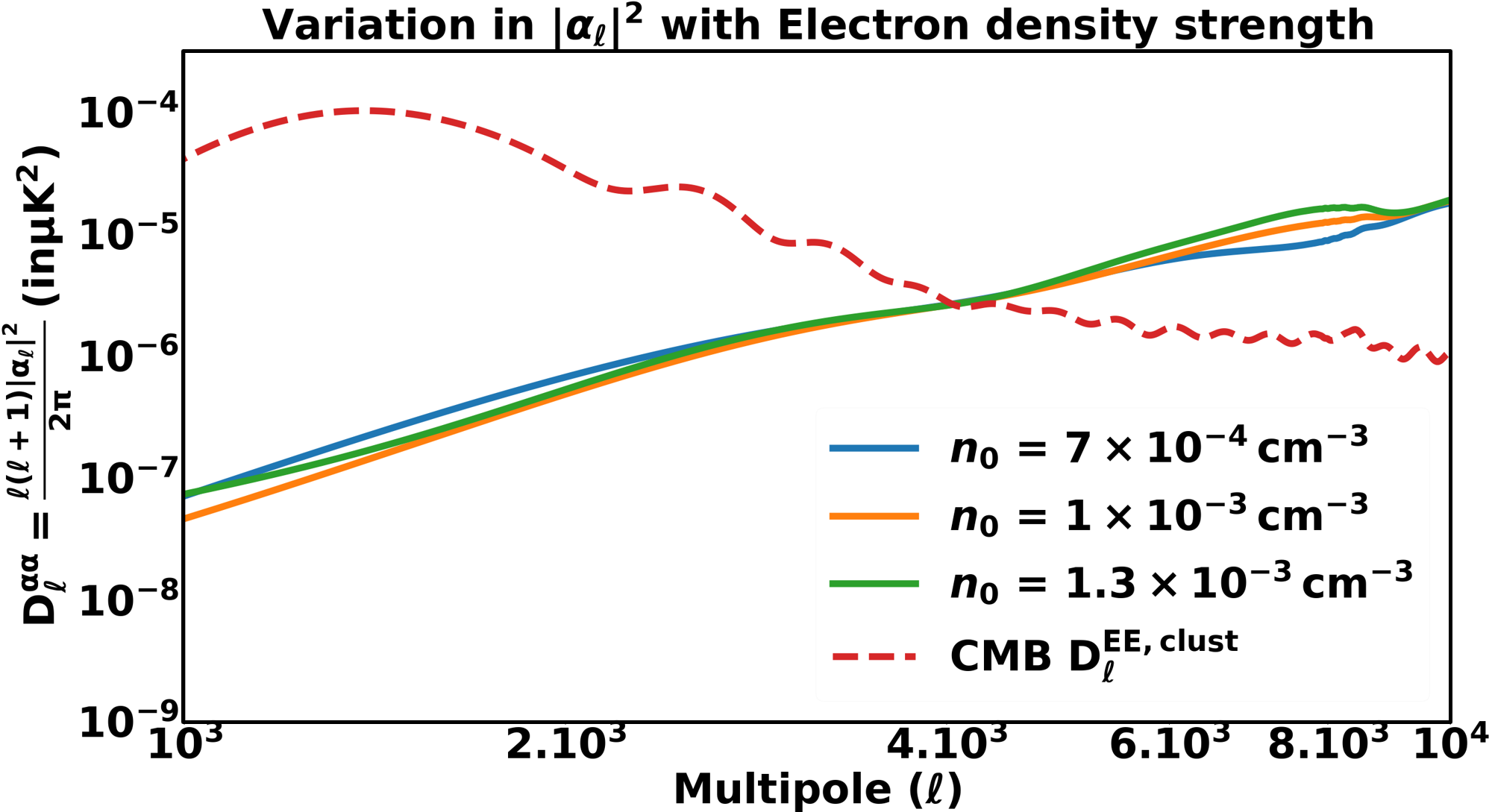}
         \caption{\textbf{Variation in $\mathrm{|\alpha_{\ell}|^2}$ with Electron density strength}}
         \label{fig:n0_alp_vary}
     \end{subfigure} 
     \hspace{0.01cm} 
     \begin{subfigure}[h]{0.45\textwidth}
         \centering
    \includegraphics[height=5cm,width=7cm]
         {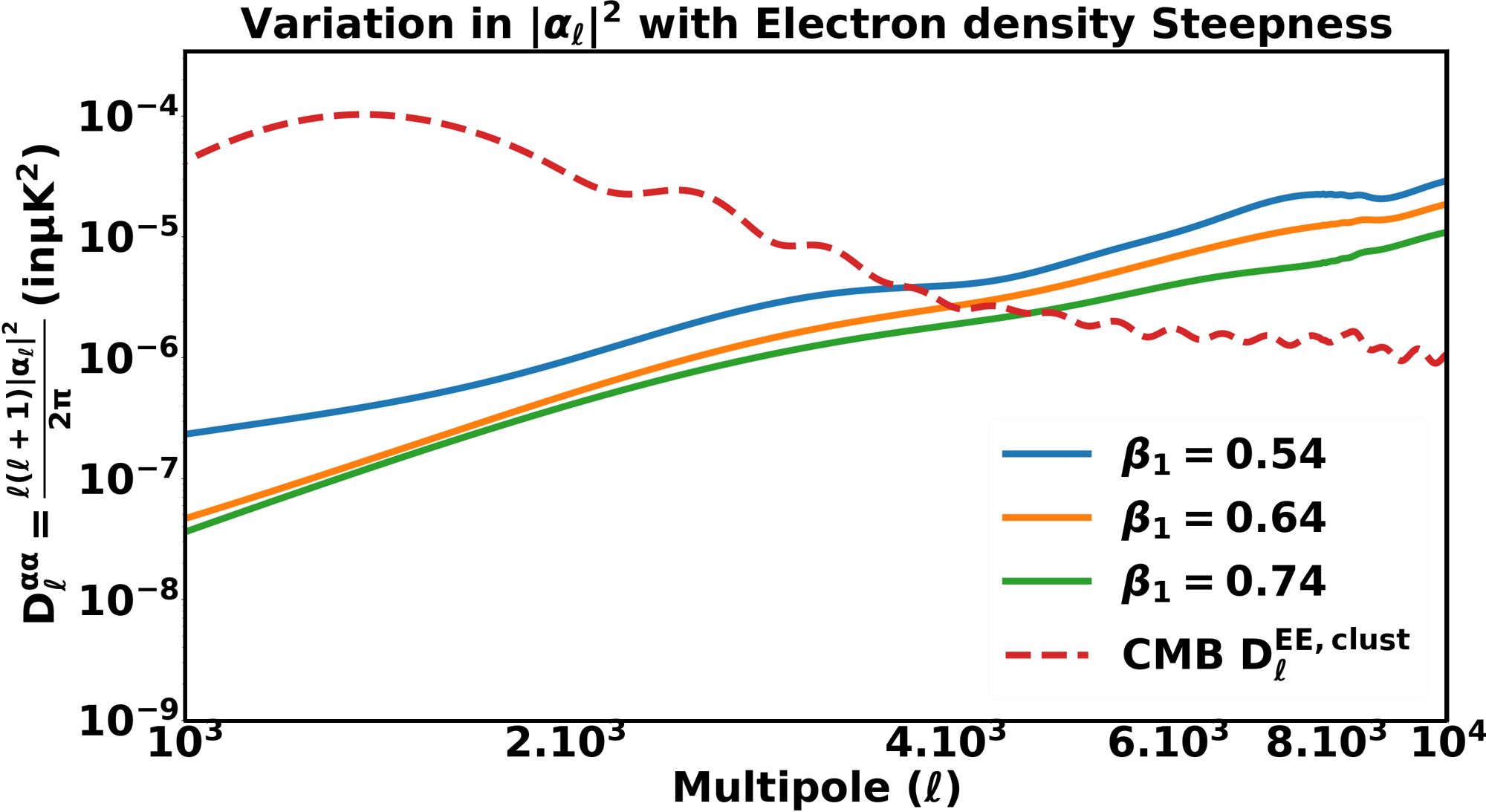}
         \caption{\textbf{Variation in $\mathrm{|\alpha_{\ell}|^2}$ with Electron density steepening}}
         \label{fig:b1_alp_vary}
     \end{subfigure}
     \begin{subfigure}[h]{0.45\textwidth}
         \centering
    \includegraphics[height=5cm,width=7cm]
         {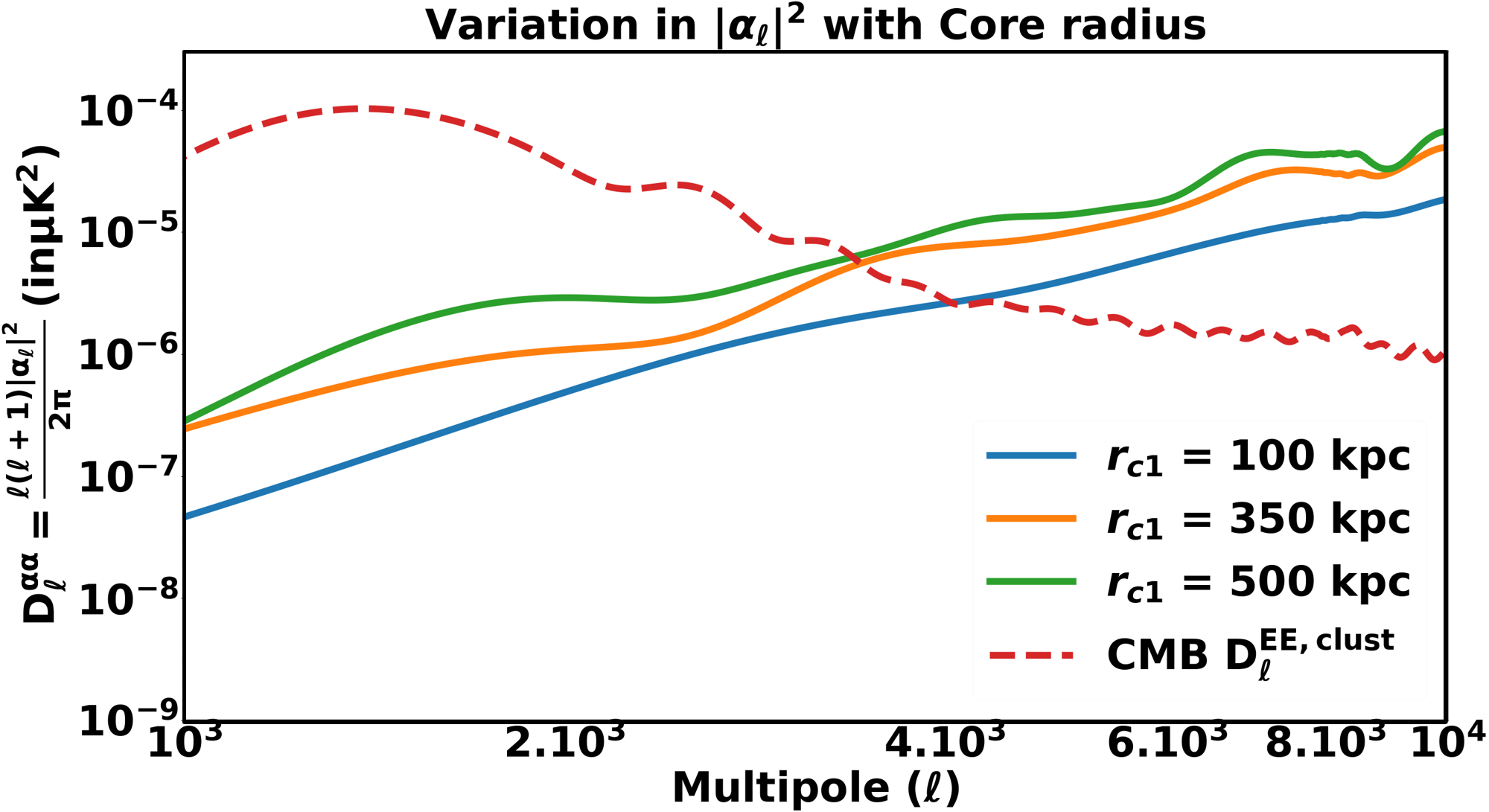}
         \caption{\textbf{Variation in $\mathrm{|\alpha_{\ell}|^2}$ with Core radius}}
         \label{fig:rc1_alp_vary}
     \end{subfigure}
     \hspace{0.01cm} 
     \begin{subfigure}[h]{0.45\textwidth}
         \centering
    \includegraphics[height=5cm,width=7cm]{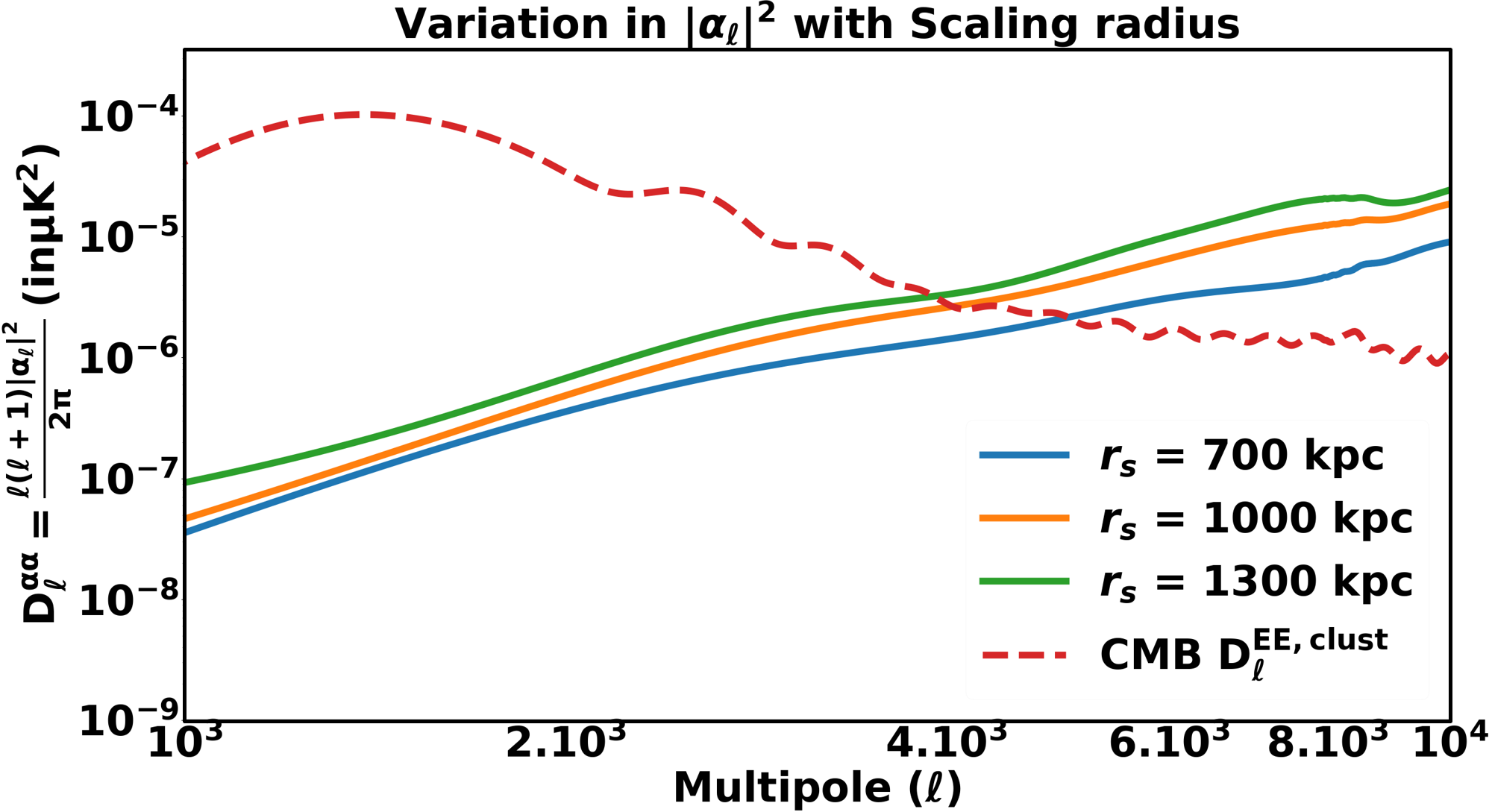}
         \caption{\textbf{Variation in $\mathrm{|\alpha_{\ell}|^2}$ with Scaling radius}}
         \label{fig:rs_alp_vary}
     \end{subfigure}
\caption{Variation in $\mathrm{|\alpha_{\ell}|^2}$ due to  electron density strength (\ref{fig:n0_alp_vary}), steepness (\ref{fig:b1_alp_vary}), core radius (\ref{fig:rc1_alp_vary}) and scaling radius  (\ref{fig:rs_alp_vary}).  Here $\mathrm{g_{a\gamma} = 10^{-11} \, GeV^{-1}}$. Also, small scales CMB power spectrum is shown around the cluster region. }
\label{fig: cluster depend}
\end{figure}

\begin{figure}[h!]
     \centering
     \begin{subfigure}[h]{0.45\textwidth}
         \centering    
\includegraphics[height=5cm,width=7cm]
         {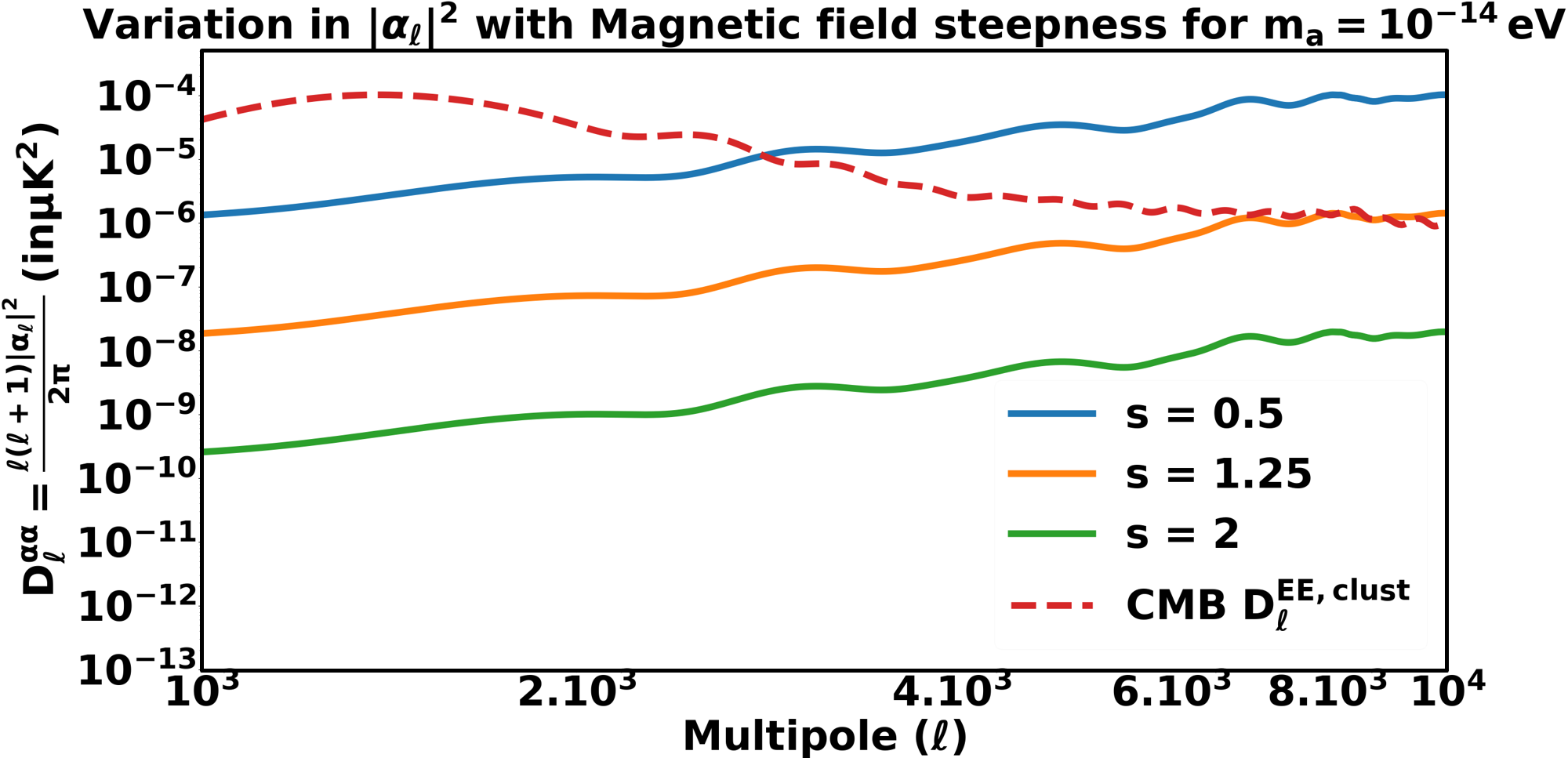}
         \caption{\textbf{Variation in $\mathrm{|\alpha_{\ell}|^2}$ for low mass ALPs (in this case, $\mathrm{m_a = 1 \times 10^{-14} \, eV}$)}}
         \label{fig:s_lowalp}
     \end{subfigure} 
     \hspace{0.01cm} 
     \begin{subfigure}[h]{0.45\textwidth}
         \centering
    \includegraphics[height=5cm,width=7cm]
         {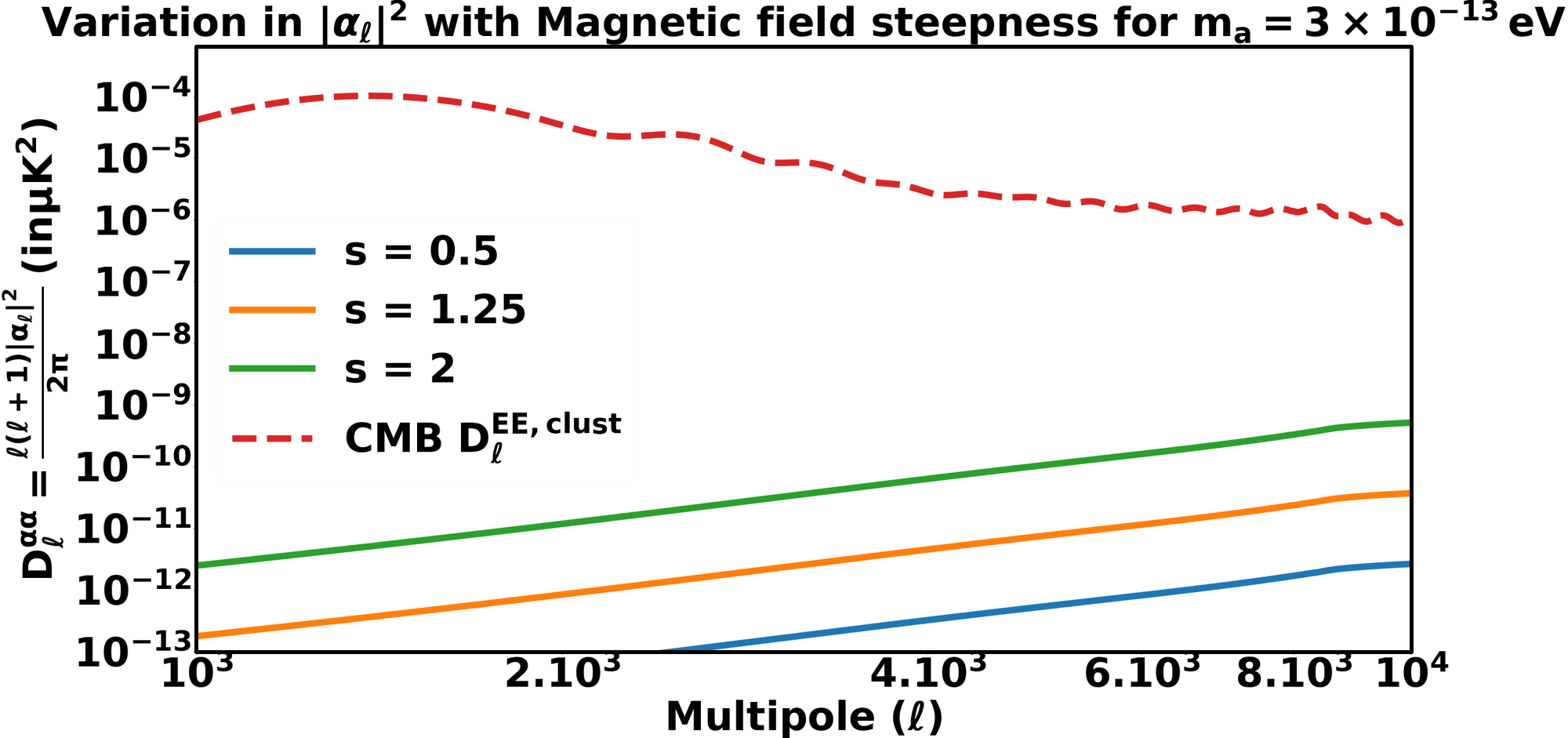}
         \caption{\textbf{Variation in $\mathrm{|\alpha_{\ell}|^2}$ for high mass ALPs (in this case, $\mathrm{m_a = 3 \times 10^{-13} \, eV}$)}}
         \label{fig:s_highalp}
     \end{subfigure} 
\caption{Variations in $\mathrm{|\alpha_{\ell}|^2}$ with Magnetic field steepness for low (a) and high (b) mass ALPs. For low mass ALPs, the power spectrum increases for the low "s" parameter. But for high mass ALPs, the magnetic field increases at low radii, when the steepness parameter "s" is high. Here $\mathrm{g_{a\gamma} = 10^{-12} \, GeV^{-1}}$. The small-scale CMB power spectrum around the cluster region is also shown. }
\label{fig: cluster depend s}
\end{figure}

\begin{figure}[h!]
     \centering
\includegraphics[height=6cm,width=12cm]
         {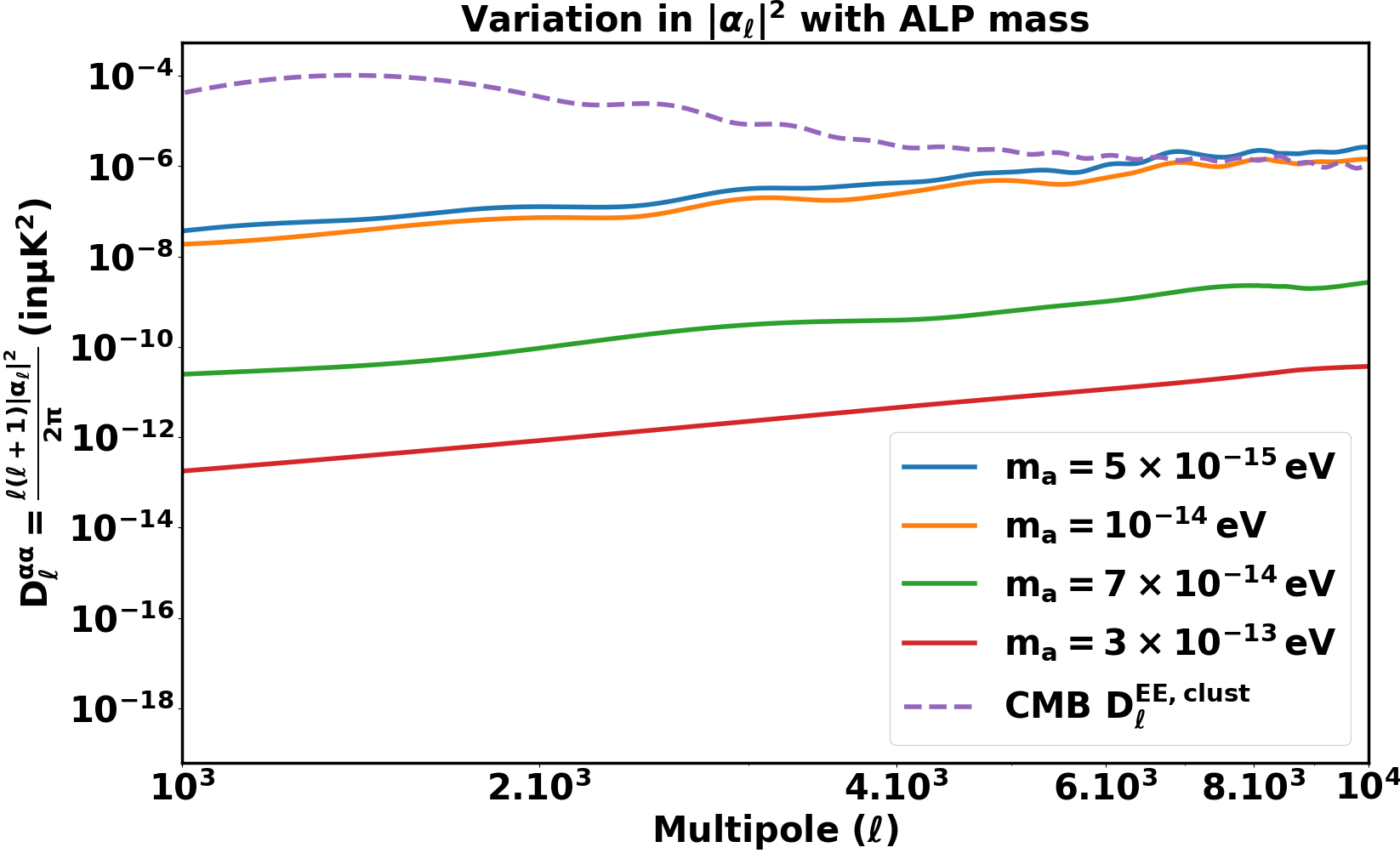}

\caption{Variations in $\mathrm{|\alpha_{\ell}|^2}$ with ALP mass. The signal generally increases for low mass ALPs as their production occurs in low electron-density regions. Here $\mathrm{g_{a\gamma} = 10^{-12} \, GeV^{-1}}$. The CMB power spectrum at small scales around the cluster region is shown as well.}

\label{fig: cluster depend m}
\end{figure}

\subsection{Change in the ALPs signal due to the change in the physical parameters}
The polarization information of the ALP distortion signal is contained in the Stokes parameters Q and U maps of the cluster region. The ALP distortion polarized intensity  $\mathrm{I_{\mathrm{pol}}}$ is given as:
\begin{equation}
\mathrm{I_{\mathrm{pol}} = \sqrt{I_{\mathrm{Q}}^2 + I_{\mathrm{U}}^2}}. 
\end{equation}
 The magnetic field direction at the location of the photon-ALP conversion in the cluster determines the contribution of the two maps to the overall signal. In our analysis, we have assumed a uniform random orientation of the cluster magnetic field. The power spectrum of the ALP distortion signal from the cluster region is given by the combined power from the Q and U maps and is what we call the ALP distortion power spectrum $|\alpha_{\ell}|^2$.

The variation of the ALP distortion power spectrum $|\alpha_{\ell}|^2$ on the profiles can be studied using the conversion probability at a location (using Eq.\ref{eq:gamma}), which scales proportionally with the square of the transverse magnetic field, while inversely with the gradient of the electron density. The outer regions of clusters have a low electron density and a gradual slope, but also a low magnetic field, as opposed to the inner regions where the strength of both these profiles increases. These factors compete to decide the strength of the ALP signal. Generally, it is the low mass ALPs that result in the generation of a stronger signal as compared to the high mass ALPs. We will discuss below the variation of the ALP signal due to variations in the astrophysical properties of clusters and due to the properties of ALPs. 

\textbf{Change in photon-ALP coupling constant: } 
The power spectrum depends on the strength of the photon-ALP coupling.
As seen in Fig. \ref{fig: g_clust_vary}, the power spectrum scales with the coupling constant as $\mathrm{g_{a\gamma}^4}$ following a $\mathrm{g_{a\gamma}^2}$ proportionality of the conversion probability. 

\textbf{Change in electron density: }As the electron density strength  (which is proportional to "$\mathrm{n_0}$") is increased, the ALPs of a certain mass will be formed away from the cluster center. This leads to the strength of the signal decreasing for low mass ALPs, which are formed in the outer regions of the cluster as the increase in strength, accompanied by a decrease in magnetic field intensity dominates over the fall in the steepness of the electron density profile in the outer regions ($\mathrm{n_e \propto n_0}$), as the gradient increases due to a steeper slope (Fig. \ref{fig:n0_alp_vary}).

The parameters that control the inner regions of galaxy clusters do not generally affect the distortion spectrum by a significant amount as the distortion signal is weak for high mass ALPs that are formed in those regions.
The dependence on the electron density steepness parameter ($\mathrm{\beta_{1}}$), outer core radius ($\mathrm{r_{c1}}$),  and scaling radius ($\mathrm{r_{s}}$) (Figs. \ref{fig:b1_alp_vary}, \ref{fig:rc1_alp_vary}, \ref{fig:rs_alp_vary}) is complicated. When one of these parameters is varied, the electron density changes as well as its gradient at different locations changes. The ALPs will be forming at a different location as compared to the case for the original parameter value. Also, there will be a different magnetic field intensity at the new location. All these factors contribute to the variation in the ALP distortion spectrum when these parameters are changed.  

\textbf{Change in magnetic field: } As the transverse magnetic field strength increases, the ALP distortion signal will increase. For our chosen magnetic field intensity profile, the parameter "s" controls the magnetic field slope at various distances from the center. The magnetic field increases at small distances from the center for high "s" values, while it decreases at large distances. Thus, with an increase in "s", the signal due to the production of high mass ALPs will increase, while it will decrease in the case of low mass ALPs (Fig. \ref{fig: cluster depend s}). We have assumed a random orientation for the transverse magnetic field at each point of the cluster. This gives  $|\alpha_{\ell}|^2$ a nearly scale-invariant shape at high multipoles. If the magnetic field orientation is not random, but follows some form of spatial trend, the variation of $|\alpha_{\ell}|^2$ with high multipoles will depend on the orientation profile of the transverse magnetic field.

\textbf{Change in cluster redshift: }The ALP distortion power spectrum also increases with an increase in the redshift of the cluster due to the conversion probability depending on the frequency of photons at the location of the cluster, which increases as $1+$z, per the expanding universe which redshifts the photon frequency as it travels from the redshift z of the cluster to being observed by us.  However, clusters at high redshifts ($> 1$) may no longer be resolved in various regions of the EM spectrum. Thus, these won't qualify for our current analysis. These are considered in a follow-up work \cite{mehta2024diffusedbackgroundaxionlikeparticles}.

\textbf{Change in ALPs mass: }The mass of ALPs that are being formed will determine not only the angular size of the signal disk, but also the strength of the signal. Fig. \ref{fig: cluster depend m} shows the variation of the ALP distortion spectrum with the mass of ALPs. The power spectrum follows the trend of decreasing with higher mass ALPs at all multipoles, as the magnitude of the electron density gradient increases in the outer regions of the cluster. In some cases, this trend may be violated for low "s" parameter values and certain multipoles as the magnetic field increases in the cluster interior. This is the case when the magnetic field increases significantly in the inner regions of the cluster and its strength dominates the variation in electron density.

\section{An estimator for ALP power spectrum from resolved galaxy clusters} \label{sec:estimator}
\subsection{Power spectrum for full sky observation} \label{sec:partial sky}
There are a variety of phenomena contributing to CMB anisotropies. The sky maps contain information about the signal at various sky locations. These signals are projected on the sky in 2d with specific angular coordinates (declination and right ascension). The power contained in these signals is quantized by taking the harmonic transform of the sky (or a certain region of it) in the angular domain. We follow the derivation given by \cite{Dodelson:2003ft,Hu_2002}.

The polarized intensity perturbations, in the CMB can be split into linear Q and U Stokes modes. These fluctuations can be treated independently at the map level. 
For any map, these fluctuations can be written using the spherical harmonic basis ($\mathrm{Y_{\ell m} (\theta, \phi)}$): 

\begin{equation}\label{eq:sphdec}
\mathrm{ \Delta^{net} = \sum_{\ell = 0}^{\infty} \sum_{m=-\ell}^{\ell} a_{\ell m}Y_{\ell m}(\theta,\phi)}. 
\end{equation} 
Here $\mathrm{a_{\ell m}}$ refers to the spherical harmonic coefficient with values of m ranging from $-\ell$ to $\ell$.
The fluctuations at the map level are encoded in the values of these spherical harmonic coefficients.  
The CMB will be affected by various signals from multiple scattering and gravitational processes. These signals add up to give the net sky signal:
\begin{equation}\label{eq:pertparts}
    \mathrm{\Delta^{net} = \Delta^{CMB} + \sum_{i \neq CMB} \Delta^{i}}.
\end{equation}
The coefficients of spherical harmonics $\mathrm{a_{\ell m}}$s depend on the origin of these fluctuations. They have a zero mean, but a non-zero variance which is called the power spectrum:
\begin{equation}\label{eq:cl_alm}
\mathrm{\langle a_{\ell m}^{net*} a_{\ell' m'}^{net} \rangle= \delta_{\ell \ell'} \delta_{m m'} C_{\ell}^{net}}.
\end{equation}
This relation comes from the orthogonality of the spherical harmonics for an isotropic universe. 

We need to estimate the power spectrum at various angular frequencies or multipoles ($\ell$) from a single sky realization (the sky map).
We consider that all the components are independent of one another and are not correlated. This enables us to write the power spectrum as:  
\begin{equation}\label{eq:clnet}
\mathrm{C_{\ell}^{net} = \sum_{j} \sum_{i} \langle a_{\ell m}^{i*} a_{\ell m}^{j} \rangle = \langle a_{\ell m}^{cmb*} a_{\ell m}^{cmb} \rangle + \langle a_{\ell m}^{fgs*} a_{\ell m}^{fgs} \rangle + \langle a_{\ell m}^{ax*} a_{\ell m}^{ax} \rangle + .....},
\end{equation}
where $\mathrm{i}$ and $\mathrm{j}$ refer to the various sky components, including CMB, ALP distortion, foregrounds (fgs), etc. There will also be the effect of limited beam resolution and instrument noise. This changes the coefficients as:
\begin{equation}\label{eq:alm_obs}
  \mathrm{a_{\ell m}^{obs} = B_{\ell}(a_{\ell m}^{cmb} + a_{\ell m}^{ax} + a_{\ell m}^{fg}) + \eta_{\ell m} },  
\end{equation}
where we take a Gaussian beam $\mathrm{B_{\ell} = exp(-\ell(\ell+1)\theta_{beam}^2 / 2)}$ and  $\mathrm{\eta_{\ell m}}$ are the spherical harmonic coefficients corresponding to instrument noise 
\begin{equation}\label{eq:Nl}
\mathrm{\langle \eta_{\ell m}* \eta_{\ell 'm'} \rangle = N_{\ell} \delta_{\ell \ell '} \delta_{mm'}}.    
\end{equation}
We use a Gaussian probability for the coefficients $\mathrm{a_{\ell m}}$'s for a given power spectrum $\mathrm{C_{\ell}}$:
\begin{equation}\label{eq:prob_alm_cl}
    \mathrm{P(a_{\ell m}^{i}|C_{\ell} ^{i} ) = \frac{1}{\sqrt{2\pi C_{\ell}^i}} exp \left( - \frac{|a_{\ell m}^{i}|^2}{2 C_{\ell}^i} \right)}.
\end{equation}
Also, a Gaussian distribution is assumed for  the $\mathrm{a_{\ell m}^{obs}}$s:
\begin{equation}\label{eq:prov_almobs_alm}
\mathrm{P(a_{{\ell}m}^{obs}|\{a_{\ell m}^i \} ) = \frac{1}{\sqrt{2\pi N_{\ell}}} exp \left( - \frac{|a_{\ell m}^{obs} - \sum_{i} B_{\ell} a_{\ell m}^i|^2}{2 N_{\ell}} \right)}.    
\end{equation}
Here $\mathrm{B_{\ell} \sum_{i}a_{\ell m}^i}$ is the mean and variance corresponds to the noise power spectrum $\mathrm{N_{\ell}}$.

Now, using the Bayes' theorem, we marginalize over the probabilities for various $\mathrm{a_{\ell m}s}$: 
\begin{equation} \label{eq:likeBayes}
    \mathrm{P(a_{\ell m}^{obs}|\{C_{\ell}^i \} ) = \prod_{m=-\ell}^{\ell} \int\int\int \prod_{i} d \, a_{\ell m}^{i}  P(a_{\ell m}^{obs} |  a_{ \ell m}^i) P(a_{\ell m}^{i}|C_{\ell}^{i} )  },
\end{equation}
where we have considered that the signals are independently affecting the observed coefficient. 

Performing the integral gives the likelihood of the observed $\mathrm{a_{\ell m}}$s to be:
\begin{equation}\label{eq:problike}
 \mathrm{\mathcal{L} = P(a_{\ell m}^{obs}|\{C_{\ell}^i \} ) = [2\pi (B_{\ell}^2 \sum_{i}C_{\ell}^{i} + N_{\ell})]^{-(2\ell + 1)/2} exp \left[ \sum_{m = -\ell}^{\ell}  -\frac{|a_{\ell m}^{obs}|^2}{2(\sum_{i}C_{\ell}^{i} B_{\ell}^2 + N_{\ell})} \right] }.    
\end{equation}
The maximum likelihood estimator of the power spectrum for component $\mathrm{i}$ is obtained by maximizing the above likelihood with respect to $\mathrm{C_{\ell}^{i}}$ :
\begin{equation}
\mathrm{\tilde{C_{\ell}^{i}} = B_{\ell}^{-2} \left[ \frac{1}{2\ell + 1} \sum_{m = -{\ell}}^{{\ell}} |a_{\ell m}^{obs}|^2 - N_{\ell} \right] - \sum_{j \neq i} C_{\ell}^{j}.}
\label{eq:estimator}
\end{equation}

There are uncertainties involved when an estimation of the power spectrum is made. The variance of the estimator is called the covariance,   
and takes into account the limited number of modes for a multipole ($\ell$) that can be used to estimate the power spectrum from a single sky map.
When a partial sky observation is made, the covariance increases as $\mathrm{1/f_{sky}}$ and is given as:
\begin{equation}
\mathrm{Cov(\tilde{C_{\ell}}) = \langle C_{\ell}^2 \rangle - \langle C_{\ell} \rangle^2 = \frac{2}{(2\ell +1)f_{sky}} \left[ \sum_i C_{\ell}^i + B_{\ell}^{-2}N_{\ell} \right]^2 .}
\label{eq:covar}
\end{equation}

The ALP fluctuations will be probed against the covariance for the fiducial (non-ALPs case) map spectrum to obtain constraints on the coupling constant.

\subsection{Power spectrum for partial sky observation} \label{sec:partial sky2}

Since galaxy clusters subtend small angular scales on the sky, the ALP distortion can be probed from the power spectrum at small angular scales, which correspond to high multipoles. This calls for partial sky observation of the cluster regions.  The features of the partial sky power spectrum do not exactly match those of the full sky power spectrum. This distinction depends on the portion and shape of the sky being considered, more generally the window function, which in our case is a step function with all pixels in the cluster region as one, and the rest zero. A partial sky imparts an
uncertainty to the observed power spectrum at different multipoles that depend on the shape and size of the window that is being observed. 
The partial sky power spectrum is obtained as \cite{Hivon_2002}
\begin{equation}\label{eq:modecouple}
  \mathrm{\bar{C_{\ell}} = \sum_{\ell'} M_{\ell \ell'}(B_{\ell'}^2 C_{\ell'} + N_{\ell'}) = \sum_{\ell'} M_{\ell \ell'}B_{\ell'}^2 C_{\ell'} + f_{sky} N_{\ell}}, 
\end{equation}
where $\mathrm{f_{sky}}$ considers the fraction of sky being considered for observation and the fluctuations are assumed to be statistically homogeneous and isotropic. Noise is assumed to be Gaussian and uncorrelated between frequencies and pixels. The masking kernel $\mathrm{M_{\ell \ell'}}$ takes into account the mode-mode coupling that is induced when partial sky observation is made. This masking kernel is a geometric quantity depending on the window function and is given as:
\begin{equation} \label{eq:maskkernel}
  \mathrm{ M_{\ell_1 \ell_2} = \frac{2 \ell_2 + 1}{4\pi}\sum_{\ell_3} W_{\ell_3} Wig(\ell_1,\ell_2,\ell_3)^2}. 
\end{equation}
Here $\mathrm{W_{\ell}}$ is the window function of the masked map in the spherical harmonics domain  and $ \mathrm{Wig(\ell_1,\ell_2,\ell_3)}$ is the Wigner 3-j symbol written as:
\begin{equation} \label{eq:wigner3j}
\mathrm{Wig(\ell_1,\ell_2,\ell_3) = \begin{pmatrix}
  \mathrm{\ell_1} & \mathrm{\ell_2} & \mathrm{\ell_3}\\ 
  0 & 0 & 0
\end{pmatrix}}.    
\end{equation}
In \cite{Hivon_2002}, the inverse of the mode coupling kernel was used to obtain the full sky power spectrum. In our case, we are interested in $\mathrm{M_{\ell_1 \ell_2}}$ itself as we need to model the power spectrum in the small cluster regions. 
Since the noise power spectrum $\mathrm{N_{\ell}}$ is assumed constant, the mode-mode coupling for a partial sky observation is assumed to be a $\mathrm{f_{sky}}$ factor being multiplied with the full sky noise power spectrum. We use the partial sky estimate in the likelihood estimation shown in Sec. \ref{sec:partial sky}. 

\section{Foreground cleaning techniques for ALPs signal}\label{sec:contaminants}
The ALP signal is contaminated by the astrophysical foregrounds (dust, synchrotron, etc.) as well as the lensed CMB anisotropies. The dust refers to the silicate and carbonaceous grains, as well as polycyclic aromatic hydrocarbons \cite{Thorne_2017}. 
 The aspherical dust grains emit preferentially along their longest axis. This axis tends to align perpendicular to the magnetic fields. This results in the thermal emission being polarized. The synchrotron emission is polarized and is a consequence of the acceleration of relativistic electrons in the presence of a transverse magnetic field \cite{1986rpa..book.....R}. The synchrotron emission peaks in the radio spectrum and decreases with frequency, mainly contributing below 70 GHz. On the other hand, thermal dust emission is in the far infrared and becomes a significant foreground for frequencies greater than 200 GHz. The CMB primordial power spectrum is independent of frequency. We thus prefer the frequencies 90 to 160 GHz for our analysis as these foregrounds weaken out in this region of the spectrum.

These foregrounds are highly dominating in the galactic plane and their effects need to be mitigated by masking the galactic region on the sky.  Masking of the galactic plane reduces the effect of these foregrounds, but there is still a significant residual impact even at high latitudes. Thus, cleaning is required to reduce the impact of the foregrounds.
 
\subsection{Interior Linear Combination for recovering the ALP signal} \label{sec: ILC}
\begin{figure}[h!]
     \centering
\includegraphics[height=5.5cm,width=10cm]{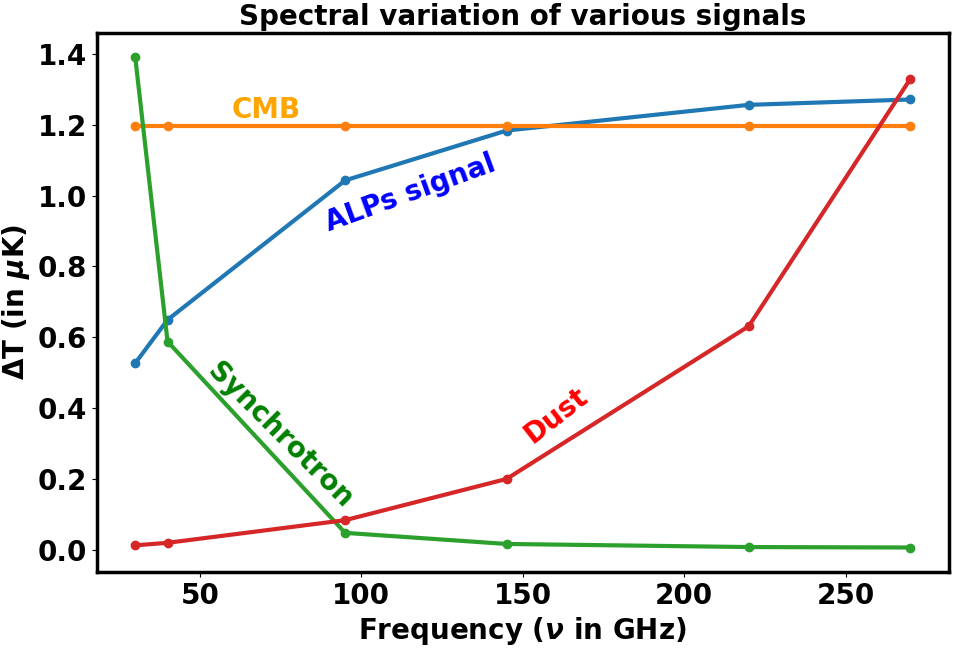}
    \caption{The ALP distortion signal ($\mathrm{g_{a\gamma} = 10^{-12} \, GeV^{-1}}$) spectral variation at a certain pixel with frequencies corresponding to the CMB-S4 bands. The signal intensity scales as $\mathrm{\Delta I \propto \nu I_{cmb}(\nu)}$.}
       \label{fig:f_signals}
\end{figure}

The ALP distortion signal ($\alpha$-distortion) follows a distinct variation with the frequency of observation ($\mathrm{\Delta I \propto \nu I_{cmb}(\nu)}$) (see Fig.\ref{fig:f_signals}). This variation can be used to clean the ALP signal, which is mainly contaminated by CMB and foregrounds.  
To reduce their impact, we employ the standard Interior Linear Combination (ILC) method (\cite{Eriksen_2004,ilc2008internal}) that extracts the ALP signal, while suppressing contamination from foregrounds and CMB. This technique uses multi-wavelength observations and assigns weights to the maps at various frequencies at a common beam resolution, such that the variance of the linearly combined map (with the corresponding weights) is minimized.  The linear combination of the various frequency maps based on their weights gives the ILC-cleaned map. The ILC weights are calculated using the covariance matrix $\mathrm{C_s}$ and the ALP spectrum ($\mathrm{f_{a\gamma}}$) which is proportional to $\mathrm{\nu I_{cmb}(\nu)}$ in intensity units: 
\begin{equation}
    \mathrm{w_{ilc } = f_{a \gamma}^T C_s^{-1} (f_{a \gamma}^T C_s^{-1} f_{a \gamma})^{-1}    
    }.
    \label{eq:weights_ilc}
\end{equation}
The ILC cleaned map ($\mathrm{S_{ilc}}$) is given as a weighted linear combination of the various frequency maps:
\begin{equation}\label{eq:ilcsum}
\mathrm{S_{ilc} = \sum_{\nu} w_{ilc}^{\nu} S_{\nu}},    
\end{equation}
where $\mathrm{S_{\nu}}$ corresponds to the map at the frequency $\nu$. In this analysis, we will primarily use the ILC technique to clean the foregrounds, as it does not make assumptions about the spectral shape of the foregrounds and relies on the spectral shape of the ALP signal. We have not considered the polarized SZ signals as these are very weak for the ALP coupling strength we are probing and the ALP signal can be separated using its spectral and spatial variation (see Appendix \ref{sec:szpol}).
We will compare the improvement in results if a foregrounds template-based technique is used, in Appendix \ref{sec:no_temp}.

\section{Demonstrating map level inference of Photon-ALP coupling constant from multi-band observations} \label{sec: bayesian}

The galaxy clusters occupy small angular regions in the sky. These correspond to high multipoles which can be probed to obtain ALP distortion signatures.  
Constraining the ALP distortion signal requires the power spectrum of the map in these cluster regions. 
We use HEALPy \cite{Zonca2019} for our analysis. 
The foreground models (dust and synchrotron) are generated using PySM \cite{Thorne_2017}. 
We use the beam resolution $\mathrm{B_{\ell}}$ and the noise power spectrum $\mathrm{N_{\ell}}$ for Simons Observatory (SO) and CMB-S4 detectors provided in Table \ref{tab:noise}. 
\begin{table}[htbp]
  \small
  \centering
  \caption{\textbf{Instrument noise and beams for the upcoming CMB experiments \cite{Ade_2019,abazajian2016cmbs4}} }
  \subfloat[\textbf{Simons observatory}]{%
    \hspace{0.5cm}%
    \begin{tabular}{|c|c|c|}
        \hline
        $\nu$ (in GHz) & Noise (in $\mu$K) & Beam \\
        \hline
        \hline
        27 & 52 & 7.4 \\
        \hline
        39 & 27 & 5.1 \\
        \hline
        93 & 5.8 & 2.2 \\
        \hline
        145 & 6.3 & 1.4 \\
        \hline
        225 & 15 & 1.0 \\
        \hline
        280 & 37 & 0.9 \\
        \hline
    \end{tabular}%
    \hspace{.5cm}%
  }\hspace{0.01cm}
  \subfloat[\textbf{CMB-S4}]{%
    \hspace{0.5cm}%
    \begin{tabular}{|c|c|c|}
        \hline
        $\nu$ (in GHz) & Noise (in $\mu$K) & Beam \\
        \hline
        \hline
        30 & 30.8 & 7.4 \\
        \hline
        40 & 17.6 & 5.1 \\
        \hline
        95 & 2.9 & 2.2 \\
        \hline
        145 & 2.8 & 1.4 \\
        \hline
        220 & 9.8 & 1.0 \\
        \hline
        270 & 23.6 & 0.9  \\
        \hline
    \end{tabular}%
    \hspace{.5cm}%
  }\label{tab:noise}
\end{table}

\subsection{Observed power spectrum}\label{sec:obs_powspec}
The CMB is generated from CAMB, while dust and synchrotron models "d-3" and "s-3" are obtained from PySM. These models are further explained in Sec.\ref{sec:no_temp}. These components are added to a generated ALPs map for a certain ALP mass range with a fixed coupling constant value (which we call $\mathrm{g_{a\gamma}^{true}}$) and smoothed, along with the instrument noise for a survey (as mentioned in the Table \ref{tab:noise}). 
We use a spatial mask as a step window function with all pixels in the cluster regions as one and the rest are set to zero.  

Using \ref{eq:modecouple}, where $\tilde{\mathrm{C_{\ell}}}$ contains contributions from the CMB, foregrounds, and the ALP signal, we get the observed map level power spectrum from which we need to obtain constraints on the ALP coupling constant. The mode coupling kernel is already taken care of when using HEALPy's anafast on a masked map for obtaining the power spectrum. This provides us with the term $\mathrm{\frac{\sum_{m=-\ell}^{\ell} |a_{\ell m}^{obs}|^2}{2\ell + 1}}$ in Eq. \ref{eq:estimator} and serves as our mock data. For our mock data, we set the ALP coupling constant as $\mathrm{g_{a\gamma}^{true} = 0}$, which corresponds to the case of the non-existence of ALPs for the applicable mass range. 

For ILC, we make mock maps at four of the frequency bands for the detectors (SO and CMB-S4), linearly combine the maps corresponding to the detector with corresponding weights (see Eq.\ref{eq:ilcsum}) obtained using the ALP spectrum ($\mathrm{f_{a\gamma}}$) and the covariance matrix ($\mathrm{C_s}$) for all maps (see Eq.\ref{eq:weights_ilc}). 
For making the ILC-cleaned maps for SO and CMB-S4, we use the frequency bands 93, 145, 225, and 280 GHz for SO and 95, 145, 220, and 270 GHz for CMB-S4. We use a common beam resolution of 2.2 arcmin to smooth the maps, which corresponds to the frequencies 93 GHz for SO and 95 GHz for CMB-S4. The power spectrum of the ILC cleaned map becomes our observed power spectrum.

\subsection{Fiducial map power spectrum}
The first term in the estimator Eq. \ref{eq:estimator} is now known from \ref{sec:obs_powspec}. The other term is a sum of power spectra of components other than the ALP distortion. We call the ensemble average of different realizations of the power spectrum of the cluster regions without ALP signal the fiducial map power spectrum, based on the assumption of the non-existence of ALPs. It is needed so that an estimation of the ALP power spectrum in the observed region can be made. As a power spectrum will always have a non-zero mean, any additional fluctuations over the covariance of the power spectrum can be used to probe ALPs. We obtain the fiducial map power spectrum as the sum of beam-smeared power spectra from different components (CMB and foregrounds)\footnote{Ultimately, we are trying to estimate the beam-smeared ALP power spectrum to find constraints on its coupling. So, the equation we are using is the estimator Eq. \ref{eq:estimator} multiplied by $\mathrm{B_{\ell}^2}$ on both sides.}.

In the case of ILC, we combine fiducial (no ALP signal) maps at all frequencies according to their weights to obtain an ILC-cleaned fiducial map. We can get the fiducial mean power spectrum by obtaining the power spectra of different fiducial ILC weighted map realizations with the same window function. The estimated fiducial mean map power spectrum will then be given as the mean of the power spectra.   
 
\begin{figure}[h!]
     \centering
\includegraphics[height=6.5cm,width=11cm]{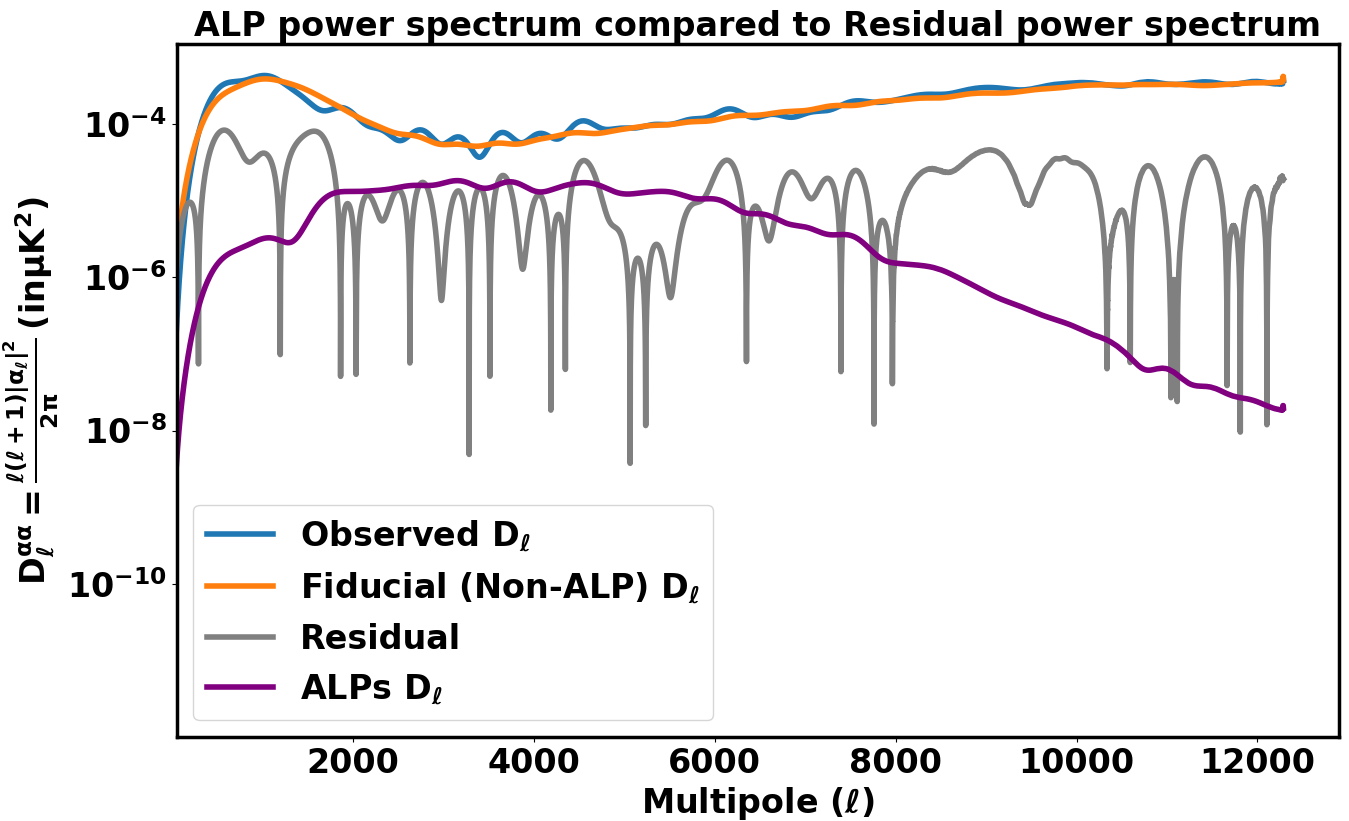}

    \caption{The observed and fiducial mean spectra provide the residual spectrum, which along with covariance sets the cutoff for possible ALP fluctuations (here $\mathrm{g_{a\gamma} = 8 \times 10^{-12} \, GeV^{-1}}$) in the observed spectrum.}
\label{fig: covar}
\end{figure}

\subsection{Modelling of the ALP power spectrum}\label{sec:makemap}
We generate the ALP signal in clusters at the required frequencies with a constant value of coupling constant for a certain ALP mass range and smooth with the instrument beam using the framework mentioned in Sec. \ref{sec:ALPs from CMB} and Sec. \ref{sec:alpha_l}. We use a standard coupling constant value of $\mathrm{g_{a\gamma}^* = 10^{-12} \, GeV^{-1}}$ (need not be equal to the value used in the observed map, which is $\mathrm{g_{a\gamma}^{true}}$) for modelling of the ALP distortion spectrum. As the power spectrum scales with the coupling strength as $\mathrm{g_{a\gamma}^4}$, we can compare the value of the ALP coupling in the observed sky with respect to a value used for modelling the signal as $\mathrm{C_{\ell}^{\rm ax,obs}= (\frac{\mathrm{g^{true}_{a\gamma}}}{\mathrm{g^{fiducial}_{a\gamma}}})^4C_{\ell}^{\rm ax,*}}$. We assume ALPs of masses in the range $10^{-13}$ eV to be forming with a tolerance of 30\%. The modelling of the ALP signal can be done for different masses separately using the information on cluster profiles and redshifts and searched from the map power spectrum.
 The power spectrum is obtained for the beam-smoothed ALP distortion map for the standard coupling constant value. 

 For ILC, we combine the beam-smoothed ALP maps at various frequencies with the weights obtained in \ref{sec:obs_powspec} and obtain the weighted ALP map, the power spectrum of which gives us the ILC-weighted ALP power spectrum.
Note that we ultimately want to estimate the beam-smeared ALP power spectrum in the cluster region ($\mathrm{B_{\ell}^2 \hat{C}^{ax,obs}_{\ell}}$).

Modelling of the ALP power spectrum requires modelling of the ALP signal from resolved clusters, which depends on the redshift, electron density profiles, and magnetic field profiles of the clusters. Here we assume typical values of these parameters as mentioned in Table. \ref{tab:params}. In reality, when applying this to the data, we plan to use multi-band observations using which we can find constraints on these quantities for resolved clusters, by studying them in different wavelengths (radio for magnetic field, X-rays and SZ observations for electron density, and optical for cluster redshifts). A multi-band Bayesian framework for cluster analysis is currently under preparation \citep{Mehta:2024:new3}. If the clusters are not resolved in any of these frequencies, then they come into the domain of unresolved clusters for which magnetic field, electron density, and redshift cannot be estimated. ALP signal from these sources would create a diffused background of the ALP distortion signal. This will be explained in a separate analysis \citep{mehta2024diffusedbackgroundaxionlikeparticles}.  

\subsection{Covariance of power spectrum}
The covariance of the estimator is mentioned in Eq. \ref{eq:covar}.
The estimated ALP power spectrum will be probed against the covariance for the fiducial mean estimated spectrum to probe any deviations arising due to the ALP distortion. The covariance decreases with the fraction of sky being observed ($\mathrm{f_{sky}}$). The covariance is high at low multipoles ($\ell$) due to a greater contribution from CMB and foregrounds, accompanied by a low number of modes for each multipole (2$\ell + 1$). The covariance is also high at very high multipoles due to the effect of the beam, which suppresses fluctuations at small angular scales. Thus, the mid multipoles provide the optimum range to probe the ALP coupling constant as the effect from the noise term ($\mathrm{B_{\ell}^2 N_{\ell}}$) as well as from the power spectra of CMB and foregrounds are weak. 

In Fig.\ref{fig: covar}, we plot the observed and fiducial spectra along with their residual for the power spectrum of a cluster region. The profile parameters are randomly assigned to the cluster within the allowed range. The fluctuations within the signal-to-noise ratio of (Residual$\mathrm{ / \sqrt{Covariance}}$) set the cutoff for probable fluctuations, while the ALP signal is scaled to get the limiting value on ALP fluctuations in the observed power spectrum. 

\section{Results} \label{sec: results}
We perform a forecast using the Bayesian framework showing the feasibility of estimating the ALP coupling constant $\mathrm{g_{a\gamma}}$ using the residual power spectrum approach for SO and CMB-S4 instrument noise.  Simons Observatory (SO) will observe about 24000 clusters \cite{Ade_2019}. We consider the clusters up to redshift z = 1 from SO for our analysis.  For CMB-S4, we consider around 70000 clusters can be detected by CMB-S4 across 50\% of the sky up to redshift of z $\approx$ 2. Thus, we use a partial sky fraction of 50\% by masking the galactic plane to minimize the impact of foregrounds. A proportion of these clusters will be resolvable in multiple EM bands, for which we can obtain information about the electron density and magnetic profiles, as well as redshift. We simulate galaxy cluster maps at different redshifts and assign random parameters for the electron density and magnetic profiles of these clusters, within the allowed cluster range to generate random profiles. The allowed range of parameters is described in Table \ref{tab:params}. These parameters are allowed to uniformly vary in the allowed range to generate random profiles for different clusters. The clusters at different redshifts are binned in widths of $\mathrm{\Delta z = 0.05}$. The number of clusters is calculated by integrating the differential cluster abundance. We have used $\mathrm{z_0} = 0.33$, (which gives a mean redshift of z = 1) and the cluster distribution given as \cite{Ade_2019} : 
\begin{equation} \label{eq:dn_gdz}
    \mathrm{\frac{dN_g}{dz} = \frac{z^2}{2z_0^3} \exp{(-z/z_0)}}.
\end{equation}
The number of clusters in each redshift bin is shown in Table \ref{tab:dndz}. 

\begin{table}[htbp]
  \small
  \centering
  \caption{\textbf{No. of clusters in various redshift bins} }
  \label{tab:dndz}
  \subfloat{
    \begin{tabular}{|c|c|}
        \hline
         Redshift bins & No. of clusters \\
        \hline
        \hline
        0.01 $\leq$ 0.05 & 38  \\
        \hline
         0.05 $\leq$ 0.10 & 236\\
        \hline
         0.10 $\leq$ 0.15 & 556 \\
        \hline
         0.15 $\leq$ 0.20 & 933\\
        \hline
         0.20 $\leq$ 0.25 & 1324 \\
        \hline
         0.25 $\leq$ 0.30 & 1699 \\
        \hline
         0.30 $\leq$ 0.35 & 2040 \\
        \hline
         0.35 $\leq$ 0.40 & 2334 \\
        \hline
         0.40 $\leq$ 0.45 & 2577 \\
        \hline
         0.45 $\leq$ 0.50 & 2766 \\
        \hline
    \end{tabular}
    }
  \subfloat{%
    \hspace{0.5cm}%
    \begin{tabular}{|c|c|}
        \hline
         Redshift bins & No. of clusters \\
        \hline
        \hline
        0.50 $\leq$ 0.55 & 2905  \\
        \hline
         0.55 $\leq$ 0.60 & 2995\\
        \hline
         0.60 $\leq$ 0.65 & 3041\\
        \hline
         0.65 $\leq$ 0.70 & 3048\\
        \hline
         0.70 $\leq$ 0.75 & 3022 \\
        \hline
         0.75 $\leq$ 0.80 & 2968 \\
        \hline
         0.80 $\leq$ 0.85 & 2891 \\
        \hline
         0.85 $\leq$ 0.90 & 2795 \\
        \hline
         0.90 $\leq$ 0.95 & 2684 \\
        \hline
         0.95 $\leq$ 1.00 & 2563 \\
        \hline
    \end{tabular}%
    \hspace{.5cm}%
  }
  
\end{table}

We use the emcee \cite{Foreman_Mackey_2013} package for Markov Chain Monte Carlo parameter estimation. We use a Gaussian likelihood ($\mathcal{L}$) with a flat prior ($\pi$) on ALP coupling constant values from $ 10^{-14}$ to $ \mathrm{10^{-11} \, GeV^{-1}}$. The posterior probability for the coupling constant is obtained using the Bayes theorem which states:
\begin{equation}\label{eq:bayes}
\mathrm{P(\theta|Data) \propto \mathcal{L}(Data|\theta)\pi(\theta) .}\end{equation}
We scale the model ALP power spectrum (obtained in \ref{sec:makemap}), with the coupling constant as:
\begin{equation} \label{Cl_g_scaling}
\mathrm{\hat{C_l}^{ax} \propto (g_{a\gamma} / g_{a\gamma}^*)^4},
\end{equation}
to obtain constraints on the coupling constant.
The bounds on the coupling constant depend highly on the covariance of the estimated spectrum.
The coupling constant has been assumed to be zero ($\mathrm{g_{a\gamma}^{true} = 0}$) for this analysis, hence the mock data is devoid of any ALP distortion signal. This corresponds to the case of the non-existence of ALPs. 
We compare the results obtained for the cases of template matching and ILC in Appendix \ref{sec:no_temp}. 
In this section, we analyze the results from ILC cleaning.

\begin{figure}[h!]
     \centering
\includegraphics[height=6cm,width=15cm]{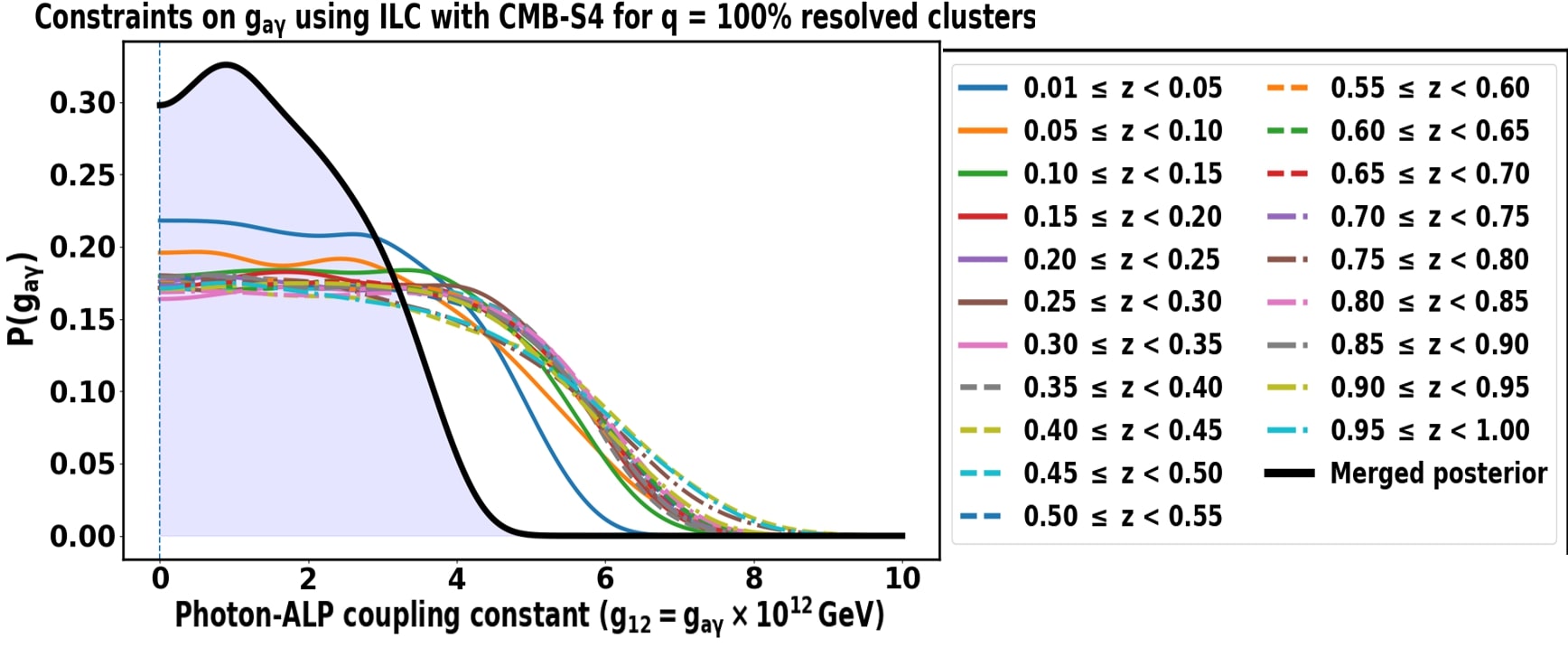}

    \caption{The constraints obtained with CMB-S4 configuration for $\mathrm{g_{a\gamma}}$ using ILC with q $= 100$\% resolved clusters for various redshift bins, with the combined posterior in bold black line. The values outside the shaded region are ruled out.}
\label{fig: ilc100cons}
\end{figure}

\begin{figure}[h!]
     \centering
\includegraphics[height=6cm,width=15cm]{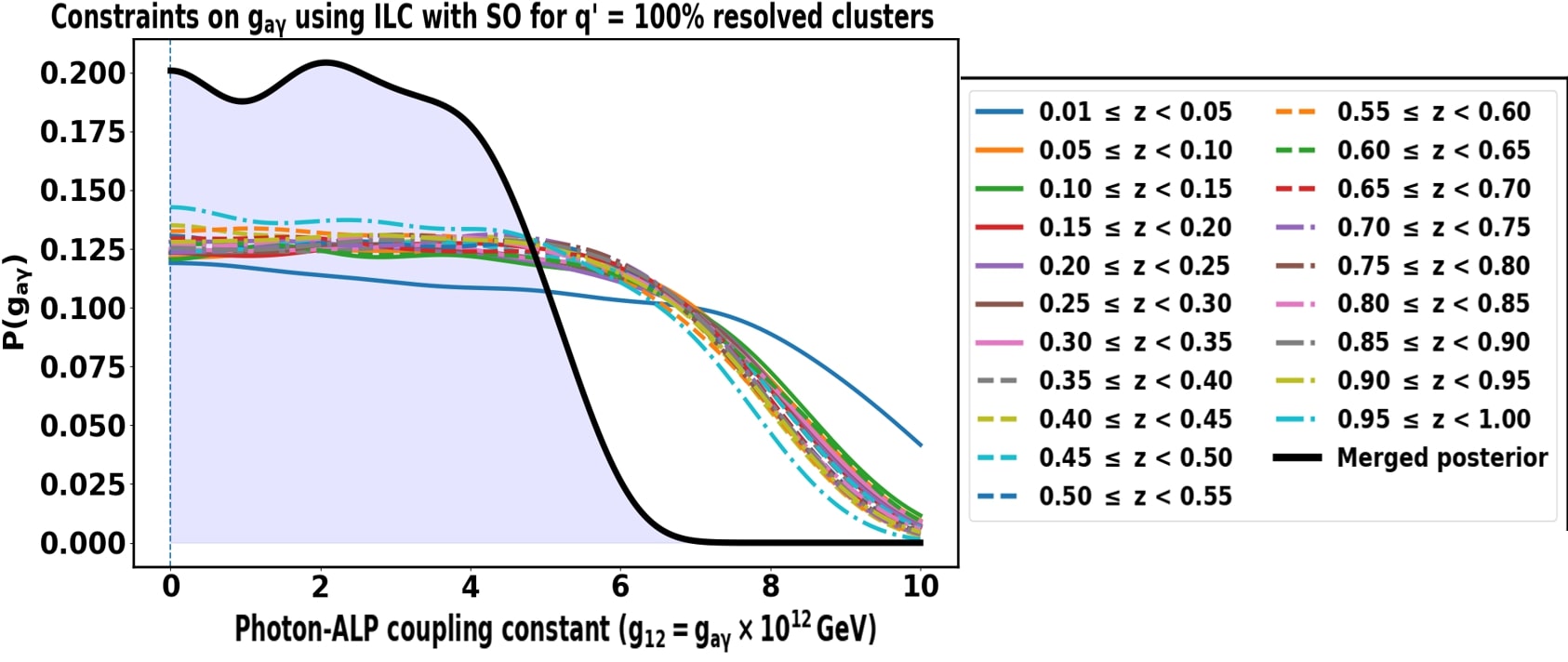}

    \caption{The constraints obtained with Simons Observatory (SO) configuration for $\mathrm{g_{a\gamma}}$ using ILC with q' = 100\% resolved clusters for various redshift bins, with the combined posterior in bold black line. The values outside the shaded region are ruled out."}
\label{fig: ilcsocons}
\end{figure} 

The Bayesian estimation is performed for a redshift bin with a window function that is one in all the cluster regions and zero elsewhere. For clusters in a redshift bin of size $\Delta$z $= 0.05$, the angular sizes of these clusters in the sky are similar. Hence, we use the minimum multipole value corresponding to the maximum cluster angular size in the redshift bin to remove any signal correlations across clusters at large angular scales or low multipoles. We use a maximum multipole value corresponding to the beam resolution i.e. $\mathrm{\ell_{max} = 3680}$ for both SO and CMB-S4 configurations.

\begin{figure}[h!]
     \centering
\includegraphics[height=6cm,width=12cm]{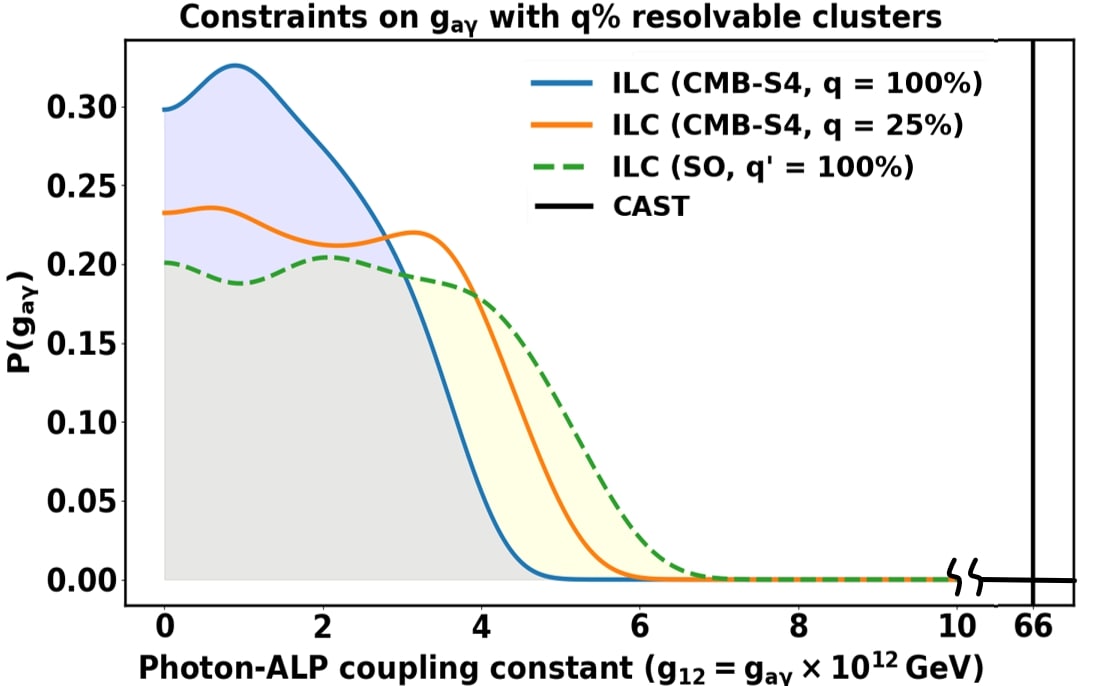}

    \caption{The combined posteriors obtained for $\mathrm{g_{a\gamma}}$ using ILC for various redshift bins. ILC is performed for different percentages of resolvable clusters as mentioned. The blue-shaded region corresponds to constraints from ILC using CMB-S4 configuration with 100\% resolvable clusters. The yellow shaded region is the constraint when ILC is applied with SO configuration. Also shown is the bound from CAST.}
\label{fig: ilcallcons}
\end{figure}
We denote the fraction of resolvable clusters with the clusters' resolvable fraction q for CMB-S4 configuration. Similarly, we use q' for SO configuration.
The posteriors with ILC for various redshifts using CMB-S4 configuration are shown in  Fig.\ref{fig: ilc100cons} when q = 100\% of the clusters up to redshift of z $= 1$ are resolvable. Similarly, with q' = 100\%, the results for Simons Observatory configuration are shown in Fig.\ref{fig: ilcsocons}. The number of clusters increases with redshift and this competes with the angular scale of the clusters to provide the constraints at these redshifts. The best constraints are obtained for the lowest redshift bin of z $= 0.01$ to z $= 0.05$ as these clusters occupy large angular scales on the sky, which in turn leads to more ALP signal fluctuations, hence providing better constraints on the coupling constant. 

The constraints at various redshifts are mainly a result of signals from bright clusters, with higher signals of ALP distortion, depending on the coupling constant values. The probability of bright clusters across a large number of clusters, as have been considered in our analysis will depend on the distribution of electron density and magnetic field profiles. If the magnetic field and electron density profiles from the clusters are not very well known, the constraints for the case of uniformly random profile parameters won't be affected much as the deviated parameters will themselves be random across a large number of clusters.
In reality, constraints will depend on the probability of occurrence of bright clusters in the cluster distribution, which itself would depend on the distribution of magnetic field and electron density profiles across various redshifts. 

The bold black line represents the combined posterior from all bins and provides the upper bounds for the ALP coupling constant. The combined posterior in the case of CMB-S4 places the upper bound on the ALP coupling constant at $\mathrm{g_{a\gamma} < 3.6 \times 10^{-12} \, GeV^{-1}}$ at 95\% confidence interval (C.I.) when ILC is used with q = 100\% of the clusters being resolvable. Similarly, SO gives the bound of $\mathrm{g_{a\gamma} < 5.2 \times 10^{-12} \, GeV^{-1}}$ at 95\% C.I for ALPs of masses $\sim 10^{-13}$ eV.

\begin{table}[h!]
    \centering
\begin{tabular}{||c c||}

\hline
        Analysis Case   & Upper bound (up to 95\% C.I.)
        \\
        \hline
        \hline
        ILC (CMB-S4, q $=$ 100\%) & $3.6\times 10^{-12} \, \mathrm{GeV^{-1}}$
        \\
        \hline
        ILC (CMB-S4, q $=$ 75\%) & $3.9\times 10^{-12} \, \mathrm{GeV^{-1}}$
        \\
        \hline 
        ILC (CMB-S4, q $=$ 50\%) & $4.2\times 10^{-12} \, \mathrm{GeV^{-1}}$
                \\
        \hline
      ILC (CMB-S4, q $=$ 25\%) & $4.6\times 10^{-12} \, \mathrm{GeV^{-1}}$
                  \\
        \hline
        ILC (CMB-S4, q $=$ 5\%) & $5.6\times 10^{-12} \, \mathrm{GeV^{-1}}$
        \\
        \hline
         ILC (SO, q' $=$ 100\%) & $5.2\times 10^{-12} \, \mathrm{GeV^{-1}}$
\\
\hline
\end{tabular}

\caption{Constraints on ALP coupling constant ($\mathrm{g_{a\gamma}}$) using ILC for different fractions of resolved clusters.}
\label{tab: gacons}
\end{table}

Next, we find constraints for different fractions of resolvable clusters up to redshift of z = 1 using ILC with CMB-S4 configuration. We consider the following percentage of resolvable clusters, with q being 5\%, 25\%, 50\%, 75\%, and 100\% (already shown in Fig.\ref{fig: ilc100cons}). The constraints weaken as the percentage of resolvable clusters decreases (see Table \ref{tab: gacons}). This is expected as the signal-to-noise ratio decreases for a low number of resolvable clusters. The combined posteriors from the cases q = 100\% and q = 25\% are shown in Fig.\ref{fig: ilcallcons}.

These results provide the upper bounds on the coupling constant for $\sim 10^{-13}$ eV mass ALPs (with 30\% tolerance) using SO and CMB-S4, if ALPs exist in nature. These bounds are much stronger than the ones from CAST ($\mathrm{6.6 \times 10^{-11} \, GeV^{-1}}$). 
The bounds will improve with detectors such as CMB-HD with a higher resolution and low noise. Information on the spectral and spatial variation of various foregrounds can further enhance the quality of constraints, as will be seen in Appendix \ref{sec:no_temp}. The constraints on the coupling constant will also depend on the mass of ALPs. The lower mass ALPs lead to stronger distortion signals (see Fig.\ref{fig: cluster depend m}). This will provide better constraints on the coupling constant as compared to high mass ALPs. Also, the broader the range over which ALPs exist, the higher will be the signal strength, leading to tighter constraints.

The clusters at higher redshifts (z > 1) may not be resolvable in different EM bands with current detectors, leading to a lack of information on their electron density and magnetic field profiles. These clusters will give rise to unresolved ALP distortion signals which will contribute to a diffused background of ALP distortion signals.  The study of ALP distortion signals from unresolved clusters is considered in a follow-up analysis \citep{mehta2024diffusedbackgroundaxionlikeparticles}. 

\section{Conclusion} \label{sec: conclusion}
In this work, we have obtained possible bounds on the photon-ALP coupling constant from CMB observations with upcoming detectors using the spectral and spatial variation of the ALP distortion signal. We have considered a power spectral approach in which the spatial fluctuations in polarization are spherical harmonics transformed to obtain their variation with angular frequencies or multipoles.  

The weak coupling of ALPs (if they exist) with photons will lead to a spectral distortion in the CMB black-body, with photons getting polarized as they pass through galaxy clusters in the presence of a transverse magnetic field (see Sec.\ref{sec:ALPs from CMB}).
These conversions will peak when the resonance condition is satisfied ($\mathrm{m_a = m_{\gamma}}$), with the probability of conversion depending on the redshift, electron density, and magnetic field profiles of galaxy clusters, the frequency of observation, ALPs mass and the ALP coupling constant (see Sec.\ref{sec:alpha_l}). This ALP distortion signal will show up as additional fluctuations in the otherwise smooth CMB, leading to a departure of the CMB power spectrum from its expected damping at high multipoles or small angular scales that galaxy clusters occupy.

We obtain bounds on the coupling constant for ALPs of masses $10^{-13}$ eV with 30\% tolerance (see Sec.\ref{sec: results}).
The bounds are stronger when the fraction of resolvable clusters increases. 
The constraints from ILC using the 4 high-frequency CMB-S4 bands place the bounds on the ALP coupling constant at 95\% confidence interval to be
$\mathrm{g_{a\gamma} < 3.6 \times 10^{-12} \, GeV^{-1}}$, when a resolvable fraction of q = 100\% is used. In case of SO, we obtain bounds of $\mathrm{g_{a\gamma} < 5.2 \times 10^{-12} \, GeV^{-1}}$ with q' = 100\%. These bounds are more than an order below the bounds from CAST ($\mathrm{g_{a\gamma} < 6.6 \times 10^{-11} \, GeV^{-1}}$). Moreover, since this method can be applied to different mass ALPs, the relation between ALP mass and coupling constant can be constrained better.

We have considered resolvable clusters up to the redshift of z $= 1$. There will be clusters resolvable in multiple bands even for z $= 1$ with improved detectors. These will improve the constraints on the coupling constant, but with the decrease in the number of clusters with redshift, accompanied by a smaller angular size on the sky, the constraints will not be much affected. Constraints on the coupling constant can also be obtained from high redshift clusters that are not well resolved in multiple bands. These will create a diffused ALP spectrum all over the sky. This will be considered in a separate work \citep{mehta2024diffusedbackgroundaxionlikeparticles}. Also, such a signal will be correlated with the large-scale structure \cite{mondino2024axioninducedpatchyscreeningcosmic}. The effect of turbulence and stochasticity in the cluster profiles have also not been considered in this work. Their effects on the distortion signal can be studied using hydrodynamical simulations like Romulus \cite{Tremmel_2017}, SIMBA \cite{Dav__2019}, etc. Also, the ellipticity of the clusters will play a role in the modelling of the ALP signal by distorting the shape of the ALP signal disk, which has not been considered.  

 The observations using future detectors such as CMB-HD, with much higher sensitivity and resolution, will enable $\sim 60\%$  better bounds on the coupling constants using the power spectral approach.  
If the foregrounds can be well modeled, we can obtain better constraints on the coupling constant. As has been shown in Appendix \ref{sec:no_temp}, better constraints can be obtained using template matching of foregrounds. Also, the shape difference and the statistical isotropy violation of foregrounds can be used to obtain better bounds on the ALP coupling constant. 

The residual power spectrum approach taken in this work will allow us to obtain better constraints on the photon-ALP coupling constants with better detectors coming up such as the SO and CMB-S4. Hence, this technique provides us with a way of obtaining the relation between coupling constants and ALP masses. The robustness of the power spectrum-based formalism, accompanied by the Bayesian approach will enable us to probe physics beyond the standard model of particle physics \cite{Marsh_2016} and string axiverse \cite{Arvanitaki_2010} from the high-resolution CMB experiments expected to be available in the near future. 
\appendix
\section{Polarized signals from galaxy clusters} 
\label{sec:szpol}

\begin{figure}[h!]
    \centering
    \includegraphics[height=7cm,width=12cm]{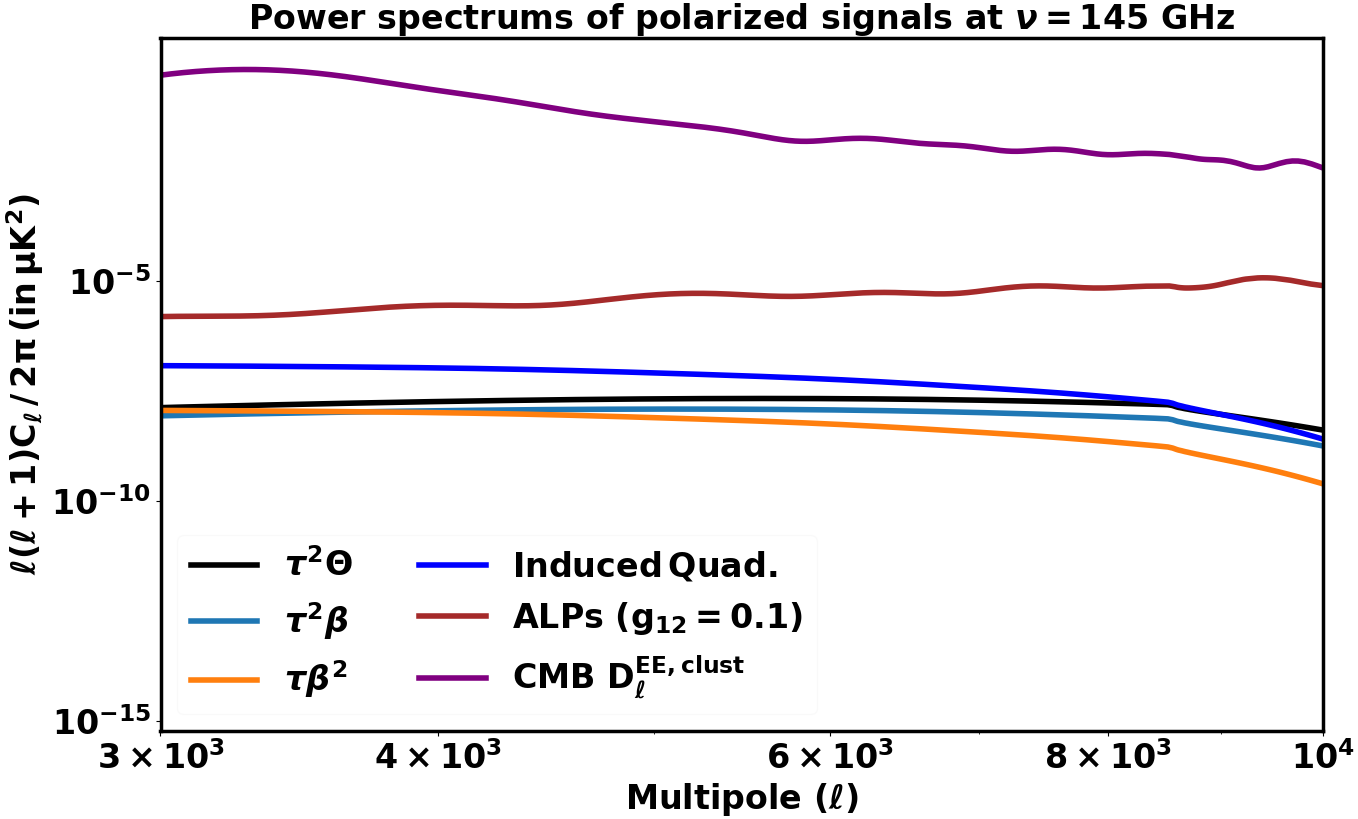}
    \caption{This figure shows the power spectrum of the ALP signal ($\mathrm{g_{a\gamma} = 10^{-13} \, GeV^{-1}}$)  against those of other polarized signals from one of the clusters. The various polarized SZ signals are difficult to separate spatially and are weak contaminants to the ALP signal. The CMB power spectrum at small scales around the cluster region is also shown. }
    \label{fig:sz_pol_all.png}
\end{figure}

\begin{figure}[h!]
    \centering
    \includegraphics[height=7cm,width=12cm]{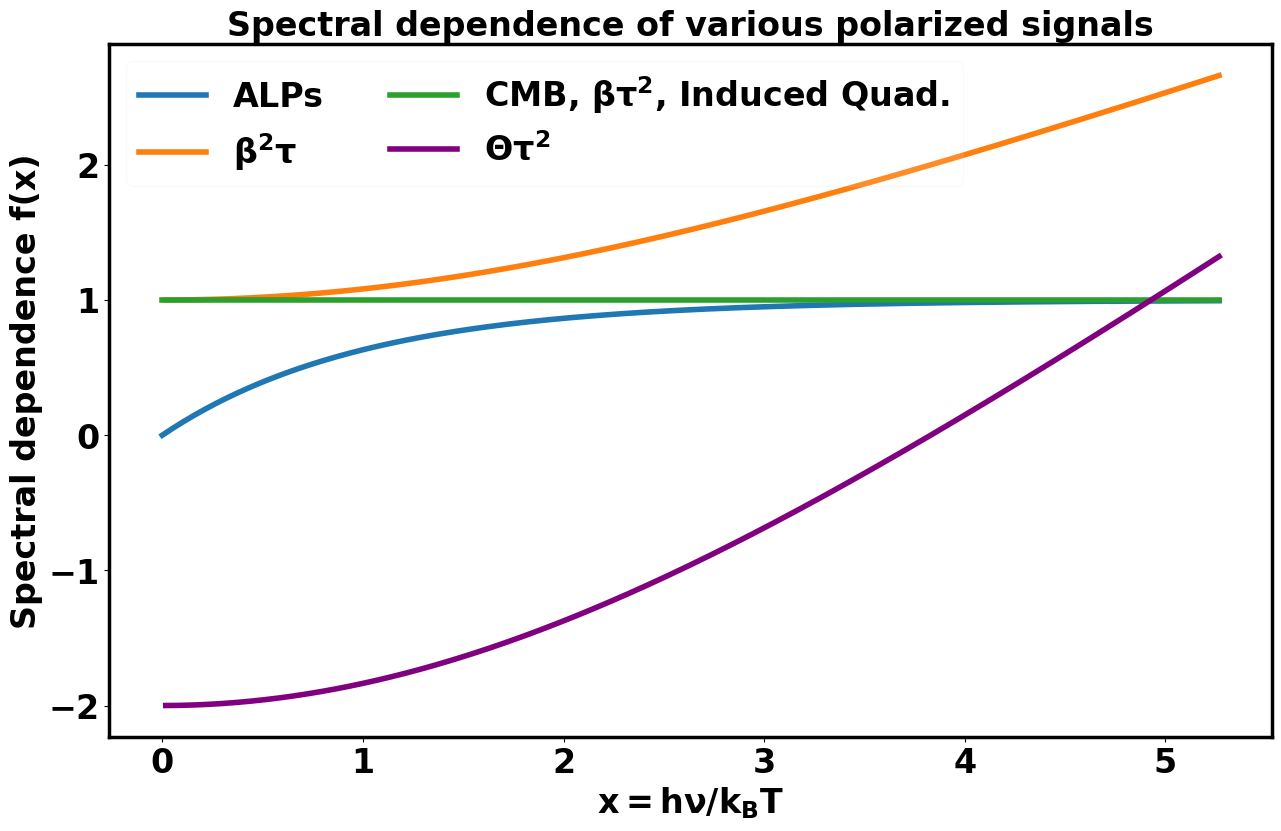}
    \caption{This figure illustrates the spectral variation of various polarized signals. The functions f(x) are mentioned in Table \ref{tab: szpol}. The distinct spectral variation of the ALP signal can distinguish it from other signals. }
    \label{fig:sz_pol_spec.png}
\end{figure}

The CMB, galactic synchrotron, and dust make the strongest contaminants to the ALP signal. 
 Among other sources of contamination, scattering by electrons in galaxy clusters will induce a polarized anisotropy due to the quadrupolar component of the local temperature distribution of CMB at the redshift of the cluster \cite{Louis_2017,Sazonov_1999,Hall_2014,aghanim2008secondary}. This contamination is proportional to the optical depth, and hence the electron density $\mathrm{n_e}$, in contrast to the ALP signal which goes as $\mathrm{n_e^{-1}}$. This signal is relevant for massive galaxy clusters and will generate anisotropies of the order of $\mathrm{10^{-4} \, to \, 10^{-2} \, \mu K}$ \cite{Birkinshaw_1999,carlstrom2002cosmology}. Also, it will follow the spectral dependence of the CMB anisotropies (independent of frequency). This will be the dominant contributor to the polarized SZ signal. There will also be higher-order SZ effects leading to polarization, caused by the cluster motion relative to the CMB and double scattering due to the finite optical depth of photons \cite{shimon2009power,challinor2000thermal,yasini2016kinetic}. The first effect will be proportional to the optical depth of photons and its degree of polarization P ($\mathrm{ = \Delta I / I}$) goes as \cite{sunyaev1980microwave}
 \begin{equation}
     \mathrm{P \propto \tau \beta_t^2 \frac{x^2 e^x(e^x + 1)}{(e^x - 1)^2}},
 \end{equation}
 where $\mathrm{x = h\nu / k_B T}$, $\tau$ is the optical depth and $\mathrm{\beta_t = v_t / c}$, with $\mathrm{v_t}$ being the transverse velocity of the cluster. The second effect will have a thermal ($\mathrm{P \propto \tau^2 k_B T_e / m_e c^2}$) or kinematic ($\mathrm{P \propto \tau^2 \beta_t }$) component, with the spectrum depending on the origin of the anisotropy \cite{aghanim2008secondary,sunyaev1980microwave,Sazonov_1999}. The spectral dependence of polarized SZ will thus be a combination of several signals. However, the spectral shape of these polarized signals is different from ALPs over the frequency band accessible from SO and CMB-S4. The spatial variation of the various polarized signals is shown in Fig.\ref{fig:sz_pol_all.png} with  $\mathrm{k_B T_e / m_e c^2}$ as $\Theta$. Spatially, the ALP signal increases at high multipoles and goes as $\ell^2$, as opposed to polarized SZ signals. At the map level, the variation of the polarized SZ signal will be different as compared to that of the ALP distortion spectrum (due to the difference in the scaling with electron density). 
 
 Spectrally, the ALP signal follows a distinctive variation with frequency and can be separated from the other polarized signals as can be seen in Fig.\ref{fig:sz_pol_spec.png}, with the spectral dependence f(x) mentioned in Table \ref{tab: szpol}. As a result, the combination of both spectral and spatial variation can be used to distinguish the ALP signal from other polarized signals. In this analysis, we have neglected the polarized SZ effect as it is not the major source of contamination to ALPs in comparison to sources such as CMB, dust, and synchrotron emission.

 \begin{table}[h!]
    \centering
\begin{tabular}{||c c||}

\hline
        Polarized Signal   & Temperature Spectral  dependence f(x)
        \\
        \hline
        \hline
        Primordial CMB anisotropy & Frequency independent ($\mathrm{x^0}$)
        \\
        \hline
        ALP distortion & $\mathrm{(\exp(x) - 1) / \exp(x)}$
        \\
        \hline 
        $\beta^2 \tau$ & $\mathrm{x(\exp(x) + 1) / 2(\exp(x) - 1) }$
                \\
        \hline
        $\beta \tau^2$ & Frequency independent ($\mathrm{x^0}$)
                  \\
        \hline
        $\Theta \tau^2$ & $\mathrm{x \coth({x/2}) - 4}$
        
        \\
        \hline
         Induced Quad. & Frequency independent ($\mathrm{x^0}$)
\\
\hline
\end{tabular}

\caption{Spectral variation of various polarized signals in temperature.}
\label{tab: szpol}
\end{table}

\section{Constraints from Template Matching of foregrounds} \label{sec:no_temp}

The constraints obtained for the case of ILC with different fractions of resolvable clusters are mentioned in Sec.\ref{sec: bayesian}. Here we consider the case of template matching of foregrounds to obtain tighter constraints using a higher beam resolution as compared to ILC.

In template matching, we assume the spectral shape of the foregrounds is known for $\mathrm{\ell > \ell_{max}}$, which corresponds to the beam resolution of the instrument.
For template matching, we fit the synchrotron power spectrum in the cluster regions at frequencies 30 and 40 GHz using Eq.\ref{eq:s3}, while we use 220 and 270 GHz frequencies to fit the dust power spectrum using Eq.\ref{eq:d3}.  We find the contribution of these foregrounds at 145 GHz, where the effect of these foregrounds is mitigated. The matched power spectrum at 145 GHz goes in the estimation of null map power spectrum from different fiducial (non-ALP) map realizations. 
 For template matching, we use the maximum multipole $\mathrm{\ell_{max} = 5728}$ corresponding to the 1.4 arcminute beam FWHM at 145 GHz. The fitted power spectrum is estimated at 145 GHz and contributes to the fiducial (non-ALPs) power spectrum. We perform the analysis for the case of 100\% resolvable clusters. The ALP distortion spectrum modeled at 145 GHz is used for the coupling constant scaling ($\sim \mathrm{g_{a\gamma}^4}$). 
 
Through template matching, we can fit the contribution of foregrounds observed power spectrum at the required frequencies (90 - 160 GHz), by modelling what it is at some other frequency. 
The dust is modeled using high-frequency maps and the low frequencies are used to model synchrotron, where their contributions are high. We have used the "s-3" model for synchrotron, for which the power spectrum can be fitted at low frequencies using:

\begin{equation}
    \mathrm{C_{\ell}^{syn}(\nu) = A_{\ell} \left( \frac{\nu}{\nu_0} \right)^{2\beta_s + 2C ln (\nu / \nu_c)}}.
    \label{eq:s3}
\end{equation}

The spectral index steepens or flattens above the frequency $\mathrm{\nu_c}$. We assume the fiducial values for the parameters with C = -0.052, $\mathrm{\nu_c = 23}$ GHz and $\mathrm{\beta_s}$ = -3.
In our analysis, we use the "s-3" model for synchrotron which considers a curved spectrum that steepens or flattens with frequency \cite{Thorne_2017}. 
 The spectrum of the thermal dust model "d-3" considers a spatially varying spectral index drawn from a Gaussian distribution. It can be fitted using the modified black-body function:
\begin{equation}
\mathrm{C_{\ell}^{dust} = A_{\ell}\nu ^{2\beta_d} B^2_{\nu}(T).
}    \label{eq:d3}
\end{equation}
We use $\mathrm{\beta_d = 1.58}$ to fit the model.
The power spectrum acquires a frequency-squared dependence.

CMB-S4 has two frequency bands above 200 GHz (at 220 and 270 GHz) which we can use to fit the dust model. 
For synchrotron, we use the 30 and 40 GHz frequencies of CMB-S4. This lets us account for the effect of foregrounds at the observation frequency of 145 GHz, which is the matched frequency.
 
\begin{figure}[h!]
    \centering
    \includegraphics[height=6cm,width=15cm]{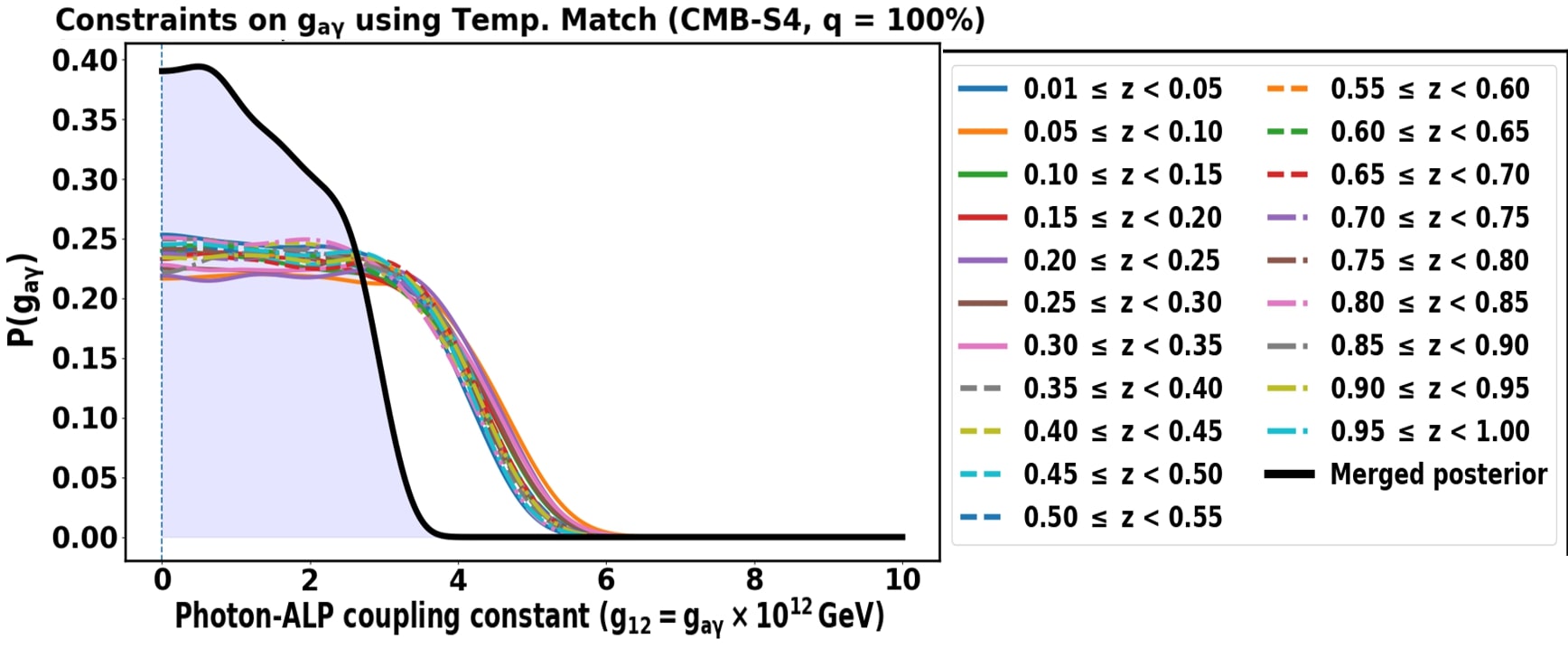}
    \caption{The constraints obtained for $\mathrm{g_{a\gamma}}$ for various redshift bins when Template matching of foregrounds is done.}
    \label{fig:g_temp}
\end{figure}

\begin{figure}[h!]
     \centering
\includegraphics[height=6cm,width=12cm]{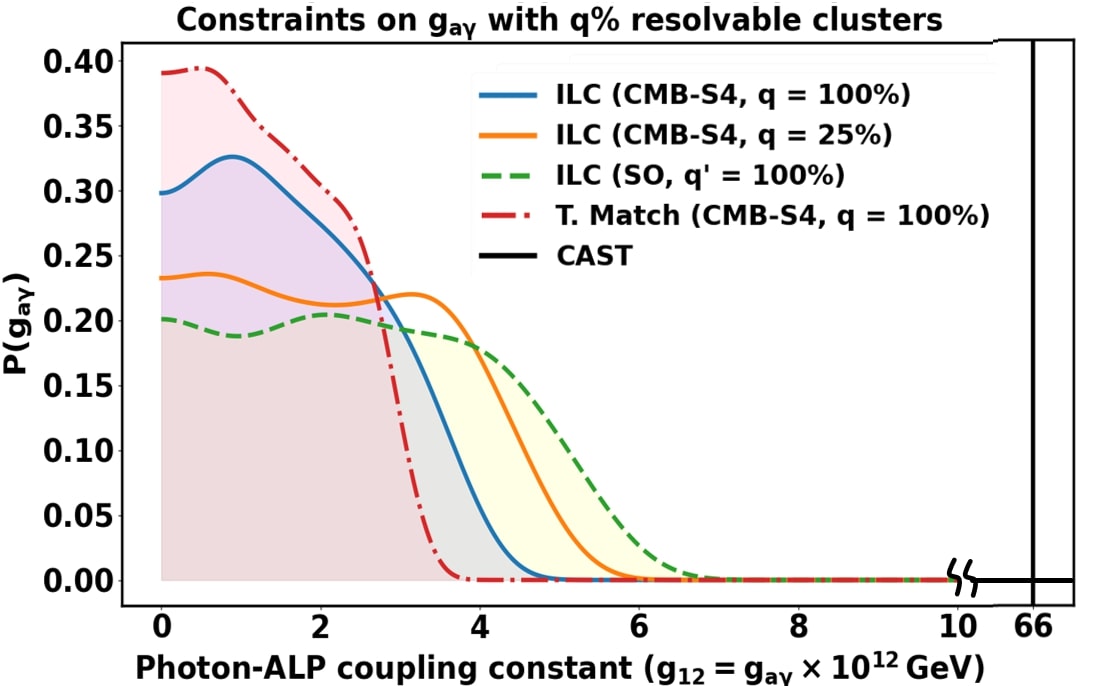}

    \caption{The posteriors obtained for $\mathrm{g_{a\gamma}}$ using ILC and Template matching for various redshift bins. ILC is performed for different percentages of resolvable clusters as mentioned. Template matching is performed for 100\% resolvable clusters case. The blue-shaded region corresponds to constraints from ILC by combining all posteriors.  The yellow shaded region is the constraint when ILC is applied with SO configuration. Also shown is the bound from CAST."}
\label{fig: tempallcons}
\end{figure}

The constraints for various redshift bins are shown in Fig.\ref{fig:g_temp}. As in the case of ILC, the angular size of the clusters and their number in a certain redshift bin compete to provide the constraints. Here again, the best constraints are provided by the redshift bin z $= 0.01$ to z $= 0.05$ due to the large angular size of these clusters in the sky.
The bold line represents the combined posterior and provides the upper bound on the coupling constant. 
We infer that template matching provides tighter results giving the bound on coupling constant as $\mathrm{g_{a\gamma} < 2.9 \times 10^{-12} \, GeV^{-1}}$ at 95\% confidence interval. 

Template matching provides about $20\%$  tighter constraints of as compared to ILC (see Table \ref{tab: gacons} and Fig.\ref{fig: tempallcons}) due to a higher beam resolution of 1.4 arcmin being used. 
The difference between the two cleaning methods comes from the beam resolution of the experiment which is 1.4 arcminutes for template matching at 145 GHz, while 2.2 arcminutes for ILC. The two methods assume the shape of the foreground power spectrum at different frequencies for $\mathrm{\ell > \ell_{max}}$ based on its shape at $\mathrm{\ell < \ell_{max}}$. Thus, template matching with a lower beam size ($\mathrm{\ell_{max} = 5782}$) provides better results as compared to ILC ($\mathrm{\ell_{max} = 3680}$). ILC is more reliable than template matching though, as there is no preassumed frequency-dependent model that is used to fit the foregrounds at various frequencies. ILC  combines data from multiple maps together to obtain a cleaned map, while template matching  assumes a frequency dependent model for foregrounds, uses data from different frequencies and estimates the contribution of foregrounds at the matched frequency. If the template matching is not precise enough, it will lead to bias in the coupling constant constraints.

\acknowledgments
    This work is a part of the $\langle \texttt{data|theory}\rangle$ \texttt{Universe-Lab}, supported by the TIFR  and the Department of Atomic Energy, Government of India. The authors express their gratitude to the TIFR CCHPC facility for meeting the computational needs. Furthermore, we would also like to thank the Simons Observatory, CMB-S4, and CMB-HD collaborations for providing the instrument noise and beam resolutions. 
 Also, the following packages were used for this work: Astropy \cite{astropy:2013,astropy:2022,astropy:2018}, NumPy \cite{harris2020array}
CAMB \cite{2011ascl.soft02026L}, emcee \cite{Foreman_Mackey_2013}, SciPy \cite{2020SciPy-NMeth}, SymPy \cite{10.7717/peerj-cs.103}, Matplotlib \cite{Hunter:2007}, HEALPix (Hierarchical Equal Area isoLatitude Pixelation of a sphere)\footnote{Link to the HEALPix website http://healpix.sf.net}\cite{2005ApJ...622..759G,Zonca2019} and PySM \cite{Thorne_2017}.

\bibliography{references.bib}

\providecommand{\href}[2]{#2}\begingroup\raggedright\begin{thebibliography}{10}

\bibitem{dine1983not}
M.~Dine and W.~Fischler, \emph{The not-so-harmless axion}, {\emph{Physics Letters B} {\bfseries 120} (1983) 137}.

\bibitem{abbott1983cosmological}
L.F.~Abbott and P.~Sikivie, \emph{A cosmological bound on the invisible axion}, {\emph{Physics Letters B} {\bfseries 120} (1983) 133}.

\bibitem{preskill1983cosmology}
J.~Preskill, M.B.~Wise and F.~Wilczek, \emph{Cosmology of the invisible axion}, {\emph{Physics Letters B} {\bfseries 120} (1983) 127}.

\bibitem{Ghosh:2022rta}
A.~Ghosh, P.~Konar and R.~Roshan, \emph{{Top-philic dark matter in a hybrid KSVZ axion framework}}, \href{https://doi.org/10.1007/JHEP12(2022)167}{\emph{JHEP} {\bfseries 12} (2022) 167} [\href{https://arxiv.org/abs/2207.00487}{{\ttfamily 2207.00487}}].

\bibitem{1992SvJNP..55.1063B}
Z.G.~{Berezhiani}, A.S.~{Sakharov} and M.Y.~{Khlopov}, \emph{{Primordial background of cosmological axions.}}, {\emph{Soviet Journal of Nuclear Physics} {\bfseries 55} (1992) 1063}.

\bibitem{khlopov1999nonlinear}
M.Y.~Khlopov, A.~Sakharov and D.~Sokoloff, \emph{The nonlinear modulation of the density distribution in standard axionic cdm and its cosmological impact}, {\emph{Nuclear Physics B-Proceedings Supplements} {\bfseries 72} (1999) 105}.

\bibitem{sakharov1994nonhomogeneity}
A.~Sakharov and M.Y.~Khlopov, \emph{The nonhomogeneity problem for the primordial axion field}, {\emph{Physics of Atomic Nuclei} {\bfseries 57} (1994) 485}.

\bibitem{sakharov1996large}
A.~Sakharov, D.~Sokoloff and M.Y.~Khlopov, \emph{Large-scale modulation of the distribution of coherent oscillations of a primordial axion field in the universe}, {\emph{Physics of Atomic Nuclei} {\bfseries 59} (1996) }.

\bibitem{Carosi:2013rla}
G.~Carosi, A.~Friedland, M.~Giannotti, M.J.~Pivovaroff, J.~Ruz and J.K.~Vogel, \emph{{Probing the axion-photon coupling: phenomenological and experimental perspectives. A snowmass white paper}},  in \emph{{Snowmass 2013}: {Snowmass on the Mississippi}}, 9, 2013 [\href{https://arxiv.org/abs/1309.7035}{{\ttfamily 1309.7035}}].

\bibitem{Mukherjee_2020}
S.~Mukherjee, D.N.~Spergel, R.~Khatri and B.D.~Wandelt, \emph{A new probe of axion-like particles: Cmb polarization distortions due to cluster magnetic fields}, \href{https://doi.org/10.1088/1475-7516/2020/02/032}{\emph{Journal of Cosmology and Astroparticle Physics} {\bfseries 2020} (2020) 032–032}.

\bibitem{Ghosh:2023xhs}
A.~Ghosh and P.~Konar, \emph{{Precision prediction at the LHC of a democratic up-family philic KSVZ axion model}},  \href{https://arxiv.org/abs/2305.08662}{{\ttfamily 2305.08662}}.

\bibitem{Fixsen_2009}
D.J.~Fixsen, \emph{The temperature of the cosmic microwave background}, \href{https://doi.org/10.1088/0004-637x/707/2/916}{\emph{The Astrophysical Journal} {\bfseries 707} (2009) 916–920}.

\bibitem{Bennett_2013}
C.L.~Bennett, D.~Larson, J.L.~Weiland, N.~Jarosik, G.~Hinshaw, N.~Odegard et~al., \emph{Nine-year wilkinson microwave anisotropy probe ( wmap ) observations: Final maps and results}, \href{https://doi.org/10.1088/0067-0049/208/2/20}{\emph{The Astrophysical Journal Supplement Series} {\bfseries 208} (2013) 20}.

\bibitem{2020}
N.~Aghanim, Y.~Akrami, F.~Arroja, M.~Ashdown, J.~Aumont, C.~Baccigalupi et~al., \emph{Planck2018 results: I. overview and the cosmological legacy ofplanck}, \href{https://doi.org/10.1051/0004-6361/201833880}{\emph{Astronomy \&; Astrophysics} {\bfseries 641} (2020) A1}.

\bibitem{Das_2011}
S.~Das, T.A.~Marriage, P.A.R.~Ade, P.~Aguirre, M.~Amiri, J.W.~Appel et~al., \emph{The atacama cosmology telescope: A measurement of the cosmic microwave background power spectrum at 148 and 218 ghz from the 2008 southern survey}, \href{https://doi.org/10.1088/0004-637x/729/1/62}{\emph{The Astrophysical Journal} {\bfseries 729} (2011) 62}.

\bibitem{benson2014spt}
B.A.~Benson, P.~Ade, Z.~Ahmed, S.~Allen, K.~Arnold, J.~Austermann et~al., \emph{Spt-3g: a next-generation cosmic microwave background polarization experiment on the south pole telescope},  in \emph{Millimeter, Submillimeter, and Far-Infrared Detectors and Instrumentation for Astronomy VII}, vol.~9153, pp.~552--572, SPIE, 2014.

\bibitem{2020_planck}
Y.~Akrami, M.~Ashdown, J.~Aumont, C.~Baccigalupi, M.~Ballardini, A.J.~Banday et~al., \emph{Planck2018 results: Vii. isotropy and statistics of the cmb}, \href{https://doi.org/10.1051/0004-6361/201935201}{\emph{Astronomy \&; Astrophysics} {\bfseries 641} (2020) A7}.

\bibitem{Austermann_2012}
J.E.~Austermann, K.A.~Aird, J.A.~Beall, D.~Becker, A.~Bender, B.A.~Benson et~al., \emph{Sptpol: an instrument for cmb polarization measurements with the south pole telescope},  in \emph{Millimeter, Submillimeter, and Far-Infrared Detectors and Instrumentation for Astronomy VI}, W.S.~Holland, ed., SPIE, Sept., 2012, \href{https://doi.org/10.1117/12.927286}{DOI}.

\bibitem{Dodelson:2003ft}
S.~Dodelson, \emph{{Modern Cosmology}}, Academic Press, Amsterdam (2003).

\bibitem{Hu_2002}
W.~Hu and S.~Dodelson, \emph{Cosmic microwave background anisotropies}, \href{https://doi.org/10.1146/annurev.astro.40.060401.093926}{\emph{Annual Review of Astronomy and Astrophysics} {\bfseries 40} (2002) 171–216}.

\bibitem{Smith_2007}
K.M.~Smith, O.~Zahn and O.~Doré, \emph{Detection of gravitational lensing in the cosmic microwave background}, \href{https://doi.org/10.1103/physrevd.76.043510}{\emph{Physical Review D} {\bfseries 76} (2007) }.

\bibitem{1972CoASP...4..173S}
R.A.~{Sunyaev} and Y.B.~{Zeldovich}, \emph{{The Observations of Relic Radiation as a Test of the Nature of X-Ray Radiation from the Clusters of Galaxies}}, {\emph{Comments on Astrophysics and Space Physics} {\bfseries 4} (1972) 173}.

\bibitem{adam2016planck}
R.~Adam, N.~Aghanim, M.~Ashdown, J.~Aumont, C.~Baccigalupi, M.~Ballardini et~al., \emph{Planck intermediate results-xlvii. planck constraints on reionization history}, {\emph{Astronomy \& Astrophysics} {\bfseries 596} (2016) A108}.

\bibitem{PhysRevD.104.022003}
{\scshape SPT-3G Collaboration} collaboration, \emph{Measurements of the $e$-mode polarization and temperature-$e$-mode correlation of the cmb from spt-3g 2018 data}, \href{https://doi.org/10.1103/PhysRevD.104.022003}{\emph{Phys. Rev. D} {\bfseries 104} (2021) 022003}.

\bibitem{osti_22525054}
M.~Schlederer and G.~Sigl, \emph{Constraining alp-photon coupling using galaxy clusters}, \href{https://doi.org/10.1088/1475-7516/2016/01/038}{\emph{Journal of Cosmology and Astroparticle Physics} {\bfseries 2016} (2016) }.

\bibitem{Mukherjee_2018}
S.~Mukherjee, R.~Khatri and B.D.~Wandelt, \emph{Polarized anisotropic spectral distortions of the cmb: galactic and extragalactic constraints on photon-axion conversion}, \href{https://doi.org/10.1088/1475-7516/2018/04/045}{\emph{Journal of Cosmology and Astroparticle Physics} {\bfseries 2018} (2018) 045–045}.

\bibitem{Mukherjee:2021jgv}
S.~Mukherjee, \emph{{Discovering Axion-Like Particles Using Cosmic Microwave Background as the Backlight}}, \href{https://doi.org/10.1134/S1063772921100243}{\emph{Astron. Rep.} {\bfseries 65} (2021) 995}.

\bibitem{2014ApJS..210....9B}
M.~{Bilicki}, T.H.~{Jarrett}, J.A.~{Peacock}, M.E.~{Cluver} and L.~{Steward}, \emph{{Two Micron All Sky Survey Photometric Redshift Catalog: A Comprehensive Three-dimensional Census of the Whole Sky}}, \href{https://doi.org/10.1088/0067-0049/210/1/9}{\emph{\apjs} {\bfseries 210} (2014) 9} [\href{https://arxiv.org/abs/1311.5246}{{\ttfamily 1311.5246}}].

\bibitem{Birkinshaw_1999}
M.~Birkinshaw, \emph{The sunyaev–zel’dovich effect}, \href{https://doi.org/10.1016/s0370-1573(98)00080-5}{\emph{Physics Reports} {\bfseries 310} (1999) 97–195}.

\bibitem{GOVONI_2004}
F.~GOVONI and L.~FERETTI, \emph{Magnetic fields in clusters of galaxies}, \href{https://doi.org/10.1142/s0218271804005080}{\emph{International Journal of Modern Physics D} {\bfseries 13} (2004) 1549–1594}.

\bibitem{Mehta:2024:new3}
H.~Mehta and S.~Mukherjee, \emph{Spectrax: A new multi-band data analysis framework to search for axion-like particles using galaxy clusters}, {\emph{(under preparation)} (2024) }.

\bibitem{Eriksen_2004}
H.K.~Eriksen, A.J.~Banday, K.M.~Gorski and P.B.~Lilje, \emph{On foreground removal from thewilkinson microwave anisotropy probedata by an internal linear combination method: Limitations and implications}, \href{https://doi.org/10.1086/422807}{\emph{The Astrophysical Journal} {\bfseries 612} }.

\bibitem{ilc2008internal}
R.~Vio and P.~Andreani, \emph{"internal linear combination" method for the separation of cmb from galactic foregrounds in the harmonic domain},  2008.

\bibitem{Hansen_2006}
F.K.~Hansen, A.J.~Banday, H.K.~Eriksen, K.M.~Gorski and P.B.~Lilje, \emph{Foreground subtraction of cosmic microwave background maps using wi‐fit (wavelet‐based high‐resolution fitting of internal templates)}, \href{https://doi.org/10.1086/506015}{\emph{The Astrophysical Journal} {\bfseries 648} (2006) 784–796}.

\bibitem{Ade_2019}
P.~Ade, J.~Aguirre and Z.e.a.~Ahmed, \emph{The simons observatory: science goals and forecasts}, \href{https://doi.org/10.1088/1475-7516/2019/02/056}{\emph{Journal of Cosmology and Astroparticle Physics} {\bfseries 2019} (2019) 056–056}.

\bibitem{abazajian2016cmbs4}
K.N.~Abazajian, P.~Adshead, Z.~Ahmed and S.W.A.~et~al., \emph{Cmb-s4 science book, first edition},  2016.

\bibitem{2016}
P.A.R.~Ade, N.~Aghanim, M.~Arnaud, M.~Ashdown and J.e.a.~Aumont, \emph{Planck2015 results: Xiii. cosmological parameters}, \href{https://doi.org/10.1051/0004-6361/201525830}{\emph{Astronomy \&; Astrophysics} {\bfseries 594} (2016) A13}.

\bibitem{Carilli_2004}
C.~Carilli and S.~Rawlings, \emph{Motivation, key science projects, standards and assumptions}, \href{https://doi.org/10.1016/j.newar.2004.09.001}{\emph{New Astronomy Reviews} {\bfseries 48} (2004) 979–984}.

\bibitem{merloni2012erosita}
A.~Merloni, P.~Predehl, W.~Becker, H.~Böhringer, T.~Boller and H.B.~et~al., \emph{erosita science book: Mapping the structure of the energetic universe},  2012.

\bibitem{Gardner_2006}
J.P.~Gardner, J.C.~Mather, M.~Clampin, R.~Doyon, M.A.~Greenhouse, H.B.~Hammel et~al., \emph{The james webb space telescope}, \href{https://doi.org/10.1007/s11214-006-8315-7}{\emph{Space Science Reviews} {\bfseries 123} (2006) 485–606}.

\bibitem{2014A&A...571A..21P}
{Planck Collaboration}, P.A.R.~{Ade}, N.~{Aghanim} and C.e.a.~{Armitage-Caplan}\href{https://doi.org/10.1051/0004-6361/201321522}{\emph{\aap} {\bfseries 571} (2014) A21} [\href{https://arxiv.org/abs/1303.5081}{{\ttfamily 1303.5081}}].

\bibitem{2014PTEP.2014fB107T}
H.~{Tashiro}, \emph{{CMB spectral distortions and energy release in the early universe}}, \href{https://doi.org/10.1093/ptep/ptu066}{\emph{Progress of Theoretical and Experimental Physics} {\bfseries 2014} (2014) 06B107}.

\bibitem{Fixsen_1996}
D.J.~Fixsen, E.S.~Cheng, J.M.~Gales, J.C.~Mather, R.A.~Shafer and E.L.~Wright, \emph{The cosmic microwave background spectrum from the fullcobefiras data set}, \href{https://doi.org/10.1086/178173}{\emph{The Astrophysical Journal} {\bfseries 473} (1996) 576–587}.

\bibitem{berezhiani1991cosmology}
Z.~Berezhiani and M.Y.~Khlopov, \emph{Cosmology of spontaneously broken gauge family symmetry with axion solution of strong cp-problem}, {\emph{Zeitschrift f{\"u}r Physik C Particles and Fields} {\bfseries 49} (1991) 73}.

\bibitem{Raffelt:1996wa}
G.G.~Raffelt, \emph{{Stars as laboratories for fundamental physics}: {The astrophysics of neutrinos, axions, and other weakly interacting particles}} (5, 1996).

\bibitem{Mirizzi_2009}
A.~Mirizzi, J.~Redondo and G.~Sigl, \emph{Constraining resonant photon-axion conversions in the early universe}, \href{https://doi.org/10.1088/1475-7516/2009/08/001}{\emph{Journal of Cosmology and Astroparticle Physics} {\bfseries 2009} (2009) 001–001}.

\bibitem{2017}
\emph{New cast limit on the axion–photon interaction}, \href{https://doi.org/10.1038/nphys4109}{\emph{Nature Physics} {\bfseries 13} (2017) 584–590}.

\bibitem{Mukherjee_2019}
S.~Mukherjee, R.~Khatri and B.D.~Wandelt, \emph{Constraints on non-resonant photon-axion conversion from the planck satellite data}, \href{https://doi.org/10.1088/1475-7516/2019/06/031}{\emph{Journal of Cosmology and Astroparticle Physics} {\bfseries 2019} (2019) 031–031}.

\bibitem{Staniszewski_2009}
Z.~Staniszewski, P.A.R.~Ade, K.A.~Aird, B.A.~Benson, L.E.~Bleem, J.E.~Carlstrom et~al., \emph{Galaxy clusters discovered with a sunyaev-zel’dovich effect survey}, \href{https://doi.org/10.1088/0004-637x/701/1/32}{\emph{The Astrophysical Journal} {\bfseries 701} (2009) 32–41}.

\bibitem{carilli2002cluster}
C.~Carilli and G.~Taylor, \emph{Cluster magnetic fields}, {\emph{Annual Review of Astronomy and Astrophysics} {\bfseries 40} (2002) 319}.

\bibitem{bonafede2010galaxy}
A.~Bonafede, L.~Feretti, M.~Murgia, F.~Govoni, G.~Giovannini and V.~Vacca, \emph{Galaxy cluster magnetic fields from radio polarized emission}, {\emph{arXiv preprint arXiv:1009.1233} (2010) }.

\bibitem{cavaliere1978distribution}
A.~Cavaliere and R.~Fusco-Femiano, \emph{The distribution of hot gas in clusters of galaxies}, {\emph{Astronomy and Astrophysics, Vol. 70, p. 677 (1978)} {\bfseries 70} (1978) 677}.

\bibitem{sarazin1986x}
C.L.~Sarazin, \emph{X-ray emission from clusters of galaxies}, {\emph{Reviews of Modern Physics} {\bfseries 58} (1986) 1}.

\bibitem{yee2001optical}
H.~Yee and M.~Gladders, \emph{Optical surveys for galaxy clusters}, {\emph{arXiv preprint astro-ph/0111431} (2001) }.

\bibitem{zehavi2011galaxy}
I.~Zehavi, Z.~Zheng, D.H.~Weinberg, M.R.~Blanton, N.A.~Bahcall, A.A.~Berlind et~al., \emph{Galaxy clustering in the completed sdss redshift survey: the dependence on color and luminosity}, {\emph{The Astrophysical Journal} {\bfseries 736} (2011) 59}.

\bibitem{burenin2018optical}
R.~Burenin, I.~Bikmaev, I.~Khamitov, I.~Zaznobin, G.~Khorunzhev, M.~Eselevich et~al., \emph{Optical identifications of high-redshift galaxy clusters from the planck sunyaev--zeldovich survey}, {\emph{Astronomy Letters} {\bfseries 44} (2018) 297}.

\bibitem{Vikhlinin_2006}
A.~Vikhlinin, A.~Kravtsov, W.~Forman, C.~Jones, M.~Markevitch, S.S.~Murray et~al., \emph{Chandrasample of nearby relaxed galaxy clusters: Mass, gas fraction, and mass‐temperature relation}, \href{https://doi.org/10.1086/500288}{\emph{The Astrophysical Journal} {\bfseries 640} (2006) 691–709}.

\bibitem{mcdonald2013growth}
M.~McDonald, B.~Benson, A.~Vikhlinin, B.~Stalder, L.~Bleem, T.~De~Haan et~al., \emph{The growth of cool cores and evolution of cooling properties in a sample of 83 galaxy clusters at 0.3< z< 1.2 selected from the spt-sz survey}, {\emph{The Astrophysical Journal} {\bfseries 774} (2013) 23}.

\bibitem{bohringer2016cosmic}
H.~B{\"o}hringer, G.~Chon and P.P.~Kronberg, \emph{The cosmic large-scale structure in x-rays (classix) cluster survey-i. probing galaxy cluster magnetic fields with line of sight rotation measures}, {\emph{Astronomy \& Astrophysics} {\bfseries 596} (2016) A22}.

\bibitem{2005ApJ...622..759G}
K.M.~{G{\'o}rski}, E.~{Hivon}, A.J.~{Banday}, B.D.~{Wandelt}, F.K.~{Hansen}, M.~{Reinecke} et~al., \emph{{HEALPix: A Framework for High-Resolution Discretization and Fast Analysis of Data Distributed on the Sphere}}, \href{https://doi.org/10.1086/427976}{\emph{\apj} {\bfseries 622} (2005) 759} [\href{https://arxiv.org/abs/arXiv:astro-ph/0409513}{{\ttfamily arXiv:astro-ph/0409513}}].

\bibitem{Zonca2019}
A.~Zonca, L.~Singer, D.~Lenz, M.~Reinecke, C.~Rosset, E.~Hivon et~al., \emph{healpy: equal area pixelization and spherical harmonics transforms for data on the sphere in python}, \href{https://doi.org/10.21105/joss.01298}{\emph{Journal of Open Source Software} {\bfseries 4} (2019) 1298}.

\bibitem{mehta2024diffusedbackgroundaxionlikeparticles}
H.~Mehta and S.~Mukherjee, \emph{A diffused background from axion-like particles in the microwave sky},  2024.

\bibitem{Hivon_2002}
E.~Hivon, K.M.~Gorski, C.B.~Netterfield, B.P.~Crill, S.~Prunet and F.~Hansen, \emph{Master of the cosmic microwave background anisotropy power spectrum: A fast method for statistical analysis of large and complex cosmic microwave background data sets}, \href{https://doi.org/10.1086/338126}{\emph{The Astrophysical Journal} {\bfseries 567} (2002) 2–17}.

\bibitem{Thorne_2017}
B.~Thorne, J.~Dunkley, D.~Alonso and S.~Næss, \emph{The python sky model: software for simulating the galactic microwave sky}, \href{https://doi.org/10.1093/mnras/stx949}{\emph{Monthly Notices of the Royal Astronomical Society} {\bfseries 469} (2017) 2821–2833}.

\bibitem{1986rpa..book.....R}
G.B.~{Rybicki} and A.P.~{Lightman}, \emph{{Radiative Processes in Astrophysics}} (1986).

\bibitem{Foreman_Mackey_2013}
D.~Foreman-Mackey, D.W.~Hogg, D.~Lang and J.~Goodman, \emph{<tt>emcee</tt>: The mcmc hammer}, \href{https://doi.org/10.1086/670067}{\emph{Publications of the Astronomical Society of the Pacific} {\bfseries 125} (2013) 306–312}.

\bibitem{mondino2024axioninducedpatchyscreeningcosmic}
C.~Mondino, D.~Pîrvu, J.~Huang and M.C.~Johnson, \emph{Axion-induced patchy screening of the cosmic microwave background},  2024.

\bibitem{Tremmel_2017}
M.~Tremmel, M.~Karcher, F.~Governato, M.~Volonteri, T.R.~Quinn, A.~Pontzen et~al., \emph{The romulus cosmological simulations: a physical approach to the formation, dynamics and accretion models of smbhs}, \href{https://doi.org/10.1093/mnras/stx1160}{\emph{Monthly Notices of the Royal Astronomical Society} {\bfseries 470} (2017) 1121–1139}.

\bibitem{Dav__2019}
R.~Davé, D.~Anglés-Alcázar, D.~Narayanan, Q.~Li, M.H.~Rafieferantsoa and S.~Appleby, \emph{simba: Cosmological simulations with black hole growth and feedback}, \href{https://doi.org/10.1093/mnras/stz937}{\emph{Monthly Notices of the Royal Astronomical Society} {\bfseries 486} (2019) 2827–2849}.

\bibitem{Marsh_2016}
D.J.~Marsh, \emph{Axion cosmology}, \href{https://doi.org/10.1016/j.physrep.2016.06.005}{\emph{Physics Reports} {\bfseries 643} (2016) 1–79}.

\bibitem{Arvanitaki_2010}
A.~Arvanitaki, S.~Dimopoulos, S.~Dubovsky, N.~Kaloper and J.~March-Russell, \emph{String axiverse}, \href{https://doi.org/10.1103/physrevd.81.123530}{\emph{Physical Review D} {\bfseries 81} (2010) }.

\bibitem{Louis_2017}
T.~Louis, E.F.~Bunn, B.~Wandelt and J.~Silk, \emph{Measuring polarized emission in clusters in the cmb s4 era}, \href{https://doi.org/10.1103/physrevd.96.123509}{\emph{Physical Review D} {\bfseries 96} (2017) }.

\bibitem{Sazonov_1999}
S.Y.~Sazonov and R.A.~Sunyaev, \emph{Microwave polarization in the direction of galaxy clusters induced by the cmb quadrupole anisotropy}, \href{https://doi.org/10.1046/j.1365-8711.1999.02981.x}{\emph{Monthly Notices of the Royal Astronomical Society} {\bfseries 310} (1999) 765–772}.

\bibitem{Hall_2014}
A.~Hall and A.~Challinor, \emph{Detecting the polarization induced by scattering of the microwave background quadrupole in galaxy clusters}, \href{https://doi.org/10.1103/physrevd.90.063518}{\emph{Physical Review D} {\bfseries 90} (2014) }.

\bibitem{aghanim2008secondary}
N.~Aghanim, S.~Majumdar and J.~Silk, \emph{Secondary anisotropies of the cmb}, {\emph{Reports on Progress in Physics} {\bfseries 71} (2008) 066902}.

\bibitem{carlstrom2002cosmology}
J.E.~Carlstrom, G.P.~Holder and E.D.~Reese, \emph{Cosmology with the sunyaev-zel’dovich effect}, {\emph{Annual Review of Astronomy and Astrophysics} {\bfseries 40} (2002) 643}.

\bibitem{shimon2009power}
M.~Shimon, Y.~Rephaeli, S.~Sadeh and B.~Keating, \emph{Power spectra of cmb polarization by scattering in clusters}, {\emph{Monthly Notices of the Royal Astronomical Society} {\bfseries 399} (2009) 2088}.

\bibitem{challinor2000thermal}
A.~Challinor, M.~Ford and A.~Lasenby, \emph{Thermal and kinematic corrections to the microwave background polarization induced by galaxy clusters along the line of sight}, {\emph{Monthly Notices of the Royal Astronomical Society} {\bfseries 312} (2000) 159}.

\bibitem{yasini2016kinetic}
S.~Yasini and E.~Pierpaoli, \emph{Kinetic sunyaev-zeldovich effect in an anisotropic cmb model: Measuring low multipoles of the cmb at higher redshifts using intensity and polarization spectral distortions}, {\emph{Physical Review D} {\bfseries 94} (2016) 023513}.

\bibitem{sunyaev1980microwave}
R.~Sunyaev and I.B.~Zeldovich, \emph{Microwave background radiation as a probe of the contemporary structure and history of the universe}, {\emph{In: Annual review of astronomy and astrophysics. Volume 18.(A81-20334 07-90) Palo Alto, Calif., Annual Reviews, Inc., 1980, p. 537-560.} {\bfseries 18} (1980) 537}.

\bibitem{astropy:2013}
{Astropy Collaboration}, T.P.~{Robitaille}, E.J.~{Tollerud}, P.~{Greenfield}, M.~{Droettboom} and E.e.a.~{Bray}, \emph{{Astropy: A community Python package for astronomy}}, \href{https://doi.org/10.1051/0004-6361/201322068}{\emph{\aap} {\bfseries 558} (2013) A33} [\href{https://arxiv.org/abs/1307.6212}{{\ttfamily 1307.6212}}].

\bibitem{astropy:2022}
{Astropy Collaboration}, A.M.~{Price-Whelan}, P.L.~{Lim} and N.e.a.~{Earl}, \emph{{The Astropy Project: Sustaining and Growing a Community-oriented Open-source Project and the Latest Major Release (v5.0) of the Core Package}}, \href{https://doi.org/10.3847/1538-4357/ac7c74}{\emph{\apj} {\bfseries 935} (2022) 167} [\href{https://arxiv.org/abs/2206.14220}{{\ttfamily 2206.14220}}].

\bibitem{astropy:2018}
{Astropy Collaboration}, A.M.~{Price-Whelan}, B.M.~{Sip{\H{o}}cz}, H.M.~{G{\"u}nther}, P.L.~{Lim} and S.M.e.a.~{Crawford}, \emph{{The Astropy Project: Building an Open-science Project and Status of the v2.0 Core Package}}, \href{https://doi.org/10.3847/1538-3881/aabc4f}{\emph{\aj} {\bfseries 156} (2018) 123} [\href{https://arxiv.org/abs/1801.02634}{{\ttfamily 1801.02634}}].

\bibitem{harris2020array}
C.R.~Harris, K.J.~Millman, S.J.~van~der Walt, R.~Gommers, P.~Virtanen and D.C.~et~al., \emph{Array programming with {NumPy}}, \href{https://doi.org/10.1038/s41586-020-2649-2}{\emph{Nature} {\bfseries 585} (2020) 357}.

\bibitem{2011ascl.soft02026L}
A.~{Lewis} and A.~{Challinor}, ``{CAMB: Code for Anisotropies in the Microwave Background}.'' Astrophysics Source Code Library, record ascl:1102.026, Feb., 2011.

\bibitem{2020SciPy-NMeth}
P.~Virtanen, R.~Gommers, T.E.~Oliphant, M.~Haberland, T.~Reddy and D.e.a.~Cournapeau, \emph{{{SciPy} 1.0: Fundamental Algorithms for Scientific Computing in Python}}, \href{https://doi.org/10.1038/s41592-019-0686-2}{\emph{Nature Methods} {\bfseries 17} (2020) 261}.

\bibitem{10.7717/peerj-cs.103}
A.~Meurer, C.P.~Smith and M.e.a.~Paprocki, \emph{Sympy: symbolic computing in python}, \href{https://doi.org/10.7717/peerj-cs.103}{\emph{PeerJ Computer Science} {\bfseries 3} (2017) e103}.

\bibitem{Hunter:2007}
J.D.~Hunter, \emph{Matplotlib: A 2d graphics environment}, \href{https://doi.org/10.1109/MCSE.2007.55}{\emph{Computing in Science \& Engineering} {\bfseries 9} (2007) 90}.

\end{thebibliography}\endgroup
\end{document}